\documentclass[preprint,pre,aps,showpacs,showkeys,amsmath]{revtex4}
\usepackage{graphicx}

\begin{document}

\title{Effect of Network Architecture on Burst and Spike Synchronization in A Scale-Free Network of Bursting Neurons}

\author{Sang-Yoon Kim}
\email{sykim@icn.re.kr}
\author{Woochang Lim}
\email{wclim@icn.re.kr}
\affiliation{Institute for Computational Neuroscience and Department of Science Education, Daegu National University of Education, Daegu 705-115, Korea}

\begin{abstract}
We investigate the effect of network architecture on burst and spike synchronization in a directed scale-free network (SFN) of bursting neurons, evolved via two independent $\alpha-$ and $\beta-$processes. The $\alpha-$process corresponds to a directed version of the Barab\'{a}si-Albert SFN model with growth and preferential attachment, while for the $\beta-$process only preferential attachments between pre-existing nodes are made without addition of new nodes. We first consider the ``pure'' $\alpha-$process of symmetric preferential attachment (with the same in- and out-degrees), and study emergence of burst and spike synchronization by varying the coupling strength $J$ and the noise intensity $D$ for a fixed attachment degree. Characterizations of burst and spike synchronization are also made by employing realistic order parameters and statistical-mechanical measures. Next, we choose appropriate values of $J$ and $D$ where only the burst synchronization occurs, and investigate the effect of the scale-free connectivity on the burst synchronization by varying (1) the symmetric attachment degree  and (2) the asymmetry  parameter (representing deviation from the symmetric case) in the $\alpha-$process, and (3) the occurrence probability of the $\beta-$process. In all these three cases, changes in the type and the degree of population synchronization are studied in connection with the network topology such as the degree distribution, the average path length $L_p$, and the betweenness centralization $B_c$. It is thus found that not only $L_p$ and $B_c$ (affecting global communication between nodes) but also the in-degree distribution (affecting individual dynamics) are important network factors for effective population synchronization in SFNs.
\end{abstract}


\maketitle

\section{Introduction}
\label{sec:INT}
We are concerned about population synchronization of bursting neurons. Bursting occurs when neuronal activity alternates, on a slow timescale, between a silent phase and an active (bursting)
phase of fast repetitive spikings \cite{Izhi1,Burst1,Burst2,Rinzel1,Rinzel2,Burst3}. This type of bursting activity occurs due to the interplay of the fast ionic currents leading to spiking activity
and the slower currents modulating the spiking activity. Hence, the dynamics of bursting neurons have two timescales: slow bursting timescale and fast spiking timescale. Thanks to a repeated sequence of
spikes in the bursting, there are many hypotheses on the importance of bursting activities in the neural information transmission \cite{Burst2,Izhi2,Burst4,Burst5,Burst6}; for example, (a) bursts are
necessary to overcome the synaptic transmission failure, (b) bursts are more reliable than single spikes in evoking responses in postsynaptic neurons, and (c) bursts can be used for selective communication
between neurons. There are several representative examples of bursting neurons such as intrinsically bursting neurons and chattering neurons in the cortex \cite{CT1,CT2}, thalamic relay neurons and thalamic
reticular neurons in the thalamus \cite{TRN1,TRN2,TR}, hippocampal pyramidal neurons \cite{HP}, Purkinje cells in the cerebellum \cite{PC}, pancreatic $\beta$-cells \cite{PBC1,PBC2,PBC3}, and respiratory
neurons in pre-Botzinger complex \cite{BC1,BC2}.

These bursting neurons exhibit two different patterns of synchronization due to the slow and the fast timescales of bursting activity. Burst synchronization (synchrony on the slow bursting timescale)
refers to a temporal coherence between the active phase (bursting) onset or offset times of bursting neurons, while spike synchronization (synchrony on the fast spike timescale) characterizes a temporal
coherence between intraburst spikes fired by bursting neurons in their respective active phases \citep{Burstsync1,Burstsync2}. For example, large-scale burst synchronization (called the sleep spindle
oscillation of 7-14 Hz) has been found to occur via interaction between the excitatory thalamic relay cells and the inhibitory thalamic reticular neurons in the thalamus during the early stage of slow-wave
sleep \cite{Spindle1,Spindle2}. These sleep spindle oscillations are involved in memory consolidation \cite{Spindle3,Spindle4}. In contrast, this kind of burst synchronization is also correlated with abnormal
pathological rhythms associated with neural diseases such as movement disorder (Parkinson's disease and essential tremor) \cite{PD1,PD2,PD3,PD4,PD5} and epileptic seizure \cite{PD5,Epilepsy}. Particularly,
for the case of the Parkinson's disease hypokinetic motor symptoms (i.e., slowness and rigidity of voluntary movement) are closely related to the burst synchronization occurring in the beta band of 10-30 Hz
range in the basal ganglia, while the hyperkinetic motor symptom (i.e., resting tremor) is associated with the burst synchronization of $\sim 5$ Hz.

In this paper, we study burst and spike synchronization of bursting neurons, associated with neural information processes in health and disease, in complex networks. Synaptic connectivity in brain
networks has been found to have complex topology which is neither regular nor completely random \cite{Sporns,Buz2,CN1,CN2,CN3,CN4,CN5,CN6,CN7}. Particularly, brain networks have been found to exhibit
power-law degree distributions (i.e., scale-free property) in the rat hippocampal networks \cite{SF1,SF2,SF3,SF4} and the human cortical functional network \cite{SF5}. Furthermore, robustness against
simulated lesions of mammalian cortical anatomical networks \cite{SF6,SF7,SF8,SF9,SF10,SF11} has also been found to be most similar to that of a scale-free network (SFN) \cite{SF12}. This kind of SFNs
are inhomogeneous ones with a few ``hubs'' (superconnected nodes), in contrast to statistically homogeneous networks such as random graphs and small-world networks \cite{BA1,BA2}. Many recent works on
various subjects of neurodynamics (e.g., coupling-induced burst synchronization, delay-induced burst synchronization, and suppression of burst synchronization) have been done in SFNs with a few percent of
hub neurons with an exceptionally large number of connections \cite{SF13,SF14,SF15,SF16,SF17,SF18}.

The main purpose of our study is to investigate the effect of scale-free connectivity on emergence of burst and spike synchronization in a directed SFN of bursting neurons, evolved via two independent
local $\alpha-$ and $\beta-$ processes which occur with probabilities $\alpha$ and $\beta$ ($\alpha + \beta=1$), respectively. The $\alpha$-process corresponds to a directed version of the
standard Barab\'{a}si-Albert SFN model (i.e., growth and preferential directed attachment) \cite{BA1,BA2}. On the other hand, for the $\beta$-process only preferential attachments between pre-existing
nodes are made without addition of new nodes (i.e., no growth) \cite{BA2,Bollobas,beta2,beta3}. Consequently, degrees of pre-existing nodes are intensified via the $\beta$-process. These $\alpha$- and
$\beta$-processes occur naturally in the evolution of communication networks (e.g., world-wide web) and social networks (e.g., collaboration graph of actors or authors) \cite{BA2,Bollobas,beta2,beta3,beta4}.
We expect that in addition to the growing $\alpha$-process, incorporation of the $\beta$-process (intensifying the internal connections between pre-existing nodes)
may be regarded as a natural extension in typical SFNs, independently of their specific nature. For details on the extended models of network evolution, refer to Refs.~\cite{BA2,Bollobas,beta2,beta3} where local processes, incorporating addition of new nodes and addition or removal of connections between pre-existing nodes, are discussed. Following this line, as our brain network of bursting neurons we employ the SFN model evolved via the $\alpha$- and the $\beta$-processes, as in our recent work on sparse synchronization of spiking neurons \cite{Kim2}. For this case, we expect that the $\alpha$- and the $\beta$-processes might be related to brain plasticity which refers to the brain's ability to change its structure and function by modifying structure and strength of synaptic connections during the development \cite{plastic}. Our SFN is composed of suprathreshold Hindmarsh-Rose (HR) neurons. The HR neurons are representative bursting neurons \cite{HR1,HR2,HR3}, and they interact through inhibitory GABAergic synapses (involving the $\rm {GABA_A}$ receptors).

We first consider the case of ``pure'' $\alpha$-process (i.e., $\alpha=1$) with symmetric preferential attachment with the same in- and out-degrees ($l_\alpha^{(in)} = l_\alpha^{(out)} \equiv l_\alpha$), and study
emergence of burst synchronization and ``complete'' synchronization (composed of both burst and spike synchronization) by varying the coupling strength $J$ and the noise intensity $D$ for a fixed attachment
degree $\widetilde{l_\alpha} (=20)$. Thus, we obtain a state diagram in the $J-D$ plane where complete synchronization occurs within a part of the region of burst synchronization. For an intensive study we fix the
value of $J$, and investigate the evolution of population states by increasing $D$. For small $D$, complete synchronization emerges. However, as $D$ passes a lower threshold $D_l^*$ the intraburst spike synchronization
breaks up, and then only the burst synchronization appears. Eventually when passing a higher threshold $D_h^*$, a transition to unsynchronization occurs due to a destructive effect of noise to spoil the synchronization.
This type of burst and (intraburst) spike synchronization may be well visualized in the raster plot of neural spikes which is a collection of spike trains of individual neurons. For the case of burst synchronization,
synchronous bursting bands appear in the raster plot, and (intraburst) spiking stripes also exist within the bursting bands for the case of complete synchronization. Such raster plots of spikes are fundamental data in
experimental neuroscience. Then, the instantaneous population firing rate (IPFR) $R(t)$ which may be directly obtained from the raster plot of spikes is often used as a collective quantity describing the whole
population behavior \cite{IPFR1,IPFR2}. Through frequency filtering, we separate the slow bursting and the fast (intraburst) spiking timescales of the bursting activity for independent characterization of burst and spike synchronization. Then, $R(t)$ can be decomposed into the instantaneous population burst rate (IPBR) $R_b(t)$ and the instantaneous population spike rate (IPSR) $R_s(t)$ which describe burst and spike synchronization separately.
For more direct visualization of bursting behavior, we also consider another raster plot of bursting onset or offset times. For the case of burst synchronization, synchronous bursting stripes appear in the raster plot. We note that, from this type of raster plot of bursting onset or offset times, one can directly obtain the IPBR $R_b^{(on)}(t)$ or $R_b^{(off)}(t)$ without frequency filtering. For characterization of burst and (intraburst) spike synchronization, we employ realistic order parameters and statistical-mechanical measures, based on the IPBRs [$R_b(t)$, $R_b^{(on)}(t)$, and $R_b^{(off)}(t)$] and the IPSR [$R_s(t)$], which were introduced in our recent work \cite{Kim1}. Then, the higher and the lower thresholds, $D^*_h$ and $D_l^*$, for the bursting and the spiking transitions may be determined in terms of the bursting and the spiking order parameters (corresponding to the time-averaged fluctuations of the IPBRs and the IPSR), respectively. Furthermore, in the region of $D < D_h^*$, the degree of burst synchronization seen in the raster plot of bursting onset times may be well measured in terms of a statistical-mechanical bursting measure $M_b^{(on)}$, introduced by considering both the occupation degree of bursting onset times (representing the density of bursting stripes) and their pacing degree (denoting the smearing of the bursting stripes) in the raster plot. Similarly, in the region of $D < D_l^*$, the degree of (intraburst) spike synchronization may also be measured in terms of a statistical-mechanical spiking measure $M_s$ which is given by the product of the occupation and the pacing degrees of the spiking times in the raster plot. Next, we choose appropriate values of $J$ and $D$ where only the burst synchronization occurs in the above pure $\alpha$-process with symmetric attachment of $l_\alpha=\widetilde{l_\alpha} (=20)$, and study the effect of the scale-free connectivity on the burst synchronization by varying (1) the degree $l_\alpha$ of symmetric attachment and (2) the ``asymmetry'' parameter $\Delta l_\alpha$ of asymmetric preferential attachment of new nodes with different in- and out-degrees ($l_\alpha^{(in)}= \widetilde{l_\alpha} + \Delta l_\alpha$ and $l_\alpha^{(out)}= \widetilde{l_\alpha} - \Delta l_\alpha$ such that $l_\alpha^{(in)} + l_\alpha^{(out)} = 2\, \widetilde{l_\alpha}=$ constant). In addition to the $\alpha-$process, as the third case of network architecture, we also study the effect of the $\beta$-process (intensifying the internal links between pre-existing nodes without adding new nodes) by (3) increasing the probability $\beta$. In theses 3 cases of varying $l_\alpha$, $\Delta l_\alpha$, and $\beta$, we investigate changes in the degree and the type of population synchronization in connection with network topology such as the average path length $L_p$ (representing typical separation between two nodes in the network) and the betweenness centralization $B_c$ (denoting the relative load of communication traffic concentrated to the head hub), both of which affect the global communication between nodes, and the in-degree distribution affecting the individual neuronal dynamics (characterized by the mean bursting/spiking rates). It is thus found that not only $L_p$ and $B_c$, but also the in-degree distribution are important network factors to determine the pacing degree of population synchronization of bursting neurons in SFNs (i.e., dispersion of mean bursting/spiking rates of individual neurons as well as effectiveness of global communication between nodes are responsible for the population synchronization). Specifically, as $l_\alpha$ is increased from $\widetilde{l_\alpha}$ and both $\Delta l_\alpha$ and $\beta$ are also increased from 0, the pacing degree of the burst synchronization increases thanks to the combined effects of the in-degree distribution, $L_p$, and $B_c$. Eventually, when passing their higher thresholds $l_{\alpha,h}^*$, $\Delta l_{\alpha,h}^*$, and $\beta^*$, complete synchronization emerges, respectively. In contrast, with decreasing $l_\alpha$ from $\widetilde{l_\alpha}$ and $\Delta l_\alpha$ from 0, the pacing degree of the burst synchronization decreases, and transitions to unsynchronization occur when paasing their lower thresholds $l_{\alpha,l}^*$ and $\Delta l_{\alpha,l}^*$, respectively.

This paper is organized as follows. In Sec.~\ref{sec:SFN}, we describe a directed SFN of inhibitory suprathreshold bursting HR neurons, evolved via two independent $\alpha-$ and $\beta-$ processes, and  then the governing equations for the population dynamics are given. Detailed explanations on methods for characterization of individual states, population states and network topology in SFNs are also given in Sec.~\ref{sec:Method}. With the characterization methods, in Sec.~\ref{sec:SFC}, we first study emergence of burst and spike synchronization for the case of pure $\alpha-$process with a fixed symmetric preferential attachment degree $\widetilde{l_\alpha} (=20)$, and then investigate the effect of scale-free connectivity on burst and spike synchronization by varying the degree $l_\alpha$ of symmetric attachment, the asymmetry parameter $\Delta l_\alpha$ of the asymmetric attachment, and the probability $\beta$ of the $\beta-$process, in relation to network topology such as the in-degree distribution, $L_p$, and $B_c$. Finally, a summary is given in Section \ref{sec:SUM}.

\section{Scale-Free Network of Inhibitory Bursting Hindmarsh-Rose Neurons}
\label{sec:SFN}
In this section, we first describe our SFN evolved via two independent $\alpha-$ and $\beta-$processes in the subsection \ref{subsec:SFN}. Then, the governing equations for the population dynamics in the SFN are given in the subsection \ref{subsec:GE}.

\subsection{Scale-Free Networks Evolved via Two Independent $\alpha-$ and $\beta-$processes}
\label{subsec:SFN}
We consider an SFN of $N$ inhibitory suprathreshold bursting neurons equidistantly placed on a one-dimensional ring of radius $N/ 2 \pi$. We employ a directed variant of the Barab\'{a}si-Albert SFN model, composed of two independent $\alpha-$ and $\beta-$processes which are performed with probabilities $\alpha$ and $\beta$ ($\alpha + \beta =1$), respectively \cite{BA1,BA2,Bollobas,beta2,beta3,Kim2}: refer to Fig. 1 of \cite{Kim2}
for the diagram of these two processes. The $\alpha$-process corresponds to a directed version of the standard Barab\'{a}si-Albert SFN model (i.e. growth and preferential directed attachment). For the $\alpha$-process, at each discrete time $t$ a new node is added, and it has $l_{\alpha}^{(in)}$ incoming (afferent) edges and $l_{\alpha}^{(out)}$ outgoing (efferent) edges via preferential attachments with $l_{\alpha}^{(in)}$ (pre-existing) source nodes and $l_{\alpha}^{(out)}$ (pre-existing) target nodes. The (pre-existing) source and target nodes $i$ (which are connected to the new node) are preferentially chosen depending on their out-degrees $d_i^{(out)}$ and in-degrees $d_i^{(in)}$ according to the attachment probabilities $\Pi_{source}(d_i^{(out)})$ and $\Pi_{target}(d_i^{(in)})$, respectively:
\begin{equation}
\Pi_{source}(d_i^{(out)})=\frac{d_i^{(out)}}{\sum_{j=1}^{N_{t -1}}d_j^{(out)}}\;\; \textrm{and} \;\; \Pi_{target}(d_i^{(in)})=\frac{d_i^{(in)}}{\sum_{j=1}^{N_{t -1}}d_j^{(in)}},
\label{eq:AP}
\end{equation}
where $N_{t-1}$ is the number of nodes at the time step $t-1$.
Hereafter, the cases of $l_{\alpha}^{(in)} = l_{\alpha}^{(out)} (\equiv l_{\alpha})$  and $l_{\alpha}^{(in)} \neq l_{\alpha}^{(out)}$ will be referred to as symmetric and asymmetric preferential attachments, respectively.
On the other hand, for the $\beta$-process, there is no addition of new nodes (i.e., no growth), and only symmetric preferential attachments with the same in- and out-degrees [$l_{\beta}^{(in)} = l_{\beta}^{(out)} (\equiv l_{\beta}$)] are made between $l_{\beta}$ pairs of (pre-existing) source and target nodes which are also preferentially chosen according to the attachment probabilities $\Pi_{source}(d_i^{(out)})$ and $\Pi_{target}(d_i^{(in)})$ of Eq.~(\ref{eq:AP}), respectively, such that self-connections (i.e., loops) and duplicate connections (i.e., multiple edges) are excluded. Through the $\beta$-process, degrees of pre-existing nodes are more intensified. For generation of an SFN with $N$ nodes, we start with the initial network at $t=0$, consisting of $N_0=50$ nodes where the node 1 is connected bidirectionally to all the other nodes, but the remaining nodes (except the node 1) are sparsely and randomly connected with a low probability $p=0.1$. Then, the $\alpha-$ and $\beta-$processes are repeated until the total number of nodes becomes $N$. For our initial network, the node 1 will be grown as the head hub with the highest degree.

\subsection{Governing Equations for The Population Dynamics}
\label{subsec:GE}
As an element in our SFN, we choose the representative bursting HR neuron model which was originally introduced to describe the time evolution of the membrane potential for the pond snails \cite{HR1,HR2,HR3}.
We consider the SFN composed of $N$ HR bursting neurons; $N=10^3$, except for the case of order parameters and spatial cross-correlation functions. The following equations (\ref{eq:CHRA})-(\ref{eq:CHRE}) govern the population dynamics in the SFN:
\begin{eqnarray}
\frac{dx_i}{dt} &=& y_i - a x^{3}_{i} + b x^{2}_{i} - z_i +I_{DC} +D \xi_{i} -I_{syn,i}, \label{eq:CHRA} \\
\frac{dy_i}{dt} &=& c - d x^{2}_{i} - y_i, \label{eq:CHRB} \\
\frac{dz_i}{dt} &=& r \left[ s (x_i - x_o) - z_i \right], \label{eq:CHRC}
\end{eqnarray}
where
\begin{eqnarray}
I_{syn,i} &=& \frac{J}{d_i^{(in)}} \sum_{j=1(\ne i)}^N w_{ij}g_j(t) (x_i - X_{syn}), \label{eq:CHRD} \\
g_j(t) &=& \sum_{f=1}^{F_j} E(t-t_f^{(j)}-\tau_l);~E(t) = \frac{1}{\tau_d - \tau_r} (e^{-t/\tau_d} - e^{-t/\tau_r}) \Theta(t). \label{eq:CHRE}
\end{eqnarray}
Here, the state of the $i$th neuron at a time $t$ (measured in units of milliseconds) is characterized by three state variables: the fast membrane potential $x_i$, the fast recovery current $y_i,$ and the slow adaptation current $z_i$. The parameter values used in our computations are listed in Table \ref{tab:Parm}. More details on the external stimulus on the single HR neuron, the synaptic currents, and the numerical integration of the governing equations are given in the following subsubsections.

\subsubsection{External Stimulus to The Single HR Neuron}
\label{subsubsec:Sti}
\begin{figure}
\includegraphics[width=0.7\columnwidth]{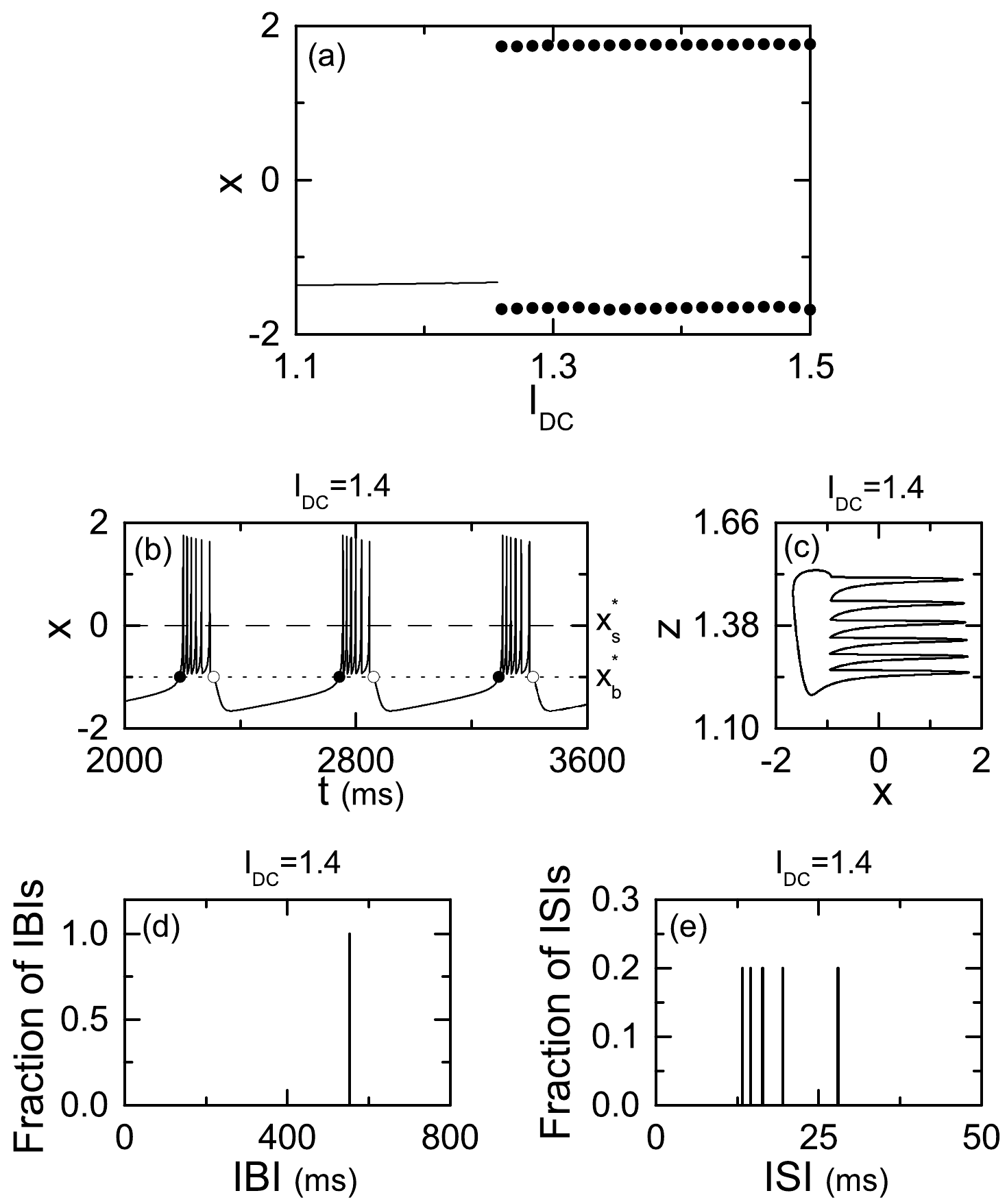}
\caption{
Single bursting HR neuron for $D=0$. (a) Bifurcation diagram in the single HR neuron. Solid line represents a stable resting state, while for the bursting state, maximum and minimum values of the membrane potential $x$ are denoted by solid circles. (b) Time series of $x(t)$ and (c) phase portrait in the $x-z$ plane for $I_{DC}=1.4$. The dotted horizontal line ($x^*_b=-1$) in (b) represents the bursting threshold (the solid and open circles denote the active phase onset and offset times, respectively), while the dashed horizontal line ($x^*_s=0$) represents the spiking threshold within the active phase. Histograms for (d) interburst intervals (IBIs) and (e) intraburst interspike intervals (ISIs) for $I_{DC}=1.4$.
}
\label{fig:Single}
\end{figure}

Each bursting HR neuron (whose parameter values are in the 1st item of Table \ref{tab:Parm}) is stimulated by a common DC current $I_{DC}$ and an independent Gaussian white noise $\xi_i$ [see the 5th and the 6th terms in Eq.~(\ref{eq:CHRA})] satisfying $\langle \xi_i(t) \rangle =0$ and $\langle \xi_i(t)~\xi_j(t') \rangle = \delta_{ij}~\delta(t-t')$, where $\langle\cdots\rangle$ denotes the ensemble average. The noise $\xi$ is a parametric one that randomly perturbs the strength of the applied current $I_{DC}$, and its intensity is controlled by the parameter $D$. As $I_{DC}$ passes a threshold $I_{DC}^* (\simeq 1.26)$ in the absence of noise (i.e., $D=0$), each single HR neuron exhibits a transition from a resting state to a bursting state [see Fig.~\ref{fig:Single}(a)]. For the suprathreshold case of $I_{DC}=1.4$, deterministic bursting occurs when neuronal activity alternates, on a slow time scale $(\simeq 552$ ms), between a silent phase and an active (bursting) phase of fast repetitive spikings. An active phase of the bursting activity begins (ends) at a bursting onset (offset) time when the membrane potential $x$ of the bursting HR neuron passes the bursting threshold of $x^*_b=-1$ from below (above). In Fig.~\ref{fig:Single}(b), the dotted horizontal line ($x^*_b=-1$) denotes the bursting threshold (the solid and open circles denote the active phase onset and offset times, respectively), while the dashed horizontal line ($x^*_s=0$) represents the spiking threshold within the active phase. As shown in Fig.~\ref{fig:Single}(c), projection of the phase flow onto the $x-z$ plane seems to be a hedgehog-like attractor. Bursting activity [alternating between a silent phase and an active (bursting) phase of fast repetitive spikings] occurs on the hedgehog-like attractor [the body (spines) of the hedgehog-like attractor corresponds to the silent (active) phase]. Figures \ref{fig:Single}(d) and \ref{fig:Single}(e) show the interburst interval (IBI) and the (intraburst) interspike interval (ISI) histograms, respectively. The average IBI is 552 ms, corresponding to the slow bursting frequency $f_b \simeq 1.8$ Hz, while the average ISI interval is 18.3 ms, corresponding to the fast spiking frequency $f_s \simeq 54.5$ Hz. In this way, the HR neuron exhibits bursting activity with the two distinct slow and fast timescales. Throughout this paper, we consider the suprathreshold case of $I_{DC} = 1.4$ (see the 2nd item of Table \ref{tab:Parm}) where each HR neuron exhibits spontaneous bursting activity without noise.

\subsubsection{Synaptic Currents}
\label{subsubsec:Syn}
The last term in Eq.~(\ref{eq:CHRA}) denotes the synaptic coupling of the network. $I_{syn,i}$ of Eq.~(\ref{eq:CHRD}) denotes a synaptic current injected into the $i$th neuron. The synaptic connectivity is given by the connection weight matrix $W$ (=$\{ w_{ij} \}$) where $w_{ij}=1$ if the neuron $j$ is presynaptic to the neuron $i$; otherwise, $w_{ij}=0$. Here, the synaptic connection is modeled by using the directed SFN (evolved via the $\alpha-$ and $\beta-$processes). Then, the in-degree of the $i$th neuron, $d_i^{(in)}$ (i.e., the number of synaptic inputs to the neuron $i$) is given by $d_i^{(in)} = \sum_{j(\ne i)}^N w_{ij}$. The fraction of open synaptic ion channels at time $t$ is represented by $g(t)$. The time course of $g_j(t)$ of the $j$th neuron is given by a sum of delayed double-exponential functions $E(t-t_f^{(j)}-\tau_l)$ [see Eq.~(\ref{eq:CHRE})], where $\tau_l$ is the synaptic delay, and $t_f^{(j)}$ and $F_j$ are the $f$th spike and the total number of spikes of the $j$th neuron at time $t$, respectively. Here, $E(t)$ [which corresponds to contribution of a presynaptic spike occurring at time $0$ to $g(t)$ in the absence of synaptic delay] is controlled by the two synaptic time constants: synaptic rise time $\tau_r$ and decay time $\tau_d$, and $\Theta(t)$ is the Heaviside step function: $\Theta(t)=1$ for $t \geq 0$ and 0 for $t <0$. The coupling strength is controlled by the parameter $J$, and $X_{syn}$ is the synaptic reversal potential. For the inhibitory GABAergic synapse (involving the $\rm{GABA_A}$ receptors), the values of $\tau_l$, $\tau_r$, $\tau_d$, and $X_{syn}$ are listed in the 3rd item of Table \ref{tab:Parm}.

\subsubsection{Numerical Integration}
\label{subsubsec:NI}
Numerical integration of stochastic differential equations (\ref{eq:CHRA})-(\ref{eq:CHRC}) is done using the Heun method \cite{SDE} (with the time step $\Delta t=0.01$ ms).
For each realization of the stochastic process, we choose a random initial point $[x_i(0),y_i(0),z_i(0)]$ for the $i$th $(i=1,\dots, N)$ neuron with uniform probability in the range of $x_i(0) \in (-1.5,1.5)$, $y_i(0) \in (-10,0)$, and $z_i(0) \in (1.2,1.5)$.

\section{Methods for Characterization of Individual and Population States and Network Topology}
\label{sec:Method}
In the following subsections, we explain methods used to characterize individual and population states of bursting neurons. Particularly, emergence of burst and spike synchronization and their degrees are characterized by employing realistic measures, based on the IPBR and the IPSR, which were introduced in our recent work \cite{Kim1}. Furthermore, network topology such as the average path length and the betweenness centralization, associated with global communication between nodes, is also explained for discussion in connection with population dynamics in the next section \ref{sec:SFC}.

\subsection{Characterization of Individual Firing Behaviors}
\label{subsec:CI}
Bursting neurons exhibit firing activity with two different timescales: slow bursting timescale and fast spiking timescale. The slow bursting behavior is characterized in terms of the interburst interval (IBI) histogram and the mean bursting rate (MBR) distribution. On the other hand, the fast spiking behavior is characterized in terms of the (intraburst) interspike interval (ISI) histogram and the mean spiking rate (MSR) distribution.
The IBI histogram and the intraburst ISI histogram are made of $10^3$ IBIs and $5 \times 10^3$ ISIs (obtained from all the neurons), respectively. The bin size for both cases is 0.5 ms.
Averaging time for the MBR of each individual neuron is  $5 \times 10^4$ ms and the bin size for the histogram is 0.1 Hz. For calculation of the MSR of each neuron, we follow 500 bursting cycles, and get the MSR in each bursting cycle. Then, through average over all bursting cycles, we obtain the bursting-averaged MSR.

\subsection{Population Variables}
\label{subsec:PV}
In computational and theoretical neuroscience, an ensemble-averaged global potential,
\begin{equation}
 X_G (t) = \frac {1} {N} \sum_{i=1}^{N} x_i(t),
\label{eq:GPOT}
\end{equation}
is often used for describing emergence of population synchronization. However, since to directly obtain $X_G$ in real experiments is practically difficult, instead of $X_G$, we employ the IPFR which is an experimentally-obtainable population quantity used in both the experimental and the computational neuroscience \cite{IPFR1,IPFR2}. The IPFR is obtained from the raster plot of neural spikes which is a collection of spike trains of individual neurons. These raster plots of spikes, where population synchronization may be well visualized, are fundamental data in experimental neuroscience (e.g. epilepsy in human \citep{EP1,EP2,EP3} and rat \citep{EP4}).
To obtain a smooth IPFR from the raster plot of spikes, we employ the kernel density estimation (kernel smoother) \citep{Kernel}. Each spike in the raster plot is convoluted (or blurred) with a kernel function $K_h(t)$ to obtain a smooth estimate of IPFR, $R(t)$:
\begin{equation}
R(t) = \frac{1}{N} \sum_{i=1}^{N} \sum_{s=1}^{n_i} K_h (t-t_{s}^{(i)}),
\label{eq:IPSRK}
\end{equation}
where $t_{s}^{(i)}$ is the $s$th spiking time of the $i$th neuron, $n_i$ is the total number of spikes for the $i$th neuron, and we use a Gaussian
kernel function of band width $h$:
\begin{equation}
K_h (t) = \frac{1}{\sqrt{2\pi}h} e^{-t^2 / 2h^2}, ~~~~ -\infty < t < \infty.
\label{eq:Gaussian}
\end{equation}
Throughout the paper, the band width of the Gaussian kernel estimate for $R(t)$ is $h=1$ ms.
This type of IPFR kernel estimate $R(t)$ of bursting neurons is a population quantity describing the ``whole'' combined collective behaviors with both slow and fast timescales. For more clear investigation of population synchronization, we separate the slow bursting timescale and the fast spiking timescale via frequency filtering, and decompose the IPFR kernel estimate $R(t)$ into the IPBR $R_b(t)$ and the IPSR $R_s(t)$.
Specifically, $R_b(t)$ and $R_s(t)$ are obtained via low-pass and band-pass filtering of $R(t)$, respectively. Then, we can study the bursting and the spiking behaviors separately in terms of $R_b(t)$ and $R_s(t)$.

\subsection{Thermodynamic Bursting and Spiking Order Parameters}
\label{subsec:Order}
We employ realistic bursting and spiking order parameters, based on the IPBR and the IPSR to characterize the bursting and the spiking transitions, respectively \cite{Kim1}.
For determination of the threshold for the bursting transition (i.e., transition from burst synchronization to unsynchronization), we employ a realistic bursting order parameter ${\cal{O}}_b$, based on the low-pass filtered IPBR $R_b(t)$, which may be applicable in both the computational and the experimental neuroscience. The mean square deviation of $R_b(t)$,
\begin{equation}
{\cal{O}}_b \equiv \overline{(R_b(t) - \overline{R_b(t)})^2},
 \label{eq:Border1}
\end{equation}
plays the role of a bursting order parameter ${\cal{O}}_b$, where the overbar represents the time average. The bursting order parameter ${\cal{O}}_b$ may be regarded as a thermodynamic measure because it concerns just the macroscopic IPBR $R_b(t)$ without any consideration between $R_b(t)$ and microscopic individual burstings. In the thermodynamic limit of $N \rightarrow \infty$, ${\cal{O}}_b$ approaches non-zero (zero) limit values for the case of burst synchronization (unsynchronization). Here, we follow a trajectory for $3 \times 10^4$ ms after a transient for $2 \times 10^3$ ms in each realization, and obtain $\langle {\cal{O}}_b \rangle_r$ via average over 20 realizations. Hereafter, $ \langle \cdots \rangle_r$ represents an average over realizations.

Similar to the case of ${\cal{O}}_b$, the mean square deviations of another IPBRs $R_b^{(on)}(t)$ and $R_b^{(off)}(t)$ (obtained directly from the raster plots of bursting onset and offset times without frequency filtering),
\begin{equation}
{\cal{O}}_b^{(on)} \equiv \overline{(R_b^{(on)}(t) - \overline{R_b^{(on)}(t)})^2}~{\rm {and}}~
{\cal{O}}_b^{(off)} \equiv \overline{(R_b^{(off)}(t) - \overline{R_b^{(off)}(t)})^2},
\label{eq:Border2}
\end{equation}
also play another bursting order parameters, used to determine the threshold for the bursting transition. As in the case of ${\cal{O}}_b$, we discard the first time steps of a trajectory as transients for $2 \times 10^3$ ms and then we compute  ${\cal{O}}_b^{(on)}$ and ${\cal{O}}_b^{(off)}$ by following the trajectory for $3 \times 10^4$ ms in each realization. Thus, we obtain $\langle{\cal{O}}_b^{(on)}\rangle_r$ and $\langle{\cal{O}}_b^{(off)}\rangle_r$ via average over 20 realizations.

Next, we also employ a realistic spiking order parameter ${\cal{O}}_s$, based on the band-passed IPSR $R_s(t)$ for determination of the threshold for the spiking transition [i.e., transitions from complete synchronization (including both burst synchronization and intraburst spike synchronization) to pure burst synchronization].
The mean square deviation of $R_s(t)$ in the $i$th global bursting cycle,
\begin{equation}
{\cal{O}}_s^{(i)} \equiv \overline{(R_s(t) - \overline{R_s(t)})^2},
\label{eq:SO}
\end{equation}
plays the role of a spiking order parameter ${\cal{O}}_s^{(i)}$ in the $i$th global bursting cycle of the IPBR $R_b(t)$ (corresponding to the $i$th bursting band in the raster plot of spikes). By averaging ${\cal{O}}_s^{(i)}$ over a sufficiently large number $N_b$ of global bursting cycles, we obtain the spiking order parameter:
\begin{equation}
{\cal{O}}_s =  {\frac {1} {N_b}} \sum_{i=1}^{N_b} {\cal{O}}_s^{(i)}.
\label{eq:SO2}
\end{equation}
Here, we follow $500$ global bursting cycles in each realization, and obtain the spiking order parameter ${\langle {\cal{O}}_s \rangle}_r$ via average over 20 realizations.

\subsection{Spatial Cross-correlation Functions}
\label{subsec:CSF}
To further understand the bursting transition, we introduce a ``microscopic'' spatial cross-correlation between neuronal pairs of bursting neurons through extension of the case of spiking neurons \cite{Kim3}.
For obtaining dynamical pair cross-correlations, each train of bursting onset times for the $i$th neuron is convoluted with the Gaussian kernel function $K_h(t)$ of band width $h$ (=50 ms) to get a smooth estimate of instantaneous individual burst rate (IIBR) $r_i^{(b,on)}(t)$:
\begin{equation}
r_i^{(b,on)}(t) = \sum_{s=1}^{n_i^{(b,on)}} K_h (t-t_i^{(b,on)}(s)),
\label{eq:IIBR}
\end{equation}
where $t_i^{(b,on)}(s)$ is the $s$th bursting onset time of the $i$th neuron, $n_i^{(b,on)}$ is the total number of bursting onset times for the $i$th neuron, and $K_h(t)$ is given in Eq.~(\ref{eq:Gaussian}). Then, the normalized temporal cross-correlation function $C_{i,j}^{(b,on)}(\tau)$ between the IIBRs $r_i^{(b,on)}(t)$ and $r_j^{(b,on)}(t)$ of the $(i,j)$ neuronal pair is given by:
\begin{equation}
C_{i,j}^{(b,on)}(\tau) = \frac{\overline{\Delta r_i^{(b,on)}(t+\tau) \Delta r_j^{(b,on)}(t)}}{\sqrt{\overline{\Delta {r_i^{(b,on)}}^2(t)}}\sqrt{\overline{\Delta {r_j^{(b,on)}}^2(t)}}},
\end{equation}
where $\Delta r_i^{(b,on)}(t) = r_i^{(b,on)}(t) - \overline{r_i^{(b,on)}(t)}$ and the overline denotes the time average.
Here, the number of data used for the calculation of each temporal cross-correlation function $C_{i,j}^{(b,on)}(\tau)$ is $2 \times 10^4$.
Then, the spatial cross-correlation $C_L^{(b,on)}$ ($L=1,...,N/2)$ between neuronal pairs separated by a spatial distance $L$ is given by the average of all the temporal cross-correlations between $r_i^{(b,on)}(t)$ and $r_{i+L}^{(b,on)}(t)$ $(i=1,...,N)$ at the zero-time lag:
\begin{equation}
C_L^{(b,on)} = \frac{1}{L} \sum_{i=1}^{N} C_{i, i+L}^{(b,on)}(0) ~~~~ {\rm for~} L=1, \cdots, N/2.
\label{eq:SCC}
\end{equation}
Here, if $i+L > N$ in Eq.~(\ref{eq:SCC}), then $i+L-N$ is considered instead of $i+L$ because neurons lie on the ring.
If the spatial cross-correlation function $C_L^{(b,on)}$ ($L=1,...,N/2)$ is non-zero in the whole range of $L$, then the spatial correlation length $\eta$ becomes $N/2$ (note that the maximal distance between neurons is $N/2$ because of the ring architecture on which neurons exist) covering the whole network. For this case, synchronization appears in the SFN; otherwise, unsynchronization occurs.

\subsection{State Diagram}
\label{subsec:SD}
Population states change depending on the synaptic coupling strength $J$ and the noise intensity $D$, which may be well shown in the state diagram in the $J-D$ plane. To get the state diagram, we first divide the $J-D$ plane into the $20 \times 10$ grids. Then, at each grid point, we calculate the bursting order parameters ${\cal{O}}_b$ for $N=10^3$ and $10^4$ to determine whether or not burst synchronization occurs at the grid points. If ${\cal{O}}_b$ for $N=10^4$ is smaller than $f \cdot {\cal{O}}_b$ for $N=10^3$ ($f$ is some appropriate factor less than unity; for convenience we set $f=0.3$), ${\cal{O}}_b$ is expected to decrease with increasing $N$. For the case of decrease in ${\cal{O}}_b$ with increasing $N$, unsynchronization occurs at the grid point; otherwise, burst synchronization appears. Next, at the grid points where burst synchronization occurs, we calculate the spiking order parameters ${\cal{O}}_s$ for $N=10^3$ and $10^4$ to determine whether or not intraburst spike synchronization occurs at the grid points. Like the case of the bursting order parameter, if ${\cal{O}}_s$ for $N=10^4$ is smaller than
$f \cdot {\cal{O}}_s$ for $N=10^3$ ($f=0.3$), ${\cal{O}}_s$ is expected to decrease with increasing $N$. For the case of decrease in ${\cal{O}}_s$ with increasing $N$, only the burst synchronization occurs at the grid point; otherwise, complete synchronization (including both the burst and the intraburst spike synchronization) appears. After determining the population states (burst or spike synchronization and unsynchronization) at all grid points, we try to obtain the threshold curves for the burst synchronization-unsynchronization transition and the burst-spike synchronization transition accurately. For this purpose, we calculate ${\cal{O}}_b$ $({\cal{O}}_s)$ in the small parameter region between the burst synchronization and the unsynchronization (the burst and the spike synchronization) grid points by varying $J$ and $D$. Moreover, to get more accurate transition curves, we divide a part of the parameter plane where the transition curves change rapidly into more minute grids and repeat the above computations.

\subsection{Statistical-Mechanical Bursting and Spiking Measures}
\label{subsec:SM}
Burst synchronization may be well visualized in the raster plot of bursting onset times. For the case of burst synchronization, bursting stripes appear successively in the raster plot. We measure the degree of burst synchronization in terms of a statistical-mechanical bursting measure $M_b^{(on)}$, based on the IPBR kernel estimates $R_b^{(on)}(t)$, which was introduced by considering the occupation pattern and the pacing pattern of bursting onset times in the bursting stripes \cite{Kim1}. The bursting measure $M_i^{(b,on)}$ of the $i$th bursting onset stripe is defined by the product of the occupation degree $O_i^{(b,on)}$ of bursting onset times (representing the density of the $i$th bursting stripe) and the pacing degree $P_i^{(b,on)}$ of bursting onset times (denoting the smearing of the $i$th bursting stripe):
\begin{equation}
M_i^{(b,on)} = O_i^{(b,on)} \cdot P_i^{(b,on)}.
\label{eq:BM1}
\end{equation}
The occupation degree $O_i^{(b,on)}$ of bursting onset times in the $i$th bursting stripe is given by the fraction of HR neurons which exhibit burstings:
\begin{equation}
   O_i^{(b,on)} = \frac {N_i^{(b,on)}} {N},
\label{eq:Occu}
\end{equation}
where $N_i^{(b,on)}$ is the number of HR neurons which exhibit burstings in the $i$th bursting onset stripe. For the full occupation $O_i^{(b,on)}=1$, while for the partial occupation $O_i^{(b,on)}<1$.
The pacing degree $P_i^{(b,on)}$ of bursting onset times in the $i$th bursting onset stripe can be determined in a statistical-mechanical way by taking into account their contributions to the macroscopic IPBR kernel estimate $R_b^{(on)}(t)$. Each global bursting onset cycle of $R_b(t)$ begins from its left minimum, passes the central maximum, and ends at the right minimum; the central maxima coincide with centers of bursting stripes in the raster
plot. An instantaneous global bursting onset phase $\Phi_b^{(on)}(t)$ of $R_b^{(on)}(t)$ is introduced via linear interpolation in the two successive subregions forming a global bursting onset cycle; for more details, refer to Fig.~4 in \cite{Kim1}. The global bursting onset phase $\Phi_b^{(on)}(t)$ between the left minimum (corresponding to the beginning point of the $i$th global bursting onset cycle) and the central maximum is given by
\begin{equation}
\Phi_b^{(on)}(t) = 2\pi(i-3/2) + \pi \left(\frac{t-t_i^{(on,min)}}{t_i^{(on,max)}-t_i^{(on,min)}} \right) {\rm~~ for~} ~t_i^{(on,min)} \leq  t < t_i^{(on,max)}
~~(i=1,2,3,\dots),
\end{equation}
and $\Phi_b^{(on)}(t)$ between the central maximum and the right minimum (corresponding to the beginning point of the $(i+1)$th global bursting onset cycle) is given by
\begin{equation}
\Phi_b^{(on)}(t) = 2\pi(i-1) + \pi \left(\frac{t-t_i^{(on,max)}}{t_{i+1}^{(on,min)}-t_i^{(on,max)}} \right)
 {\rm~~ for~} ~t_i^{(on,max)} \leq  t < t_{i+1}^{(on,min)}~~(i=1,2,3,\dots),
\end{equation}
where $t_i^{(on,min)}$ is the beginning time of the $i$th global bursting onset cycle (i.e., the time at which the left minimum of $R_b^{(on)}(t)$ appears in the $i$th global bursting onset cycle) and $t_i^{(on,max)}$ is the time at which the maximum of $R_b^{(on)}(t)$ appears in the $i$th global bursting onset cycle. Then, the contribution of the $k$th microscopic bursting onset time in the $i$th bursting onset stripe occurring at the time $t_k^{(b,on)}$ to $R_b^{(on)}(t)$ is given by $\cos \Phi_k^{(b,on)}$, where $\Phi_k^{(b,on)}$ is the global bursting onset phase at the $k$th bursting onset time [i.e., $\Phi_k^{(b,on)} \equiv \Phi_b^{(on)}(t_k^{(b,on)})$]. A microscopic bursting onset time makes the most constructive (in-phase) contribution to $R_b^{(b,on)}(t)$ when the corresponding global phase $\Phi_k^{(b,on)}$ is $2 \pi n$ ($n=0,1,2, \dots$), while it makes the most destructive (anti-phase) contribution to $R_b^{(b,on)}(t)$ when $\Phi_k^{(b,on)}$ is $2 \pi (n-1/2)$. By averaging the contributions of all microscopic bursting onset times in the $i$th bursting onset stripe to $R_b^{(on)}(t)$, we obtain the pacing degree of bursting onset times in the $i$th bursting onset stripe:
\begin{equation}
 P_i^{(b,on)} = {\frac {1} {B_i^{(on)}}} \sum_{k=1}^{B_i^{(on)}} \cos \Phi_k^{(b,on)},
\label{eq:Pacing}
\end{equation}
where $B_i^{(on)}$ is the total number of microscopic bursting onset times in the $i$th bursting onset stripe.
By averaging $M_i^{(b,on)}$ of Eq.~(\ref{eq:BM1}) over a sufficiently large number $N_b^{(on)}$ of bursting onset stripes, we obtain the statistical-mechanical bursting measure $M_b^{(on)}$:
\begin{equation}
M_b^{(on)} =  {\frac {1} {N_b^{(on)}}} \sum_{i=1}^{N_b^{(on)}} M_i^{(b,on)}.
\label{eq:BM2}
\end{equation}
Here, we follow 500 global bursting onset cycles (corresponding to the bursting onset stripes in the raster plot) in each realization and obtain $\langle O_b^{(on)} \rangle_r$, $\langle P_b^{(on)} \rangle_r$, and
$\langle M_b^{(on)} \rangle_r$ via average over 20 realizations.

Next, we consider the intraburst spike synchronization which may be well visualized in the raster plot of spikes. Spiking stripes (composed of intraburst spikes and indicating intraburst spike synchronization) appear within the  bursting bands of the raster plot. Similar to the case of burst synchronization, we also measure the degree of spike synchronization by employing both a statistical-mechanical spiking measure $M_s$, which is explained in the main section \ref{sec:SFC}.

\subsection{Network Topology}
\label{subsec:NT}
We explain the network topology such as the average path length $L_p$ and the betweenness centralization $B_c$, associated with global communication between nodes. The average path length $L_p$, representing typical separation between two nodes in the network, is obtained through the average of the shortest path lengths of all nodal pairs \cite{BA2}:
\begin{equation}
L_p = \frac{1}{N(N-1)} \sum_{i=1}^{N}\sum_{j=1 (j \ne i)}^{N} l_{ij},
\label{eq:APL}
\end{equation}
where $l_{ij}$ is the shortest path length from the node $i$ to the node $j$.
Next, we consider the betweenness centrality $B_i$ of the node $i$ which represents the fraction of all the shortest paths between any two other nodes that pass through the node $i$ \cite{BTC1,BTC2}:
\begin{equation}
B_i = \sum_{j=1(j \ne i)}^{N} \sum_{k=1(k \ne j ~ \& ~ k \ne i)}^{N} \frac{\sigma_{jk}(i)}{\sigma_{jk}},
\label{eq:BC1}
\end{equation}
where $\sigma_{jk}(i)$ is the number of shortest paths from the node $j$ to the node $k$ passing through the node $i$ and $\sigma_{jk}$ is the total number of shortest paths from the node $j$ to the node $k$.
To examine how evenly the betweenness centrality (representing the load of communication traffic) is distributed among nodes, we consider the group betweenness centralization $B_c$, denoting the degree to which the maximum betweenness centrality $B_{max}$ of the head hub exceeds the betweenness centralities of all the other nodes. We obtain $B_c$ through the sum of differences between the maximum betweenness centrality $B_{max}$ of the head hub and the betweenness centrality $B_i$ of other node $i$, and normalization by dividing the sum of differences with its maximum possible value \cite{BTC1,BTC2}:
\begin{equation}
B_c=\frac{\sum_{i=1}^{N} (B_{max} - B_i)}{{\rm max}{\sum_{i=1}^{N} (B_{max} - B_i)}};~ {\rm max}{\sum_{i=1}^{N} (B_{max} - B_i)}= \frac {(N-1)(N^2 - 3 N +2)} {2},
\label{eq:BC2}
\end{equation}
where the maximum sum of differences in the denominator corresponds to that for the star network.
Association of $L_p$ and $B_c$ with burst and spike synchronization is discussed in the subsection \ref{subsec:SFC}.

\section{Effect of Scale-Free Connectivity on Burst and Spike Synchronization of Bursting HR Neurons}
\label{sec:SFC}
In this section, we study the effect of scale-free connectivity on burst and spike synchronization in a directed SFN of inhibitory suprathreshold bursting HR neurons, evolved via two independent local $\alpha-$ and $\beta-$processes. In the subsection \ref{subsec:DBA}, we study emergence of burst and spike synchronization in the directed Barab\'{a}si-Albert SFN (corresponding to the pure $\alpha-$process) of HR neurons  for a fixed symmetric preferential attachment degree. Characterization of burst and spike synchronization is made in terms of realistic order parameters and statistical-mechanical measures,
explained in Sec.~\ref{sec:Method}. In the next subsection \ref{subsec:SFC}, we investigate the effect of scale-free connectivity on burst and spike synchronization by varying the degree of symmetric attachment $l_\alpha$, the asymmetry parameter $\Delta l_\alpha$ of the asymmetric attachment, and the probability $\beta$ of the $\beta-$process. Changes in the degree and the type of population synchronization are also discussed in association with  network topology such as the in-degree distribution, the average path length $L_p$, and the betweenness centralization $B_c$.

\subsection{Burst and Spike Synchronization in the Directed  Barab\'{a}si-Albert SFN}
\label{subsec:DBA}
\begin{figure}
\includegraphics[width=0.7\columnwidth]{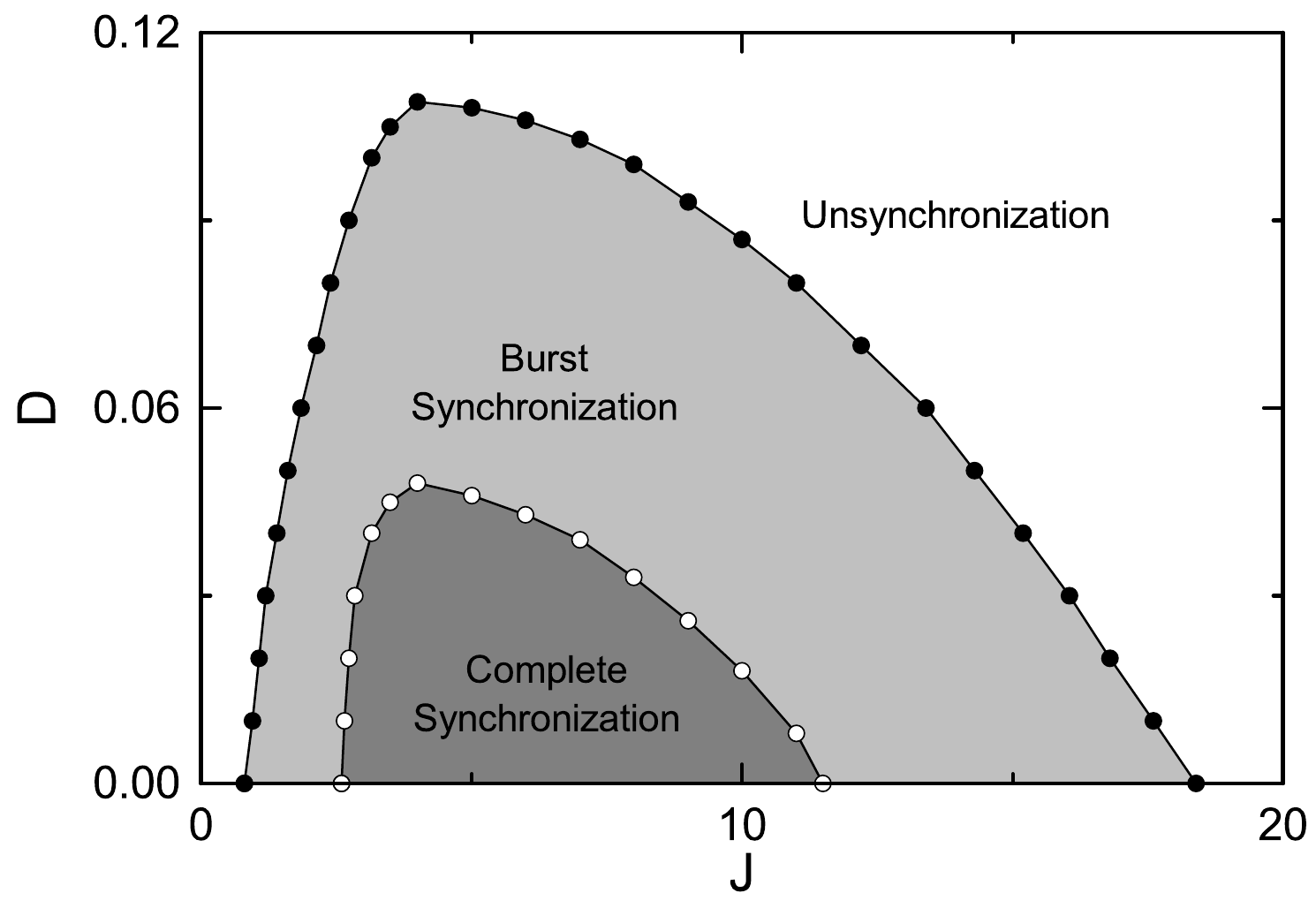}
\caption{
State diagram in the $J-D$ plane in the pure $\alpha-$process ($\alpha=1$) with symmetric preferential attachment of $\widetilde{l_\alpha} (=20)$. Complete synchronization (including both burst and intraburst spike synchronization) occurs in the dark gray region, while in the gray region only the burst synchronization appears.
}
\label{fig:SD}
\end{figure}

We first consider the directed Barab\'{a}si-Albert SFN (i.e., pure $\alpha-$process) of $N$ bursting HR neurons, equidistantly placed on a one-dimensional ring of radius $N/ 2 \pi$. The HR neurons are suprathreshold ones which can fire spontaneously, and they are coupled via inhibitory synapses. We investigate emergence of burst and spike synchronization by varying the coupling strength $J$ and the noise intensity $D$ for a fixed symmetric attachment degree $\widetilde{l_\alpha} (=20)$ (see the fourth item in Table \ref{tab:Parm}). Figure \ref{fig:SD} shows the state diagram in the $J-D$ plane. Complete synchronization (including both the burst and (intraburst) spike synchronization) emerges in the dark gray region, while in the gray region only the burst synchronization (without spike synchronization) appears. In the absence of noise (i.e., $D=0$), unsynchronization occurs for sufficiently small $J$. However, when passing a lower bursting threshold $J_{b,l}^* (\simeq 0.8)$ a transition to burst synchronization occurs due to a constructive role of $J$ for the population synchronization. With increasing $J$ from $J_{b,l}^*$, the degree of burst synchronization increases, and eventually complete synchronization emerges as $J$ passes a lower spiking threshold $J_{s,l}^* (\simeq 2.6)$. However, with further increase in $J$, the degree of burst synchronization decreases because of a destructive role of $J$ to spoil the synchronization. Thus, the (intraburst) spike synchronization first breaks up when passing a higher spiking threshold $J_{s,h}^* (\simeq 11.5)$. Then, only the burst synchronization persists. Eventually, as $J$ passes a higher bursting threshold $J_{b,h}^* (\simeq 18.4)$, a transition to unsynchronization occurs.

\begin{figure}
\includegraphics[width=0.8\columnwidth]{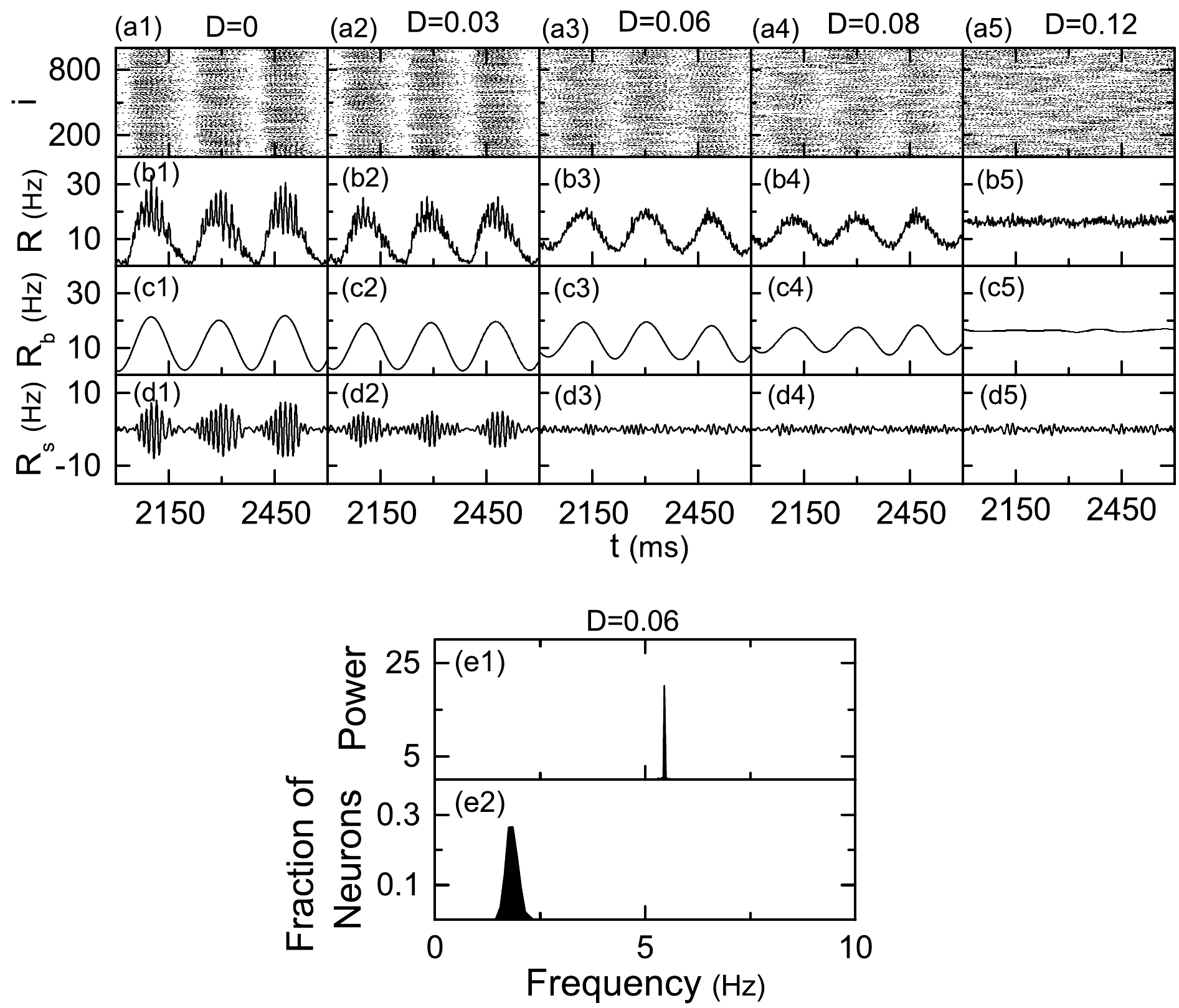}
\caption{
Complete and burst synchronization in the pure $\alpha-$process ($\alpha=1$) with symmetric preferential attachment of $\widetilde{l_\alpha} (=20)$: complete synchronization (including both burst and intraburst spike synchronization) for $D=0$ and 0.03, burst synchronization (without intraburst spike synchronization) for $D=0.06$ and 0.08, and unsynchronization for $D=0.12$. (a1)-(a5) Raster plots of spikes, (b1)-(b5) plots of IPFR kernel estimates $R(t)$, (c1)-(c5) plots of low-pass filtered IPBRs $R_b(t)$, and (d1)-(d5) plots of band-pass filtered IPSRs $R_s(t)$ for various values of $D=0$, 0.03, 0.06, 0.08, and 0.12. (e1) One-sided power spectrum of $\Delta R_b(t) [=R_b(t) - \overline{R_b(t)}]$ (the overbar represents the time average) with mean-squared amplitude normalization and (e2) distribution of mean bursting rates (MBRs) of individual neurons for $D=0.06$. Power spectrum is obtained from $2^{16} (=65536)$ data points.
}
\label{fig:CBS}
\end{figure}

For further intensive study we fix the value of $J$ at $J=4$, and investigate the evolution of population states by increasing $D$. As examples of population states, Figs.~\ref{fig:CBS}(a1)-\ref{fig:CBS}(a5) show the raster plots of neural spikes for various values of noise intensity $D$: complete synchronization for $D=0$ and 0.03, burst synchronization for $D=0.06$ and 0.08, and unsynchronization for $D=0.12$. From these raster plots, we obtain smooth IPFR kernel estimates $R(t)$ of Eq.~(\ref{eq:IPSRK}) in Figs.~\ref{fig:CBS}(b1)-\ref{fig:CBS}(b5). We note that $R(t)$ describes the whole population behavior with both the slow bursting and the fast spiking timescales. For more clear investigation of population synchronization, we separate the slow and the fast timescales via frequency filtering, and decompose the IPFR kernel estimate $R(t)$ into the IPBR $R_b(t)$ and the IPSR $R_s(t)$. Through low-pass filtering of $R(t)$ with cut-off frequency of 10 Hz, we obtain the slowly-oscillating IPBR $R_b(t)$ (containing only the bursting behavior without spiking) in  Figs.~\ref{fig:CBS}(c1)-\ref{fig:CBS}(c5). On the other hand, via band-pass filtering of $R(t)$ with lower and higher cut-off frequencies of 30 Hz (high-pass filter) and 90 Hz (low-pass filter), we obtain the fast-oscillating IPSR $R_s(t)$ (including only the intraburst spiking behavior) in Figs.~\ref{fig:CBS}(d1)-\ref{fig:CBS}(d5). For $D=0$, ``bursting bands,'' each of which is composed of ``spiking stripes,'' appear successively at nearly regular time intervals, as shown in Fig.~\ref{fig:CBS}(a1); a magnification of the 1st bursting band is given in Fig.~\ref{fig:ST}(a1) (where the spiking stripes are well seen). For this case, in addition to burst synchronization, (intraburst) spike synchronization also occurs in each bursting band. As a result of this complete synchronization, the IPFR kernel estimate $R(t)$ exhibits a bursting activity [i.e., fast spikes appear on a slow wave in $R(t)$], as shown in Fig.~\ref{fig:CBS}(b1). Through frequency filtering of $R(t)$, the IPBR $R_b(t)$ and the IPSR $R_s(t)$ show the slow bursting and the fast (intraburst) spiking oscillations in Figs.~\ref{fig:CBS}(c1) and \ref{fig:CBS}(d1), respectively. However, with increasing $D$, bursting bands become smeared in the raster plot, and loss of spike synchronization also begins to occur in each bursting band due to smearing of spiking stripes. As an example, see the case of $D=0.03$: the raster plot of spikes, the IPFR kernel estimate $R(t)$, the IPBR $R_b(t)$, and the IPSR $R_s(t)$ are shown in Figs.~\ref{fig:CBS}(a2), \ref{fig:CBS}(b2), \ref{fig:CBS}(c2) and \ref{fig:CBS}(d2), respectively. As a result of smearing, the amplitudes of $R(t)$, $R_b(t)$, and $R_s(t)$ decrease (i.e., the degrees of both burst and spike synchronization decrease). When passing a lower spiking threshold $D^*_l$ $(\simeq 0.048)$, complete loss of spike synchronization occurs in each bursting band (i.e., intraburst spikes become incoherent within each bursting band). Consequently, only the burst synchronization (without spike synchronization) persists. As an example, see the case of $D=0.06$ in Figs.~\ref{fig:CBS}(a3), \ref{fig:CBS}(b3), \ref{fig:CBS}(c3), and \ref{fig:CBS}(d3). For this case of burst synchronization, $R(t)$ shows a slow-wave oscillation without spikes. Hence, the IPBR $R_b(t)$ exhibits slowly-oscillating behavior, while the IPSR $R_s(t)$ with small fluctuations becomes nearly stationary. As $D$ is further increased, such ``incoherent'' bursting bands (where intraburst spikes are incoherent) become more and more smeared, and thus the degree of burst synchronization decreases [e.g., see Figs.~\ref{fig:CBS}(a4), \ref{fig:CBS}(b4), \ref{fig:CBS}(c4), and \ref{fig:CBS}(d4) for $D=0.08$]. Consequently, the amplitudes of both $R(t)$ and $R_b(t)$ are further decreased, and $R_s(t)$ becomes more nearly stationary. Eventually, as $D$ passes a higher bursting threshold $D_h^* (\simeq 0.109)$, bursting bands begin to overlap, which leads to complete loss of burst synchronization. In this way, completely unsynchronized states with nearly stationary $R(t)$, $R_b(t)$, and $R_s(t)$ appear, as shown in Figs.~\ref{fig:CBS}(a5), \ref{fig:CBS}(b5),  \ref{fig:CBS}(c5) and \ref{fig:CBS}(d5) for $D=0.12$. We also note that only some fraction of HR neurons make burstings in each bursting band (i.e., burst skipping occurs). Figures \ref{fig:CBS}(e1) and \ref{fig:CBS}(e2) show the power spectrum of $\Delta R_b(t) [=R_b(t) - \overline{R_b(t)}]$ and distribution of mean bursting rates (MBRs) $f_b^{(i)}$ of individual neurons for $D=0.06$. The population bursting frequency $f_b^{(p)}$ [i.e., corresponding to the frequency of $R_b(t)$] is 5.4 Hz, while the ensemble-averaged MFR of individual neurons $\langle f_b^{(i)} \rangle$ is 1.8 Hz. Hence, in an average sense, only one third of the whole HR neurons exhibit burstings in each bursting band.

\begin{figure}
\includegraphics[width=0.7\columnwidth]{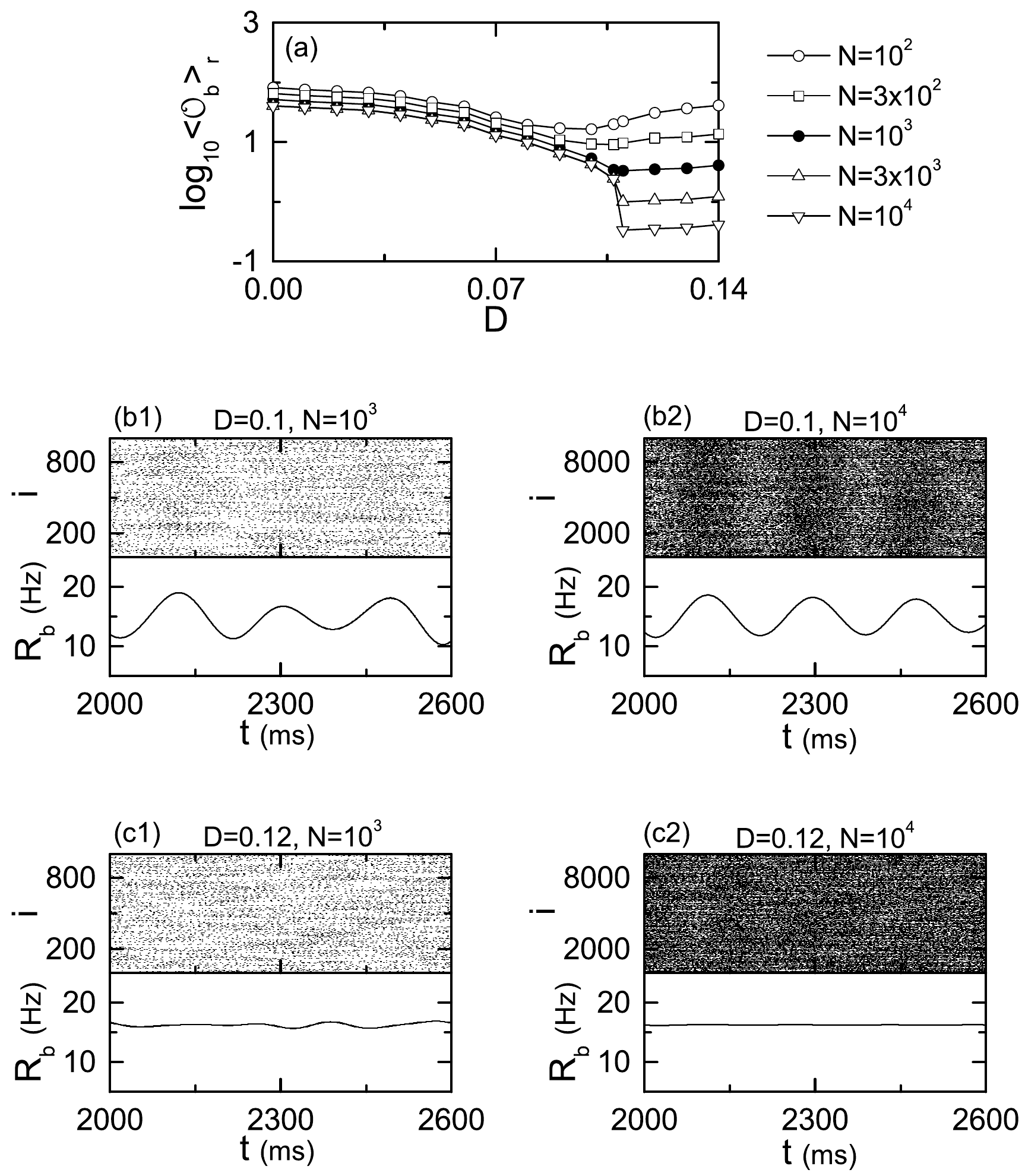}
\caption{
Determination of the higher bursting noise threshold $D^*_h$ for the bursting transition in terms of a realistic thermodynamics bursting order parameter $\langle {\cal{O}}_b \rangle_r$ in the pure $\alpha-$process ($\alpha=1$) with symmetric preferential attachment of $\widetilde{l_\alpha} (=20)$. (a) Plots of the bursting order parameter $\langle {\cal{O}}_b \rangle_r$ versus $D$. Burst synchronization for $D=0.1$: raster plots of spikes and plots of low-pass filtered IPBRs $R_b(t)$ for $N=$ (b1) $10^3$  and (b2) $10^4$. Unsynchronization for $D=0.12$: raster plots of spikes and plots of low-pass filtered IPBRs $R_b(t)$ for $N=$ (c1) $10^3$ and (c2) $10^4$.
}
\label{fig:BT1}
\end{figure}

For determination of the higher bursting threshold $D_h^*$ for the bursting transition (i.e., transition from burst synchronization to unsynchronization), we employ a realistic bursting order parameter ${\cal{O}}_b$ of
Eq.~(\ref{eq:Border1}), corresponding to the time-averaged fluctuation of the low-pass filtered IPBR $R_b(t)$. Figure \ref{fig:BT1}(a) shows the plot of $\langle {\cal{O}}_b \rangle_r$ ($\langle \cdots \rangle_r:$ realization-average) versus $D$. For $D < D^*_h$ $(\simeq 0.109$), burst synchronization appears because the values of $\langle {\cal{O}}_b \rangle_r$ become saturated to non-zero limit values as $N$ is increased. However, as $D$ passes the higher bursting threshold $D^*_h$, the bursting order parameter $\langle {\cal{O}}_b \rangle_r$ tends to zero as $N \rightarrow \infty$, and hence a transition to unsynchronization occurs due to a destructive role of noise spoiling the burst synchronization. We consider examples for $D=0.1$ and $D=0.12$. Figures \ref{fig:BT1}(b1) and \ref{fig:BT1}(b2) show the raster plots of spikes and the IPBRs $R_b(t)$ for $N=10^3$ and $10^4$, respectively, in the case of burst synchronization of $D=0.1$. For this case, with increasing $N$, more clear bursting bands appear in the raster plot, and $R_b(t)$ shows more regular oscillation with nearly same amplitudes. On the other hand, for the case of unsynchronization of $D=0.12$, spikes in the raster plots become completely scattered (without forming any bursting bands) and $R_b(t)$ becomes more and more nearly
stationary (i.e., noisy fluctuations of $R_b(t)$ becomes reduced) as $N$ is increased [see Figs.~\ref{fig:BT1}(c1) and \ref{fig:BT1}(c2)].

\begin{figure}
\includegraphics[width=0.8\columnwidth]{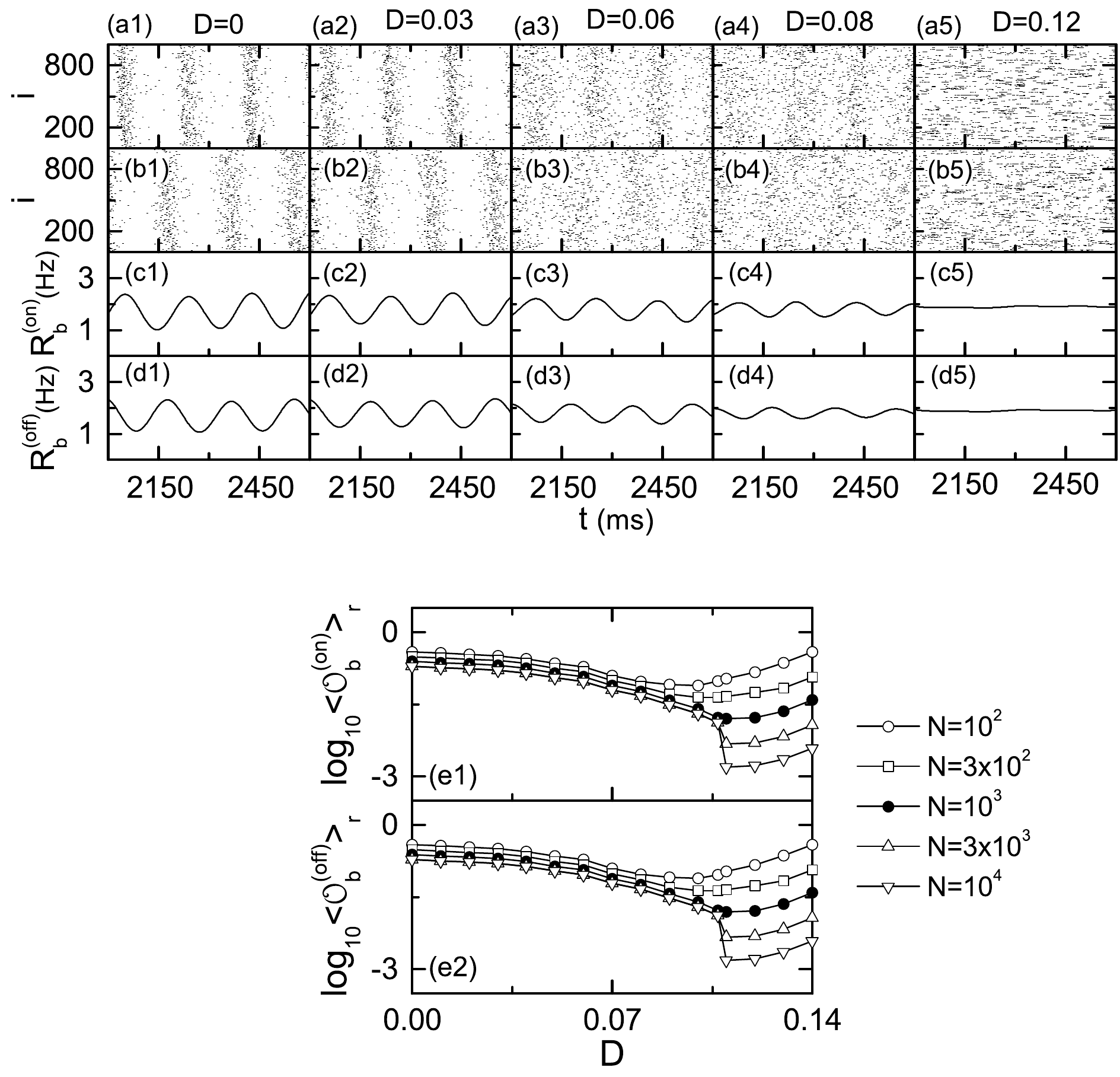}
\caption{
Representation of bursting states in terms of active phase (bursting) onset and offset times and determination of the higher bursting noise threshold $D^*_h$ for the bursting transition in terms of realistic bursting order parameters $\langle {\cal{O}}_b^{(on)} \rangle_r$ and $\langle {\cal{O}}_b^{(off)} \rangle_r$ in the pure $\alpha-$process ($\alpha=1$) with symmetric preferential attachment of $\widetilde{l_\alpha} (=20)$: burst synchronization for $D=0$, 0.03, 0.06, and 0.08, and unsynchronization for $D=0.12$. (a1)-(a5) Raster plots of bursting onset times, (b1)-(b5) raster plots of bursting offset times, (c1)-(c5) time series of IPBR kernel estimates $R_b^{(on)}(t)$, and (d1)-(d5) time series of IPBR kernel estimates $R_b^{(off)}(t)$. Plots of realistic thermodynamic bursting order parameters (e1) $\langle {\cal{O}}_b^{(on)} \rangle_r$ [based on $R_b^{(on)}(t)$] and (e2) $\langle {\cal{O}}_b^{(off)} \rangle_r$ [based on $R_b^{(off)}(t)$] versus $D$.
}
\label{fig:BT2}
\end{figure}

For more direct visualization of bursting behavior, we also consider another raster plots of bursting onset and offset times, from which we can directly obtain the IPBR kernel estimates of band width $h=50$ ms, $R_b^{(on)}(t)$ and $R_b^{(off)}(t)$, without frequency filtering. Figures \ref{fig:BT2}(a1)-\ref{fig:BT2}(a5) show the raster plots of bursting onset times for various values of $D$, and the corresponding IPBR kernel estimates $R_b^{(on)}(t)$ are shown in Figs.~\ref{fig:BT2}(c1)-\ref{fig:BT2}(c5). Similarly, the raster plots of bursting offset times are given in Figs.~\ref{fig:BT2}(b1)-\ref{fig:BT2}(b5), and Figs.~\ref{fig:BT2}(d1)-\ref{fig:BT2}(d5) show the corresponding IPBR kernel estimates $R_b^{(off)}(t)$. For the case of burst synchronization, bursting stripes (composed of bursting onset or offset times and representing burst synchronization) appear successively in the raster plot. The bursting onset and offset stripes are time-shifted [e.g., compare Figs.~\ref{fig:BT2}(a1) and \ref{fig:BT2}(b1) for $D=0$]. For this synchronous case, the corresponding IPBR kernel estimates, $R_b^{(on)}(t)$ and $R_b^{(off)}(t)$, show slow-wave oscillations with the same population bursting frequency $f_b^{(p)} \simeq 5.4$ Hz, although they are phase-shifted [e.g., compare Figs.~\ref{fig:BT2}(c1) and \ref{fig:BT2}(d1) for $D=0$]. On the other hand, in the case of unsynchronization, bursting onset or offset times are scattered completely without forming any bursting stripes in the raster plots, and the corresponding IPBR kernel estimates $R_b^{(on)}(t)$ and $R_b^{(off)}(t)$ become nearly stationary. Then, like the case of the bursting order parameter ${\cal{O}}_b$ of Eq.~(\ref{eq:Border1}), the mean square deviations of $R_b^{(on)}(t)$ and $R_b^{(off)}(t)$ also play another bursting order parameters
of Eq.~(\ref{eq:Border2}), used to determine the higher bursting threshold $D^*_h$ for the bursting transition. Figures \ref{fig:BT2}(e1) and \ref{fig:BT2}(e2) show plots of the bursting order parameters $\langle{\cal{O}}_b^{(on)}\rangle_r$ and $\langle{\cal{O}}_b^{(off)}\rangle_r$ versus $D$, respectively. As in the case of $\langle{\cal{O}}_b\rangle_r$, in the same region of $D < D^*_h (\simeq 0.108$), burst synchronization appears because the values of $\langle{\cal{O}}_b^{(on)}\rangle_r$ and $\langle{\cal{O}}_b^{(off)}\rangle_r$ become saturated to non-zero limit values as $N \rightarrow \infty$. On the other hand, when passing the higher threshold $D^*_h$, $\langle{\cal{O}}_b^{(on)}\rangle_r$ and $\langle{\cal{O}}_b^{(off)}\rangle_r$ tend to zero in the thermodynamic limit of $N \rightarrow \infty$, and hence transition to unsynchronization occurs. In this way, the higher bursting threshold $D^*_h$ for the bursting transition may be well determined through calculation of each of the three realistic bursting order parameters, $\langle {\cal{O}}_b \rangle_r$, $\langle {\cal {O}}_b^{(on)} \rangle_r$ and $\langle {\cal {O}}_b^{(off)} \rangle_r$. Particularly, $\langle {\cal {O}}_b^{(on)} \rangle_r$ and $\langle {\cal {O}}_b^{(off)} \rangle_r$ are more direct ones than $\langle {\cal {O}}_b \rangle_r$ because they are based on the IPBR kernel estimates $R_b^{(on)}(t)$ and $R_b^{(off)}(t)$ which are directly obtained from the raster plots of the bursting onset and offset times without frequency filtering, respectively. Hereafter, for convenience we consider only the raster plot of bursting onset times for characterization of burst synchronization, because both the raster plots of bursting onset and offset times show the same bursting behaviors.

\begin{figure}
\includegraphics[width=0.8\columnwidth]{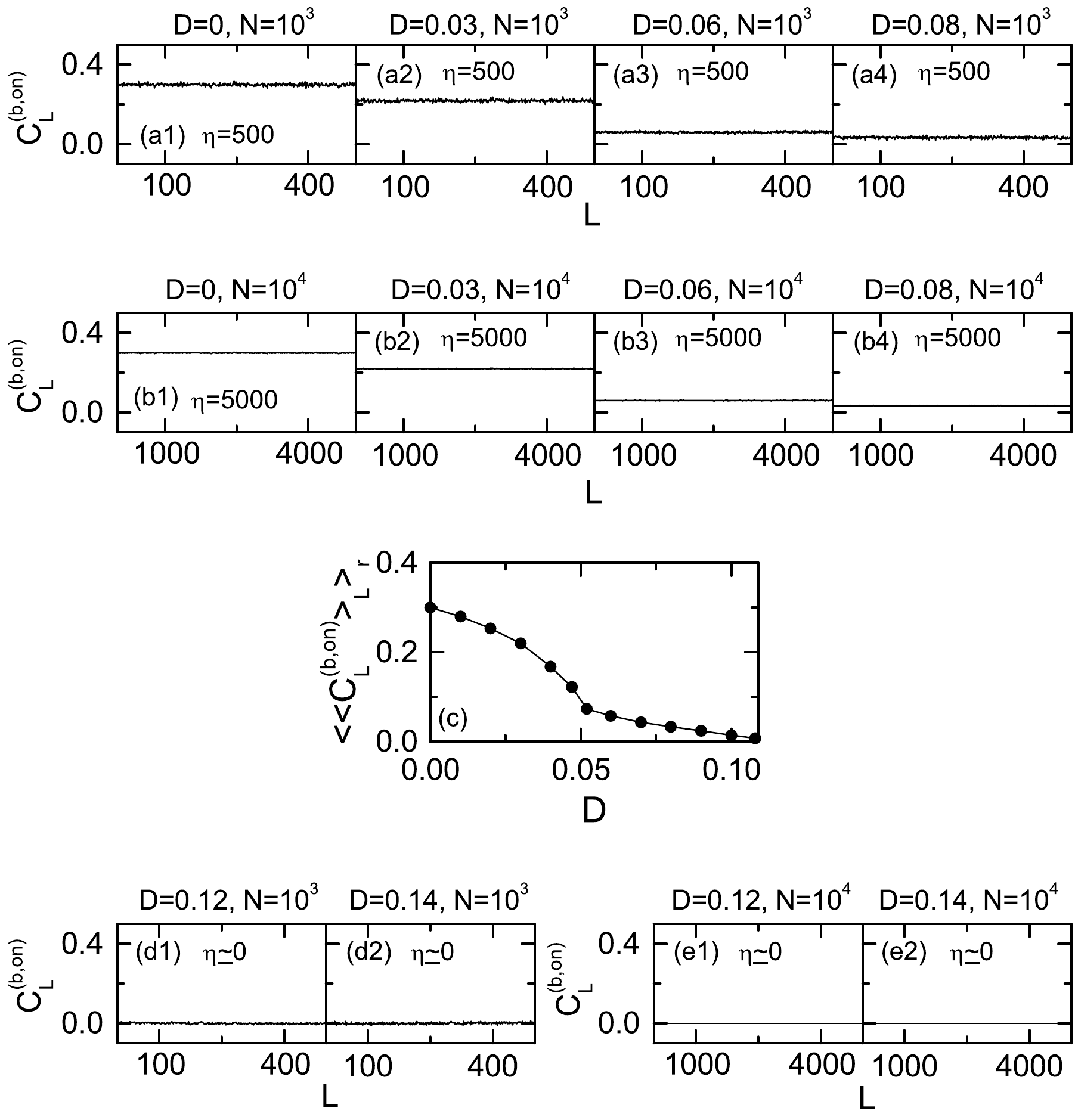}
\caption{
Characterization of bursting transition in terms of spatial cross-correlations in the pure $\alpha-$process ($\alpha=1$) with symmetric preferential attachment of $\widetilde{l_\alpha} (=20)$.
Plots of the spatial correlation function $C_L^{(b,on)}$ between neuronal pairs versus spatial distance $L$ for the synchronized cases of $D=0$, 0.03, 0.06, and 0.08 when $N=$ (a1)-(a4) $10^3$  and (b1)-(b4) $10^4$. (c) Plot of the average spatial-correlation degree $\langle C_L^{(b,on)} \rangle_L$  versus $D$. Plots of the spatial correlation function $C_L^{(b,on)}$ versus $L$ for the unsynchronized cases of $D=0.12$ and 0.14 when  $N=$ (d1)-(d2) $10^3$ and (e1)-(e2) $10^4$.
}
\label{fig:BT3}
\end{figure}

To further understand the bursting transition, we investigate the effect of the noise intensity $D$ on the ``microscopic'' dynamical cross-correlations between neuronal pairs of bursting neurons. To this end, we introduce the spatial cross-correlation $C_L^{(b,on)}$ ($L=1,...,N/2)$ between neuronal pairs separated by a spatial distance $L$ in Eq.~(\ref{eq:SCC}) which corresponds to the average of all the temporal cross-correlations between the IIBRs of Eq.~(\ref{eq:IIBR}) $r_i^{(b,on)}(t)$ and $r_{i+L}^{(b,on)}(t)$ $(i=1,...,N)$ at the zero-time lag. Figures \ref{fig:BT3}(a1)-\ref{fig:BT3}(a4) show plots of the spatial cross-correlation function $C_L^{(b,on)}$ versus $L$ in the case of $N=10^3$ for various values of $D$ in the region of burst synchronization. The spatial cross-correlation functions $C_L^{(b,on)}$ are nearly non-zero constants in the whole range of $L$, and hence the correlation length $\eta$ for all cases of (a1)-(a5) becomes $N/2$ (=500) covering the whole network (note that the maximal distance between neurons is $N/2$ because of the ring architecture on which HR neurons exist). Consequently, the whole network is composed of just one single synchronized block. For $N=10^4$, the non-zero flatness of $C_L^{(b,on)}$ in Figs.~\ref{fig:BT3}(b1)-\ref{fig:BT3}(b4) also extends to the whole range ($L=N/2=5000$) of the network, and the correlation length becomes $\eta=5000$, which also covers the whole network. Then, the normalized correlation length $\tilde{\eta}$ ($= \frac {\eta} {N}$), representing the ratio of the correlation length $\eta$ to the network size $N$ (i.e., denoting the relative size of synchronized blocks when compared to the whole network size), has a non-zero limit value, $1/2$, and consequently burst synchronization emerges in the whole network. The degree of burst synchronization may also be measured in terms of the average spatial cross-correlation degree $\langle C_L^{(b,on)} \rangle_L$ ($\langle \cdots \rangle_L$: length-average) given by averaging of $C_L^{(b,on)}$ over all lengths $L$. For each $D$, we obtain $\langle \langle C_L^{(b,on)} \rangle_L \rangle_r$ via average over 20 realizations. Figure \ref{fig:BT3}(c) shows the plot of $\langle \langle C_L^{(b,on)} \rangle_L \rangle_r$. In the region of complete synchronization for $D< D^*_l$, $\langle \langle C_L^{(b,on)} \rangle_L \rangle_r$ drops rapidly, and then it slowly decreases to zero in the region of ``pure'' burst synchronization for $D^*_l < D < D^*_h$. In contrast to the case of complete and burst synchronization, the spatial cross-correlation functions $C_L^{(b,on)}$ for $D=0.12$ and 0.14 are nearly zero for both cases of $N=10^3$ and $10^4$, as shown in Figs.~\ref{fig:BT3}(d1)-\ref{fig:BT3}(d2) and Figs.~\ref{fig:BT3}(e1)-\ref{fig:BT3}(e2). For theses cases, due to a destructive role of noise spoiling the pacing between bursting onset times, the correlation lengths $\eta$ become nearly zero, independently of $N$, and hence no population synchronization occurs in the network.

\begin{figure}
\includegraphics[width=0.8\columnwidth]{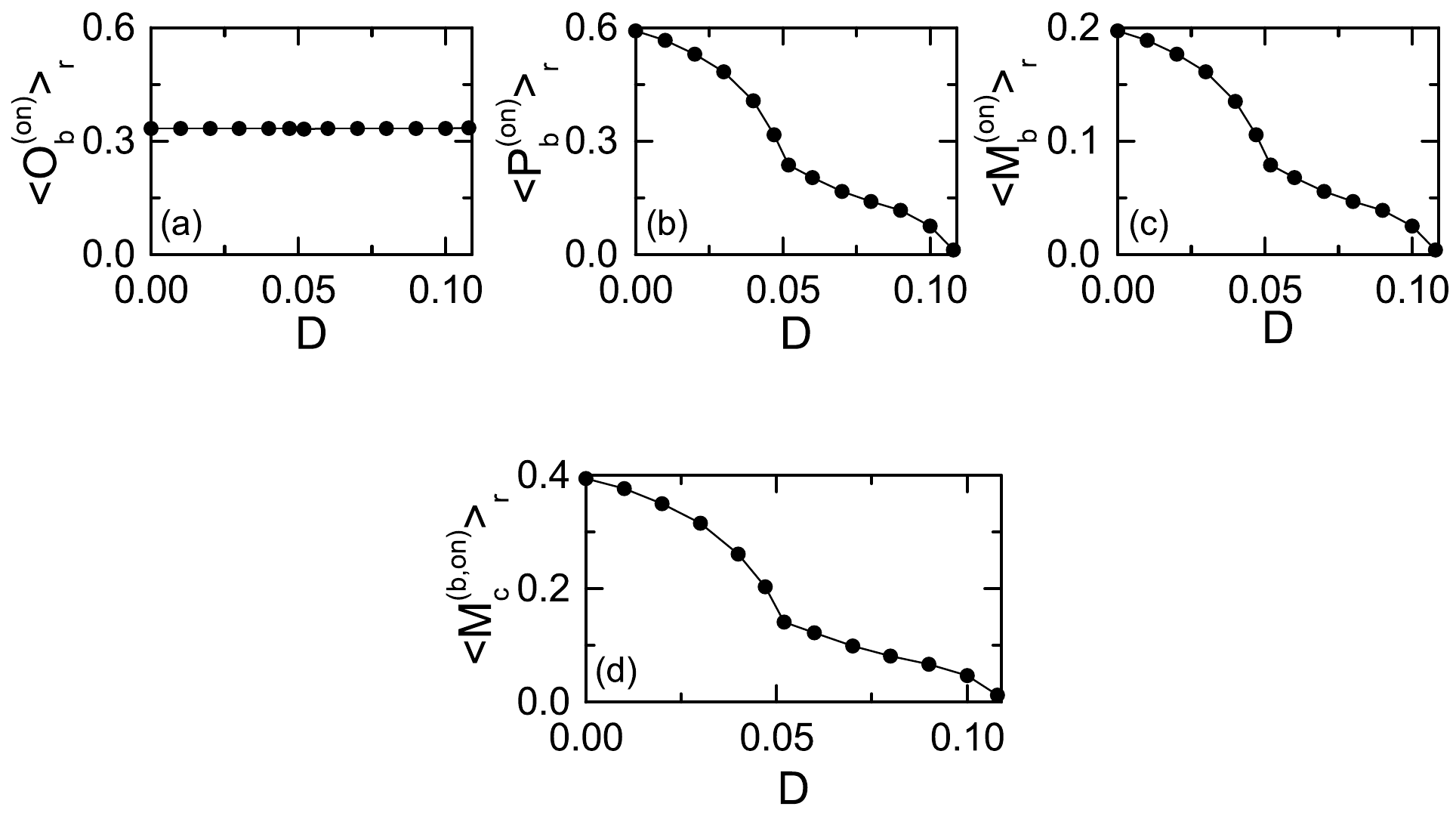}
\caption{
Measurement of the degree of burst synchronization in the pure $\alpha-$process ($\alpha=1$) with symmetric preferential attachment of $\widetilde{l_\alpha} (=20)$. Plots of (a) the average occupation degree $\langle O_b^{(on)} \rangle_r$ of bursting onset times, (b) the average pacing degree $\langle P_b^{(on)} \rangle_r$ of bursting onset times, and (c) the statistical-mechanical bursting measure $\langle M_b^{(on)} \rangle_r$ versus $D$. (d) Plot of the statistical-mechanical bursting correlation measure $\langle M_c^{(b,on)} \rangle_r$ versus $D$.
}
\label{fig:SBM}
\end{figure}

\begin{figure}
\includegraphics[width=0.8\columnwidth]{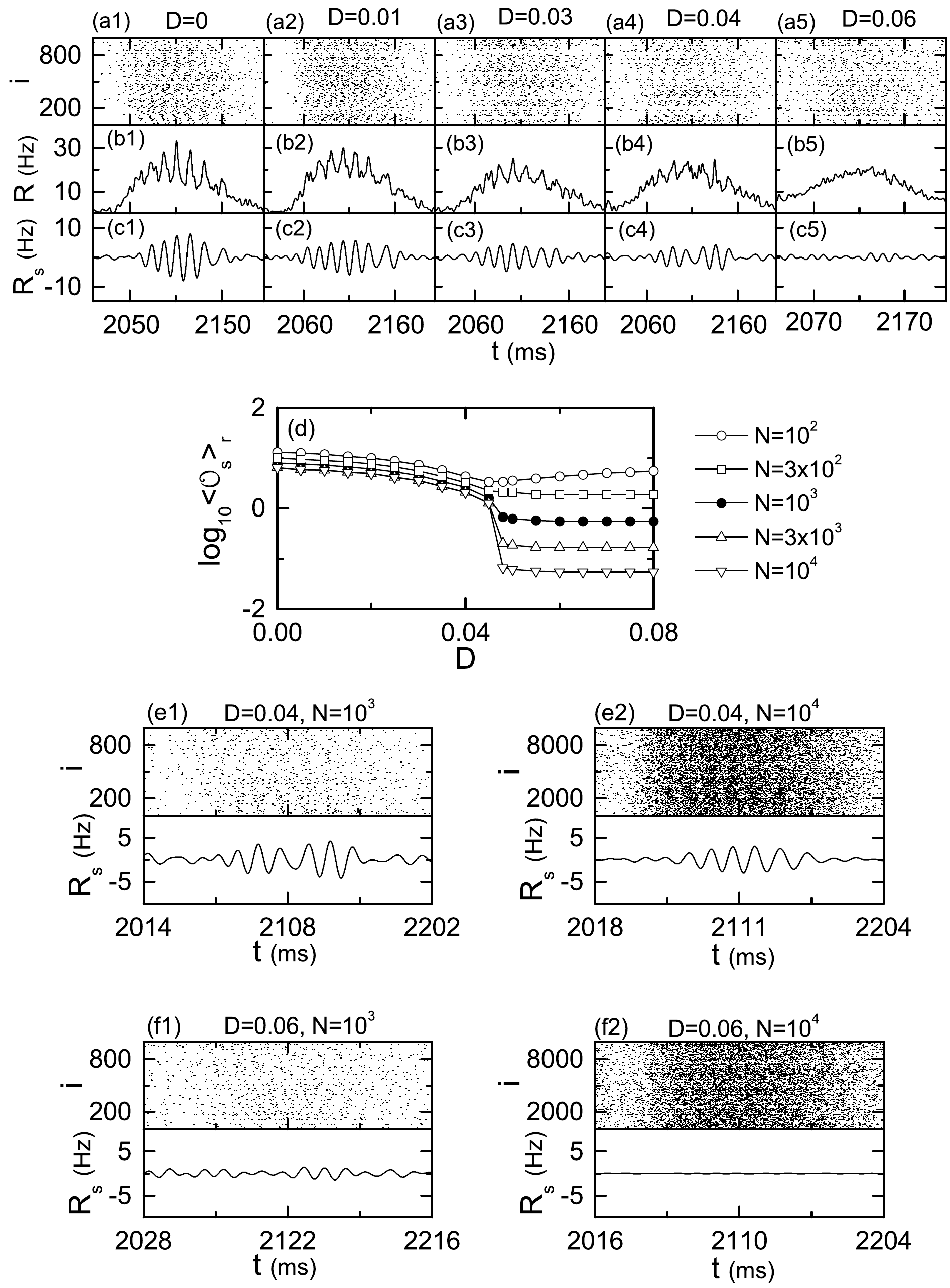}
\linespread{1.2}
\caption{
Intraburst spike synchronization and determination of the lower spiking noise threshold $D^*_l$ for the intraburst spiking transition in terms of a realistic thermodynamics spiking order parameter $\langle {\cal{O}}_s \rangle_r$
in the pure $\alpha-$process ($\alpha=1$) with symmetric preferential attachment of $\widetilde{l_\alpha} (=20)$. (a1)-(a5) Raster plots of spikes, (b1)-(b5) IPFR kernel estimates $R(t)$, and (c1)-(c5) band-pass filtered IPSRs $R_s(t)$ in the 1st global bursting cycle of the low-pass filtered IPBR $R_b(t)$ (after the transient time of $ 2 \times 10^3$ ms) for various values of $D=0$, 0.01, 0.03, 0.04, and 0.06. (d) Plots of the spiking order parameter $\langle {\cal{O}}_s \rangle_r$ versus $D$. Intraburst spike synchronization for $D=0.04$: raster plots of spikes and plots of IPSRs $R_s(t)$ for $N=$ (e1) $10^3$ and (e2) $10^4$. Intraburst spike unsynchronization for $D=0.06$: raster plots of spikes and plots of IPSRs $R_s(t)$ for $N=$ (f1) $10^3$ and (f2) $10^4$.
}
\label{fig:ST}
\end{figure}

We now measure the degree of burst synchronization in the synchronized region of $0 < D < D^*_h$. As shown in Figs.~\ref{fig:BT2}(a1)-\ref{fig:BT2}(a4), burst synchronization may be well visualized in the raster plots of bursting onset times. For $D=0$ clear bursting stripes (composed of bursting onset times and indicating burst synchronization) appear in the raster plot. As $D$ is increased, bursting stripes become more and more smeared. Eventually, when passing the higher bursting threshold $D^*_h$, bursting onset times are completely scattered without forming any bursting stripes, as shown in Fig.~\ref{fig:BT2}(a5). For this case of burst synchronization, the IPBR kernel estimates $R_b^{(on)}(t)$ exhibit slow-wave oscillations, as shown in Figs.~\ref{fig:BT2}(c1)-\ref{fig:BT2}(c4). With increasing $D$, the amplitude of $R_b(t)$ decreases, and it becomes nearly stationary as $D$ passes $D^*_h$. We measure the degree of burst synchronization seen in the raster plot of bursting onset times in Figs.~\ref{fig:BT2}(a1)-\ref{fig:BT2}(a4) in terms of a statistical-mechanical bursting measure $M_b^{(on)}$, based on $R_b^{(on)}(t)$, introduced by considering the occupation pattern (representing the density of the bursting onset stripes) and the pacing pattern (denoting the smearing of the bursting onset stripes) of bursting onset times in the bursting stripes as explained in the subsection \ref{subsec:SM}. Figures \ref{fig:SBM}(a)-\ref{fig:SBM}(c) show $\langle O_b^{(on)} \rangle_r$ of Eq.~(\ref{eq:Occu}) (average occupation degree), $\langle P_b^{(on)} \rangle_r$ of Eq.~(\ref{eq:Pacing}) (average pacing degree), and the statistical-mechanical bursting measure $\langle M_b^{(on)} \rangle_r$ of Eq.~(\ref{eq:BM2}) for 13 values of $D$ in the synchronized region. In the whole region of $D$, $\langle O_b^{(on)} \rangle_r \simeq 0.33$. Hence, about one third of total HR neurons make burstings in each global bursting cycle, as shown in Figs.~\ref{fig:CBS}(e1) and \ref{fig:CBS}(e2). Similar to the case of the average spatial correlation degree $\langle \langle C_L^{(b,on)} \rangle_L \rangle_r$ in Fig.~\ref{fig:BT3}(c), $\langle P_b^{(on)}  \rangle_r$ decreases rapidly in the region of complete synchronization (i.e., region of $D< D^*_l$), while it decreases slowly to zero in the region of ``pure'' burst synchronization (i.e., region of $D^*_l < D < D^*_h$). Since $ \langle O_b^{(on)} \rangle_r \simeq 0.33$, $\langle M_b^{(on)} \rangle_r \simeq \langle P_b^{(on)} \rangle_r /3$.

For characterization of the burst synchronization, we also introduce another statistical-mechanical bursting correlation measure $M_c^{(b,on)}$, based on the cross-correlations between the kernel estimate IPBR $R_b^{(on)}(t)$ and the IIBR kernel estimates $r_i^{(b,on)}(t)$ ($i=1, ..., N$) through extension of the case of spiking neurons \cite{Kim4}. This correlation-based measure $M_c^{(b,on)}$ may also be regarded as a statistical-mechanical measure because it quantifies the average contribution of (microscopic) IIBRs $r_i^{(b,on)}(t)$ to the (macroscopic) IPBR $R_b^{(on)}(t)$. The normalized cross-correlation function $C_i^{(b,on)}(\tau)$ between $R_b^{(on)}(t)$ and $r_i^{(b,on)}(t)$ is given by
\begin{equation}
C_i^{(b,on)}(\tau) = \frac{\overline{\Delta R_b^{(on)}(t+\tau) \Delta r_i^{(b,on)}(t)}}{\sqrt{\overline{\Delta {R_b^{(on)}}^2(t)}}\sqrt{\overline{\Delta {r_i^{(b,on)}}^2(t)}}},
\label{eq:CCFi}
\end{equation}
where $\tau$ is the time lag, $\Delta R_b^{(on)}(t) = R_b^{(on)}(t) - \overline{R_b^{(on)}(t)}$, $\Delta r_i^{(b,on)}(t) = r_i^{(b,on)}(t)-\overline{r_i^{(b,on)}(t)}$, and the overline denotes the time average.
Then, the statistical-mechanical bursting correlation measure $M_c^{(b,on)}$ is given by the ensemble-average of $C_i^{(b,on)}(0)$ at the zero-time lag:
\begin{equation}
M_c^{(b,on)} = \frac{1}{N} \sum_{i=1}^{N} C_i^{(b,on)}(0).
\label{eq:CM}
\end{equation}
Here, the number of data used for the calculation of temporal cross-correlation function is $2 \times 10^4$ in each realization, and we obtain $\langle M_c^{(b,on)} \rangle_r$ via average over 20 realizations.
Figure \ref{fig:SBM}(d) shows the plot of $\langle M_c^{(b,on)} \rangle_r$ versus $D$.
As $D$ is increased, $\langle M_c^{(b,on)} \rangle_r$ drops rapidly in the region of complete synchronization for $D< D^*_l$, while it slowly decreases to zero in the region of ``pure'' burst synchronization
for $D^*_l < D < D^*_h$, like the case of $M_b^{(b,on)}$.

In addition to the above burst synchronization, we also investigate intraburst spike synchronization of bursting HR neurons by varying the noise intensity $D$ for $J=4$. Figures \ref{fig:ST}(a1)-\ref{fig:ST}(a5) and \ref{fig:ST}(b1)-\ref{fig:ST}(b5) show the raster plots of intraburst spikes and the corresponding IPFR kernel estimates $R(t)$ during the 1st global bursting cycle, respectively. For $D=0$, the 1st bursting band is composed of (somewhat clear) spiking stripes, and hence the corresponding $R(t)$ exhibits a bursting activity [i.e., fast spikes appear on a slow wave in $R(t)$]. Through band-pass filtering of $R(t)$, we obtain the fast-oscillating IPSR $R_s(t)$ (showing only the intraburst spiking behavior without a slow wave) in Fig.~\ref{fig:ST}(c1). As $D$ is increased, intraburst spiking stripes become more and more smeared due to a destructive role of noise spoiling the spike synchronization, and hence the amplitudes of $R_s(t)$ decrease [see Figs.~\ref{fig:ST}(c2)-\ref{fig:ST}(c5)], although the underlying slow-wave oscillations of $R(t)$ persist. Eventually, when passing a lower spiking threshold $D^*_l$, complete loss of the intraburst spike synchronization occurs, and then only pure burst synchronization persists.

For determination of $D_l^*$ for the spiking transition [i.e., transitions from complete synchronization (including both burst synchronization and intraburst spike synchronization) to pure burst synchronization], we employ a realistic spiking order parameter ${\cal{O}}_s$ of Eq.~(\ref{eq:SO2}), corresponding to the time-averaged fluctuation of the IPSR $R_s(t)$. Figure \ref{fig:ST}(d) shows plots of ${\langle {\cal{O}}_s \rangle}_r$ versus $D$. When passing the lower spiking threshold $D^*_l (\simeq 0.048$), a transition from intraburst spike synchronization to intraburst spike unsynchronization occurs because the values of ${\langle {\cal{O}}_s \rangle}_r$ tend to zero in the thermodynamic limit of $N \rightarrow \infty$. Consequently, only for $D < D^*_l$ intraburst spike synchronization appears. In this way, $D^*_l$ for the intraburst spiking transition may be well determined in terms of the spiking order parameter ${\langle {\cal{O}}_s \rangle}_r$. We consider two examples for $D=0.04$ and 0.06. Figures \ref{fig:ST}(e1) and \ref{fig:ST}(e2) show the raster plots of intraburst spikes and the IPSRs $R_s(t)$ for $N=10^3$ and $10^4$, respectively, in the case of intraburst spike synchronization of $D=0.04$. For this case, with increasing $N$, more clear spiking stripes appear in the raster plot, and $R_s(t)$ shows more regular oscillation. On the other hand, for the case of intraburst unsynchronization of $D=0.06$, intraburst spikes in the raster plot become completely scattered (without forming any spiking stripes) and $R_s(t)$ becomes more and more nearly stationary [i.e., noisy fluctuations of $R_s(t)$ becomes reduced] as $N$ is increased [see Figs.~\ref{fig:ST}(f1) and \ref{fig:ST}(f2)].

\begin{figure}
\includegraphics[width=0.8\columnwidth]{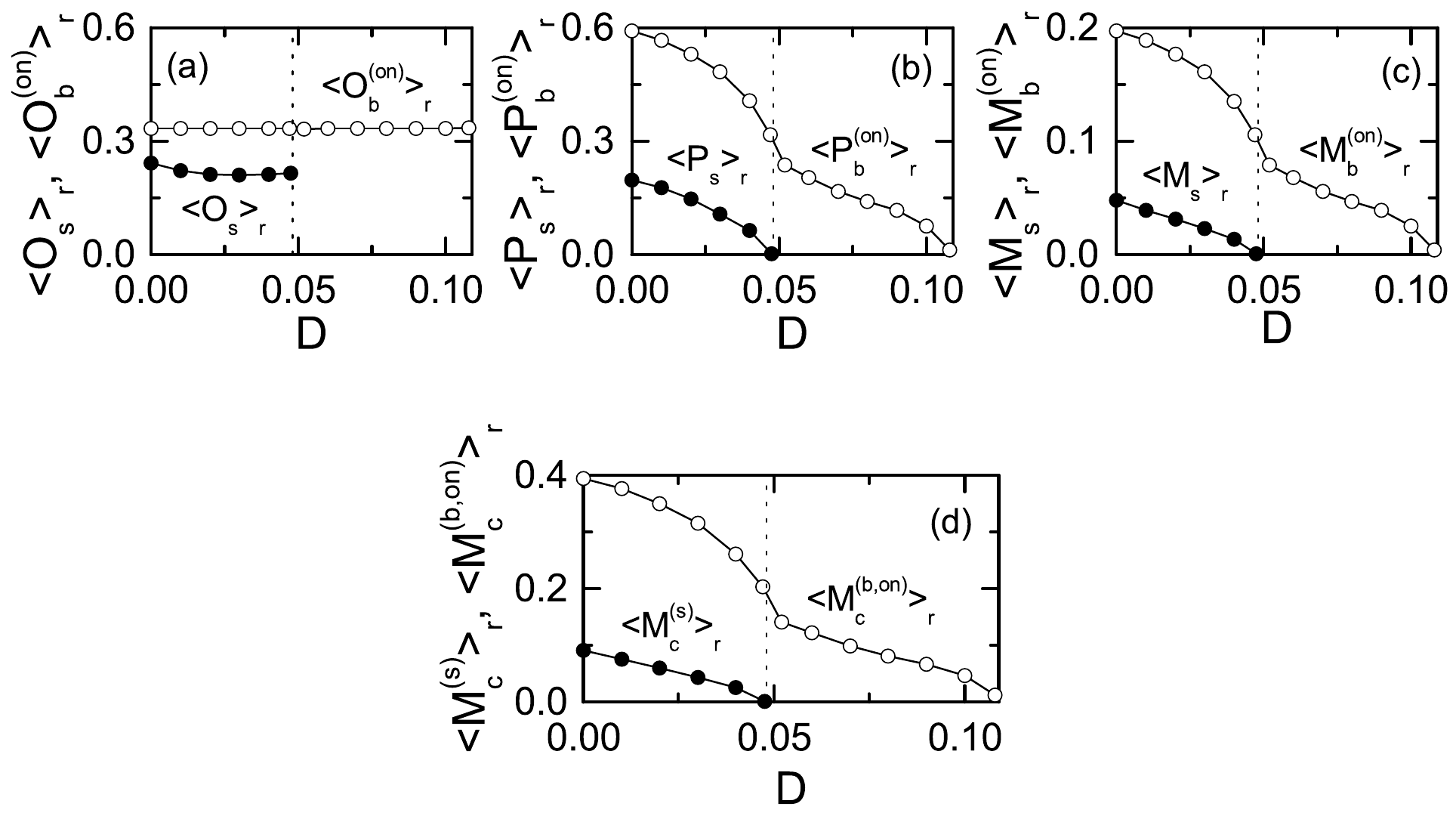}
\caption{
Measurement of the degree of intraburst spike synchronization in the pure $\alpha-$process ($\alpha=1$) with symmetric preferential attachment of $\widetilde{l_\alpha} (=20)$. Plots of (a) the average occupation degree $\langle O_s \rangle_r$ of spikes, (b) the average pacing degree $\langle P_s \rangle_r$ of spikes, (c) the statistical-mechanical spiking measure $\langle M_s \rangle_r$, and (d) the statistical-mechanical spiking correlation measure $\langle M_c^{(s)} \rangle_r$ versus $D$. Data for $\langle O_s \rangle_r$, $\langle P_s \rangle_r$, $\langle M_s \rangle_r$, and  $\langle M_c^{(s)} \rangle_r$ are denoted by solid circles. For comparison, data for $\langle O_b^{(on)} \rangle_r$, $\langle P_b^{(on)} \rangle_r$, $\langle M_b^{(on)} \rangle_r$, and $\langle M_c^{(b,on)} \rangle_r$ represented by open circles and denoting the degree of burst synchronization, are also given. The vertical dotted line represents the lower spiking threshold $D^*_l (\simeq 0.048)$.
}
\label{fig:SSM}
\end{figure}

Within the whole region of intraburst spike synchronization ($D<D^*_l$), we also measure the degree of intraburst spike synchronization by employing a statistical-mechanical spiking measure $M_s$, based on the IPSR $R_s(t)$  \cite{Kim1}. As shown in Figs.~\ref{fig:ST}(a1)-\ref{fig:ST}(a4), spike synchronization may be well visualized in the raster plot of spikes; spiking stripes (composed of intraburst spikes and indicating intraburst spike synchronization) appear in the 1st bursting band [corresponding to the 1st bursting cycle of $R_b(t)$] of the raster plot. Like the case of bursting cycles of $R_b(t)$, spiking cycles (corresponding to the spiking stripes)
of $R_s(t)$ may also be introduced: for more details, refer to Fig.~7 in \cite{Kim1}. Then, similar to the case of burst synchronization, we measure the degree of intraburst spike synchronization seen in the raster plot in terms of a statistical-mechanical spiking measure, based on $R_s(t)$, by considering the occupation and the pacing patterns of intraburst spikes in the spiking stripes. The spiking measure $M_{1,j}^{(s)}$ of the $j$th spiking cycle in the 1st bursting cycle is defined by the product of the occupation degree $O_{1,j}^{(s)}$ of spikes (denoting the density of the $j$th spiking stripe) and the pacing degree $P_{1,j}^{(s)}$ of spikes (representing the smearing of  the $j$th spiking stripe). For the 1st bursting cycle, we obtain the spiking-averaged occupation degree $O_1^{(s)}$ (=${\langle O_{1,j}^{(s)} \rangle}_s$), the spiking-averaged pacing degree $P_1^{(s)}$ (=${\langle P_{1,j}^{(s)} \rangle}_s$), and the spiking-averaged statistical-mechanical spiking measure $M_1^{(s)}$ (=${\langle M_{1,j}^{(s)} \rangle}_s$), where ${\langle \cdots \rangle}_s$ represents the average over the spiking cycles. In each realization, we follow $500$ bursting cycles and get $O_i^{(s)}$, $P_i^{(s)}$, and $M_i^{(s)}$ in each $i$th bursting cycle. Then, through average over all bursting cycles, we obtain the bursting-averaged occupation degree $O_s$ (=${\langle O_i^{(s)} \rangle}_b$), the bursting-averaged pacing degree $P_s$ (=${\langle P_i^{(s)} \rangle}_b$), and the bursting-averaged statistical-mechanical spiking measure $M_s$ (=${\langle M_i^{(s)} \rangle}_b$). We note that $O_s$, $P_s$, and $M_s$ are obtained through double-averaging $[{\langle {\langle \cdots \rangle}_s \rangle}_b]$ over the spiking and bursting cycles. For each $D$, we repeat the above process to get $O_s$, $P_s$, and $M_s$ for multiple realizations. Thus, we obtain ${\langle O_s \rangle}_r$ (average occupation degree of spikes), ${\langle P_s \rangle}_r$ (average pacing degree of spikes), and ${\langle M_s \rangle}_r$ (average statistical-mechanical spiking measure) through average over 20 realizations. The data of ${\langle O_s \rangle}_r$, ${\langle P_s \rangle}_r$, and ${\langle M_s \rangle}_r$ are denoted by solid circles in Figs.~\ref{fig:SSM}(a)-\ref{fig:SSM}(c), respectively. For comparison, data of $\langle O_b^{(on)} \rangle_r$, $\langle P_b^{(on)} \rangle_r$, and $\langle M_b^{(on)} \rangle_r$ for the case of burst synchronization are also represented by open circles in the region of $0 \leq D < D^*_h$. In the whole region of intraburst spike synchronization, ${\langle O_s \rangle}_r \sim 0.22$. When compared with $\langle O_b \simeq 0.33 \rangle_r$ for the case of burst synchronization, only a fraction (about 2/3) of the HR neurons exhibiting the bursting active phases fire spikings in the spiking cycles. Unlike the nearly constant ${\langle O_s \rangle}_r$, ${\langle P_s \rangle}_r$ decreases monotonically to zero. We also note that  ${\langle P_s \rangle}_r$ is much less than $\langle P_b^{(on)} \rangle_r$ (e.g., for $D=0$, ${\langle P_s \rangle}_r \simeq 0.20$ and $\langle P_b^{(on)} \rangle_r \simeq 0.59$). Since ${\langle O_s \rangle}_r$ is nearly constant, ${\langle M_s \rangle}_r$ is given approximately by $0.22 ~{\langle P_s \rangle}_r$. Consequently, the degree of intraburst spike synchronization, ${\langle M_s \rangle}_r$, is much less than the degree of burst synchronization, ${\langle M_b^{(on)} \rangle}_r$.

Moreover, we also introduce another statistical-mechanical spiking correlation measure $M_c^{(s)}$, based on the cross-correlations between the IPSR $R_s(t)$ and the instantaneous individual spike rates (IISRs) $r_i^{(s)}(t)$ ($i=1, ..., N$) \cite{Kim4}. Like the case of the IIBR $r_i^{(b,on)}(t)$ of Eq.~(\ref{eq:IIBR}), the IISR $r_i^{(s)}(t)$ may also be obtained through convolution of the spike train of the $i$th neuron with the Gaussian kernel function $K_h(t)$ of band width $h$ (=1 ms). We may get the normalized cross-correlation function $C_i^{(s)} (\tau)$ between $R_s(t)$ and $r_i^{(s)}(t)$, as in Eq.~(\ref{eq:CCFi}) for the burst synchronization. Then, the statistical-mechanical spiking correlation measure $M_c^{(s)}$ is given by the ensemble-average of the normalized cross-correlation function of $C_i^{(s)}(0)$ at the zero-time lag. For calculation of $M_c^{(s)}$, the average number of data used for the calculation of temporal cross-correlation function is 186 for each global bursting cycle, and we follow 500 global bursting cycles in each realization. We obtain $\langle M_c^{(s)} \rangle_r$ via average over 20 realizations. Figure \ref{fig:SSM}(d) shows the plot of $\langle M_c^{(s)} \rangle_r$ versus $D$. As $D$ is increased, $\langle M_c^{(s)} \rangle_r$ also decreases to zero monotonically, like the case of $\langle M_s \rangle_r$.

\subsection{Effect of The Scale-Free Connectivity on Burst and Spike Synchronization}
\label{subsec:SFC}
In this subsection, we fix the values of $J$ and $D$ at $J=4$ and $D=0.06$ (see the fifth item in Table \ref{tab:Parm}) where only the burst synchronization (without intraburst spike synchronization) occurs in the above pure $\alpha$-process with symmetric attachment of $l_\alpha=\widetilde{l_\alpha} (=20)$ [see Figs.~\ref{fig:CBS}(a3), \ref{fig:CBS}(b3), \ref{fig:CBS}(c3), and \ref{fig:CBS}(d3)], and then investigate the effect of scale-free connectivity on the burst synchronization in the following three cases by varying (1) the degree of symmetric attachment $l_\alpha$ and (2) the asymmetry parameter $\Delta l_\alpha$ of the asymmetric attachment in the pure $\alpha-$process (i.e., the standard Barab\'{a}si-Albert SFN model with growth and preferential directed attachment) and by changing (3) the probability $\beta$ of the $\beta-$process (intensifying the internal connections between pre-existing nodes without addition of new nodes). The results for these 3 cases are given in the following subsubsections.

\subsubsection{1st Case of Network Architecture: Varying The Degree of Symmetric Attachment $l_\alpha$}
\label{subsubsec:1st}
\begin{figure}
\includegraphics[width=0.8\columnwidth]{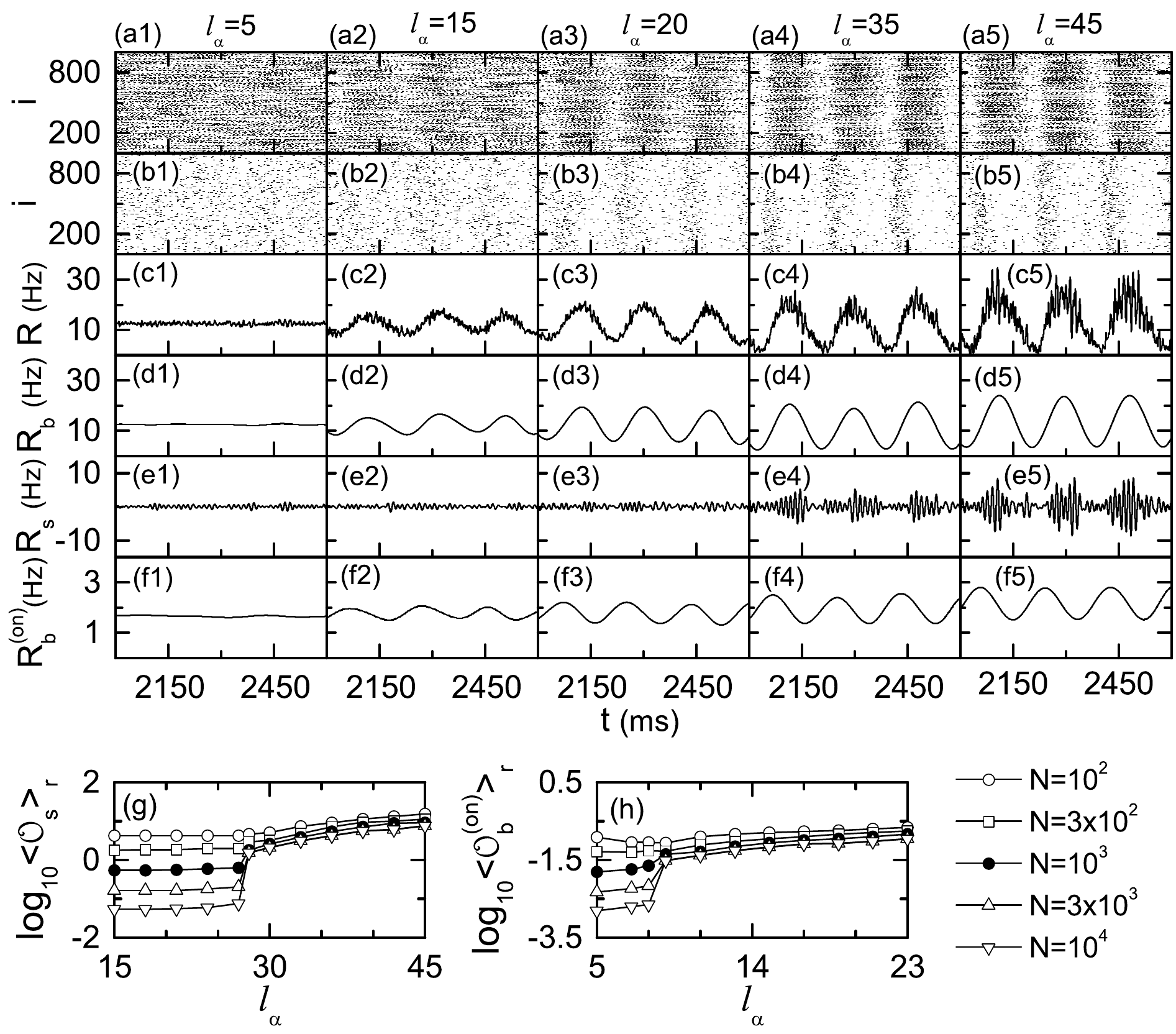}
\caption{
Emergence of burst and spike synchronization for various values of symmetric attachment degree $l_\alpha$ in the pure $\alpha-$process ($\alpha=1$): unsynchronization for $l_\alpha=5$, burst synchronization for $l_\alpha=15$ and 20, and complete synchronization (including both burst and intraburst spike synchronization) for $l_\alpha=35$ and 45. (a1)-(a5) Raster plots of spikes, (b1)-(b5) raster plots of bursting onset times, (c1)-(c5) plots of IPFR kernel estimates $R(t)$, (d1)-(d5) plots of low-pass filtered IPBRs $R_b(t)$, (e1)-(e5) plots of band-pass filtered IPSRs $R_s(t)$, and (f1)-(f5) plots of IPBR kernel estimates $R_b^{(on)}(t)$ for various values of $l_\alpha=$5, 15, 20, 35, and 45. Plots of (g) spiking order parameter $\langle {\cal{O}}_s \rangle_r$ and (h) bursting order parameter $\langle {\cal{O}}_b^{(on)} \rangle_r$  versus $l_\alpha$.
}
\label{fig:SA1}
\end{figure}

As the first case of network architecture, we consider the effect of the degree $l_{\alpha}$ of the symmetric preferential attachment ($l_{\alpha}^{(in)}$ = $l_{\alpha}^{(out)} \equiv l_{\alpha})$.
Figures \ref{fig:SA1}(a1)-\ref{fig:SA1}(a5) show raster plots of spikes for $l_\alpha=5$, 15, 20, 35, and 45, respectively. Their corresponding IPFR kernel estimates $R(t)$ are also given in Figs.~\ref{fig:SA1}(c1)-\ref{fig:SA1}(c5). We note that $R(t)$ exhibit the whole combined behaviors (including both burst and intraburst spike synchronization). To see the bursting and spiking behaviors separately, we obtain the IPBR $R_b(t)$ and the IPSR $R_s(t)$ through frequency-filtering of $R(t)$, which are shown in Figs.~\ref{fig:SA1}(d1)-\ref{fig:SA1}(d5) for $R_b(t)$ and Figs.~\ref{fig:SA1}(e1)-\ref{fig:SA1}(e5) for $R_s(t)$.
Moreover, for more direct visualization of bursting behavior, we also get the raster plots of bursting onset times [see Figs.~\ref{fig:SA1}(b1)-\ref{fig:SA1}(b5)] for various values of $l_\alpha$, and the corresponding
IPBR kernel estimates $R_b^{(on)}(t)$ are shown in Figs.~\ref{fig:SA1}(f1)-\ref{fig:SA1}(f5). As $l_\alpha$ is increased from $l_\alpha=\widetilde{l_\alpha} (=20)$ (studied above in the subsection \ref{subsec:DBA}), bursting bands in the raster plots of spikes and bursting stripes in the raster plots of bursting onset times become more clear. Hence, the amplitudes of $R_b(t)$ and $R_b^{(on)}(t)$ for $l_\alpha = 35$ and 45 become larger than that for $l_\alpha = \widetilde{l_\alpha}$. Furthermore, when passing a higher spiking threshold $l_{\alpha,h}^*$, a transition to intraburst spike-synchronized states occurs. This type of spike synchronization is well shown in the fast-oscillating IPSR $R_s(t)$ for $l_\alpha=35$ and 45. We also note that the spiking amplitude of $R_s(t)$ (i.e., the degree of spike synchronization) for $l_\alpha=45$ is larger than that for $l_\alpha=35$. In this way, with increasing $l_\alpha$ from $l_\alpha=\widetilde{l_\alpha}$ the population synchronization becomes better. On the other hand, as $l_\alpha$ is decreased from $l_\alpha=\widetilde{l_\alpha}$, the amplitudes of $R_b(t)$ and $R_b^{(on)}(t)$ decrease, as shown in the case of $l_\alpha=15$. Eventually, when passing a lower bursting threshold $l_{\alpha,l}^*$, a transition to unsynchronization occurs. As an example of unsynchronization, see the case of $l_\alpha=5$ where $R_b(t)$ and $R_b^{(on)}(t)$ are nearly stationary. To determine the higher and the lower thresholds, $l_{\alpha,h}^*$ and $l_{\alpha,l}^*$, for the spiking and the bursting transitions, we employ the spiking
and the bursting order parameters, $\langle {\cal{O}}_s \rangle_r$ of Eq.~(\ref{eq:SO2}) [representing the time-averaged fluctuation of $R_s(t)$] and  $\langle {\cal{O}}_b^{(on)} \rangle_r$ of Eq.~(\ref{eq:Border2}) [denoting the time-averaged fluctuation of $R_b^{(on)}(t)$]. Figures \ref{fig:SA1}(g) and \ref{fig:SA1}(h) show plots of $\langle {\cal{O}}_s \rangle_r$ and $\langle {\cal{O}}_b^{(on)} \rangle_r$ versus $l_\alpha$, respectively. When passing the higher spiking threshold $l_{\alpha,h}^* (\simeq 28)$, a transition to intraburst spike synchronization occurs because ${\langle {\cal{O}}_s \rangle}_r$ goes to non-zero limit values in the thermodynamic limit of $N \rightarrow \infty$. Consequently, for $l_\alpha > l_{\alpha,h}^*$ complete synchronization (including both burst and spike synchronization) appears. On the other hand, as $l_\alpha$ is decreased and passes the lower threshold $l_{\alpha,l}^* (\simeq 9)$, a transition from burst synchronization to unsynchronization takes place because the values of $\langle {\cal{O}}_b^{(on)} \rangle_r$ tend to zero as $N$ is increased to $\infty$.

\begin{figure}
\includegraphics[width=0.75\columnwidth]{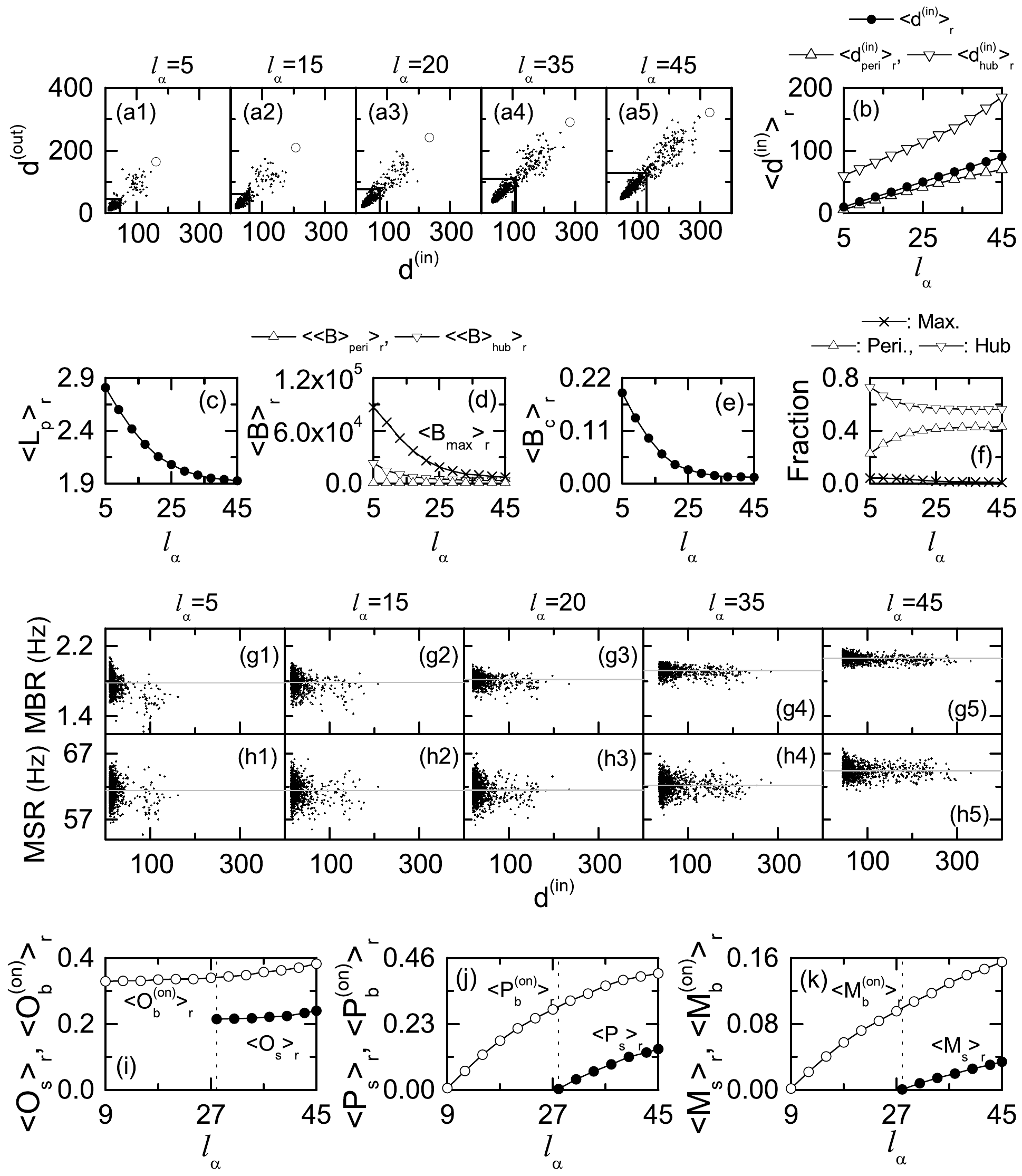}
\linespread{1.1}
\caption{
Effect of the symmetric attachment degree $l_\alpha$ on the degree of burst and spike synchronization in the pure $\alpha-$process ($\alpha=1$). Plots of the out-degree $d^{(out)}$ versus the in-degree $d^{(in)}$ for $l_\alpha=$  (a1) 5, (a2) 15, (a3) 20, (a4) 35, and (a5) 45. (b) Plots of the average in-degree $\langle d^{(in)} \rangle_r$ in the whole population, the average in-degree $\langle d^{(in)}_{peri} \rangle_r$ in the group of peripheral nodes, and the average in-degree $\langle d^{(in)}_{hub} \rangle_r$ in the group of hubs versus $l_\alpha$. (c) Plot of the average path length $\langle L_p \rangle_r$ versus $l_\alpha$. (d) Plots of the maximum betweenness centrality $\langle B_{max} \rangle_r$ of the head hub, the average betweenness centrality $\langle \langle B \rangle_{hub} \rangle_r$ of secondary hubs, and the average betweenness centrality $\langle \langle B \rangle_{peri} \rangle_r$ of peripheral nodes versus $l_\alpha$. (e) Plot of  the betweenness centralization $\langle B_c \rangle_r$  versus  $l_\alpha$. (f) Plots of fractions $\langle B_{max} \rangle_r / \langle B_{tot} \rangle_r$, $\langle B_{tot}^{(hub)} \rangle_r / \langle B_{tot} \rangle_r$, and $\langle B_{tot}^{(peri)} \rangle_r / \langle B_{tot} \rangle_r$ versus $l_\alpha$. Here, quantities in (b)-(f) are obtained via 20 realizations. Plots of MBRs of individual neurons for $l_\alpha=$ (g1) 5, (g2) 15, (g3) 20, (g4) 35, and (g5) 45. Plots of MSRs of individual neurons for $l_\alpha=$ (h1) 5, (h2) 15, (h3) 20, (h4) 35, and (h5) 45. Plots of (i) the average occupation degrees of bursting onset times $\langle O_b^{(on)} \rangle_r$  and spikings $\langle O_s \rangle_r$, (j) the average pacing degrees of bursting onset times $\langle P_b^{(on)} \rangle_r$ and spikings $\langle P_s \rangle_r$, and (k) the statistical-mechanical bursting measure $\langle M_b^{(on)} \rangle_r$ and spiking measure $\langle M_s \rangle_r$ versus $l_\alpha$. Data for $\langle O_b^{(on)} \rangle_r$, $\langle P_b^{(on)} \rangle_r$, and $\langle M_b^{(on)} \rangle_r$ are denoted by open circles, while those for $\langle O_s \rangle_r$, $\langle P_s \rangle_r$, and $\langle M_s \rangle_r$ are represented by solid circles. The vertical dotted line represents the higher spiking threshold $l_{\alpha,h}^* (\simeq 28)$.
}
\label{fig:SA2}
\end{figure}

As in our recent work on spiking neurons in SFNs \cite{Kim2}, for characterization of the effect of $l_\alpha$ on network topology, we study the local property of the SFN in terms of the in- and out-degrees. Plots of the out-degree $d^{(out)}$ versus the in-degree $d^{(in)}$ for $l_\alpha=5$, 15, 20, 35, and 45 are shown in Figs.~\ref{fig:SA2}(a1)-\ref{fig:SA2}(a5), respectively. The nodes are classified into the hub group (composed of the head hub with the highest degree and the secondary hubs with higher degrees) and the peripheral group (consisting of a majority of nodes with lower degrees). For visualization, the peripheral groups are enclosed by rectangles.
Hereafter, boundaries of the rectangles are determined by the thresholds $d^{(in)}_{th}$ and $d^{(out)}_{th}$ where fraction of nodes is 0.2 $\%$. Then, the hub groups lie outside the rectangles, where the node 1 (denoted by the open circle) corresponds to the head hub with the highest degree and the other ones are called as secondary hubs. This type of degree distribution is a ``comet-shaped'' one; the peripheral and the hub groups correspond to the coma (surrounding the nucleus) and the tail of the comet, respectively. The in- and out-degrees are distributed nearly symmetrically around the diagonal, and with increasing $l_\alpha$ they are shifted upward. Figure \ref{fig:SA2}(b) shows plots of the average in-degree $\langle d^{(in)} \rangle_r$ (solid circles) in the whole population, the average in-degree $\langle d^{(in)}_{peri} \rangle_r$ (triangles) in the peripheral group, and the average in-degree $\langle d^{(in)}_{hub} \rangle_r$ (inverted triangles) in the hub group versus $l_\alpha$. As $l_\alpha$ is increased, both $\langle d^{(in)}_{peri} \rangle_r$ and $\langle d^{(in)}_{hub} \rangle_r$ increase in a similar rate. Since the peripheral group is a majority one, $\langle d^{(in)} \rangle_r$ lies a little above $\langle d^{(in)}_{peri} \rangle_r$. In this way, with increasing $l_\alpha$ the total number of connections in the SFN increases.

In addition to the degree distribution of individual nodes, we study the group property of the SFN in terms of the average path length $L_p$ and the betweenness centralization $B_c$ by varying $l_{\alpha}$. The average path length $L_p$ of Eq.~(\ref{eq:APL}), representing typical separation between two nodes in the network, is obtained through the average of the shortest path lengths of all nodal pairs. We note that $L_p$ characterizes the global efficiency of information transfer between distant nodes. Next, we consider the betweenness centrality $B_i$ of the node $i$, given in Eq.~(\ref{eq:BC1}), which denotes the fraction of all the shortest paths between any two other nodes that pass through the node $i$. This betweenness centrality $B_i$ characterizes the potentiality in controlling communication between other nodes in the rest of the network. In our SFN, the head hub (i.e., node 1) with the highest degree also has the maximum betweenness centrality $B_{max}$, and hence the head hub has the largest load of communication traffic passing through it. To examine how much the load of communication traffic is concentrated on the head hub, we obtain the group betweenness centralization $B_c$ of Eq.~(\ref{eq:BC2}), representing the degree to which the maximum betweenness centrality $B_{max}$ of the head hub exceeds the betweenness centrality of all the other nodes. Large $B_c$ implies that load of communication traffic is much concentrated on the head hub, and hence the head hub tends to become overloaded by the communication traffic passing through it. As a result, it becomes difficult to obtain efficient communication between nodes because of destructive interference between many signals passing through the head hub \cite{BTC3}. Figures \ref{fig:SA2}(c) shows the plot of the average path length $\langle L_p \rangle_r$ versus $l_{\alpha}$. As $l_\alpha$ is increased, $\langle L_p \rangle_r$ decreases monotonically due to increase in the total number of connections. Such decrease in $\langle L_p \rangle_r$ leads to reduction in intermediate mediation of nodes controlling the communication in the whole network. Hence, with increasing $l_\alpha$ the total centrality $B_{tot}$, given by the sum of betweenness centralities $B_i$ of all nodes [i.e., $B_{tot} = \sum_i^N B_i = B_{max} + B_{tot}^{(hub)} + B_{tot}^{(peri)},$ where $B_{tot}^{(hub)}$ ($B_{tot}^{(peri)}$) is the total centrality in the group of secondary hubs (peripheral nodes)] is reduced. How the total betweenness $B_{tot}$ decreases with increase in $l_\alpha$ is shown in Fig.~\ref{fig:SA2}(d). As $l_\alpha$ is increased, the maximum betweenness $\langle B_{max} \rangle_r$ (crosses) of the head hub is much more reduced than the average centralities of the secondary hubs and the peripheral nodes, $\langle {\langle B \rangle}_{hub} \rangle_r$ (inverted triangles) and $\langle {\langle B \rangle}_{peri} \rangle_r$ (triangles), which results in decrease in differences between $B_{max}$ of the head hub and $B_i$ of other nodes. As a result, with increasing $l_\alpha$ the betweenness centralization $\langle B_c \rangle_r$ decreases, as shown in Fig.~\ref{fig:SA2}(e). Figure \ref{fig:SA2}(f) also shows plots of fractions $\langle B_{max} \rangle_r / \langle B_{tot} \rangle_r$ (crosses), $\langle B_{tot}^{(hub)} \rangle_r / \langle B_{tot} \rangle_r$ (inverted triangles), and $\langle B_{tot}^{(peri)} \rangle_r / \langle B_{tot} \rangle_r$ (triangles) versus $l_\alpha$. These fractions represent how the total centrality $B_{tot}$ is distributed in the head hub, the secondary hub group, and the peripheral group. As already shown in $\langle B_c \rangle_r$, with increasing $l_\alpha$, the fraction $\langle B_{max} \rangle_r / \langle B_{tot} \rangle_r$ for the head hub (i.e., relative load of communication traffic for the head hub) decreases. More than half load of total communication traffic is given to the group of secondary hubs because of their large average in-degree. However, with increasing $l_\alpha$, the role of peripheral nodes, controlling communication between nodes, also increases due to increase in their average in-degree. Hence, as $l_\alpha$ is increased, the fraction $\langle B_{tot}^{(hub)} \rangle_r / \langle B_{tot} \rangle_r$ for the secondary hub group decreases due to increase in the fraction $\langle B_{tot}^{(peri)} \rangle_r / \langle B_{tot} \rangle_r$ for the peripheral group. Thanks to the effect of $l_\alpha$ on $\langle L_p \rangle_r$ and $\langle B_c \rangle_r$, with increasing $l_\alpha$, typical separation between two nodes in the network becomes shorter and load of communication traffic becomes less concentrated on the head hub, which results from increase in the total number of connections. Consequently, with increase in $l_\alpha$, efficiency of global communication between nodes becomes better, which may lead to increase in the degree of burst and spike synchronization.

We note that individual dynamics vary depending on the synaptic inputs with the in-degree $d^{(in)}$ of Eq.~(\ref{eq:CHRD}). Hence, the in-degree distribution affects MBRs (mean bursting rates) and MSRs (mean spiking rates) of individual neurons. Figures \ref{fig:SA2}(g1)-\ref{fig:SA2}(g5) show plots of MBRs of individual neurons versus $d^{(in)}$ for $l_\alpha=5$, 15, 20, 35, and 45, respectively. We first consider the case of $l_\alpha=5$. Since the in-degree of a peripheral neuron is small, its pre-synaptic neurons belong to a small subset of the whole population. MBRs of the peripheral neurons change depending on the average MBR of pre-synaptic neurons in the small subset. If MBRs of the pre-synaptic neurons are fast (slow) on average, then the post-synaptic peripheral neuron receives more (less) synaptic inhibition, and hence its  MBR becomes slow (fast). Consequently, MBRs of the peripheral neurons are broadly distributed around the ensemble-averaged horizontal gray line of $\langle f_i^{(b)} \rangle \simeq 1.78$ Hz. The average MBR $\langle f_i^{(b)} \rangle_{peri} (\simeq 1.80$ Hz) of peripheral neurons is a little faster than $\langle f_i^{(b)} \rangle$ because MBRs of the peripheral neurons are distributed a little more above the gray line. On the other hand, the pre-synaptic neurons of a hub neuron with higher in-degree belong to a relatively larger sub-population of the whole network, which results in reduced variation in the synaptic inhibitions received by the hub neurons. Hence, the distribution of MBRs of the hub neurons becomes a little reduced.  Moreover, since $\langle f_i^{(b)} \rangle_{peri} > \langle f_i^{(b)} \rangle$, the average MBR $\langle f_i^{(b)} \rangle_{hub} (\simeq 1.62$ Hz) of hub neurons becomes slower than the ensemble-averaged MBR $\langle f_i^{(b)} \rangle$. However, as $l_\alpha$ is increased from 5, the total number of inward connections increases, which favors the pacing between neurons. Then, the ensemble-averaged MBRs $\langle f_i^{(b)} \rangle$ increase, while variations in MBRs from $\langle f_i^{(b)} \rangle$ decrease [see Figs.~\ref{fig:SA2}(g1)-\ref{fig:SA2}(g5)]. For the case of MSRs of individual neurons, with increasing $l_\alpha$ similar tendency for changes in both the ensemble-averaged MSRs $\langle f_i^{(s)} \rangle$ and variations in MSRs of neurons from $\langle f_i^{(s)} \rangle$ also occurs, as shown in Fig.~\ref{fig:SA2}(h1)-\ref{fig:SA2}(h5). Particularly, this kind of changes appear distinctly for both cases of burst [$l_\alpha > l_{\alpha,l}^* (\simeq 9)$] and (intraburst) spike [$l_\alpha > l_{\alpha,h}^* (\simeq 28)$] synchronization because individual neurons receive more coherent inputs; for both cases of burst unsynchronization and intraburst spike unsynchronization only a little changes occur. Figures \ref{fig:SA2}(i)-\ref{fig:SA2}(k) show the average occupation degrees of bursting onset times $\langle O_b^{(on)} \rangle_r$  and spikings $\langle O_s \rangle_r$, the average pacing degrees of bursting onset times $\langle P_b^{(on)} \rangle_r$  and spikings $\langle P_s \rangle_r$, and the statistical-mechanical bursting measure $\langle M_b^{(on)} \rangle_r$ and spiking measure $\langle M_s \rangle_r$ versus $l_\alpha$; open circles represent data for burstings, while solid circles denote data for spikings. For the case of burst synchronization, with increasing $l_\alpha$ the variation in MBRs decreases, and hence $\langle P_b^{(on)} \rangle_r$ increases distinctly. On the other hand, $\langle O_b^{(on)} \rangle_r$  makes a little increase around 0.33 because of a slight increase in the ensemble-averaged MBR $\langle f_i^{(b)} \rangle$. Then, $\langle M_b^{(on)} \rangle_r$ (given by the product of the occupation and the pacing degrees) also increases markedly like the case of $\langle P_b^{(on)} \rangle_r$. For the case of (intraburst) spike synchronization, $\langle O_s \rangle_r$ also makes a little increase around 0.22, and hence only a fraction (about 2/3) of neurons showing bursting activity fire spikings in the intraburst spiking cycles. On the other hand, with increasing $l_\alpha$ $\langle P_s \rangle_r$ makes a distinct increase. But, it is much less than $\langle P_b^{(on)} \rangle_r$, because both the ensemble-averaged MSR $\langle f_i^{(s)} \rangle$ and the variation in MSRs from $\langle f_i^{(s)} \rangle$ are much larger than those for the bursting case. Consequently, the degree of spike synchronization ${\langle M_s \rangle_r} (\sim 0.22~ {\langle P_s \rangle_r})$ is also much less than the degree of burst synchronization $\langle M_b^{(on)} \rangle_r (\sim 0.33~{\langle P_b^{(on)} \rangle_r})$.

\subsubsection{2nd Case of Network Architecture: Varying The Asymmetry Parameter $\Delta l_\alpha$}
\label{subsubsec:2nd}
\begin{figure}
\includegraphics[width=0.8\columnwidth]{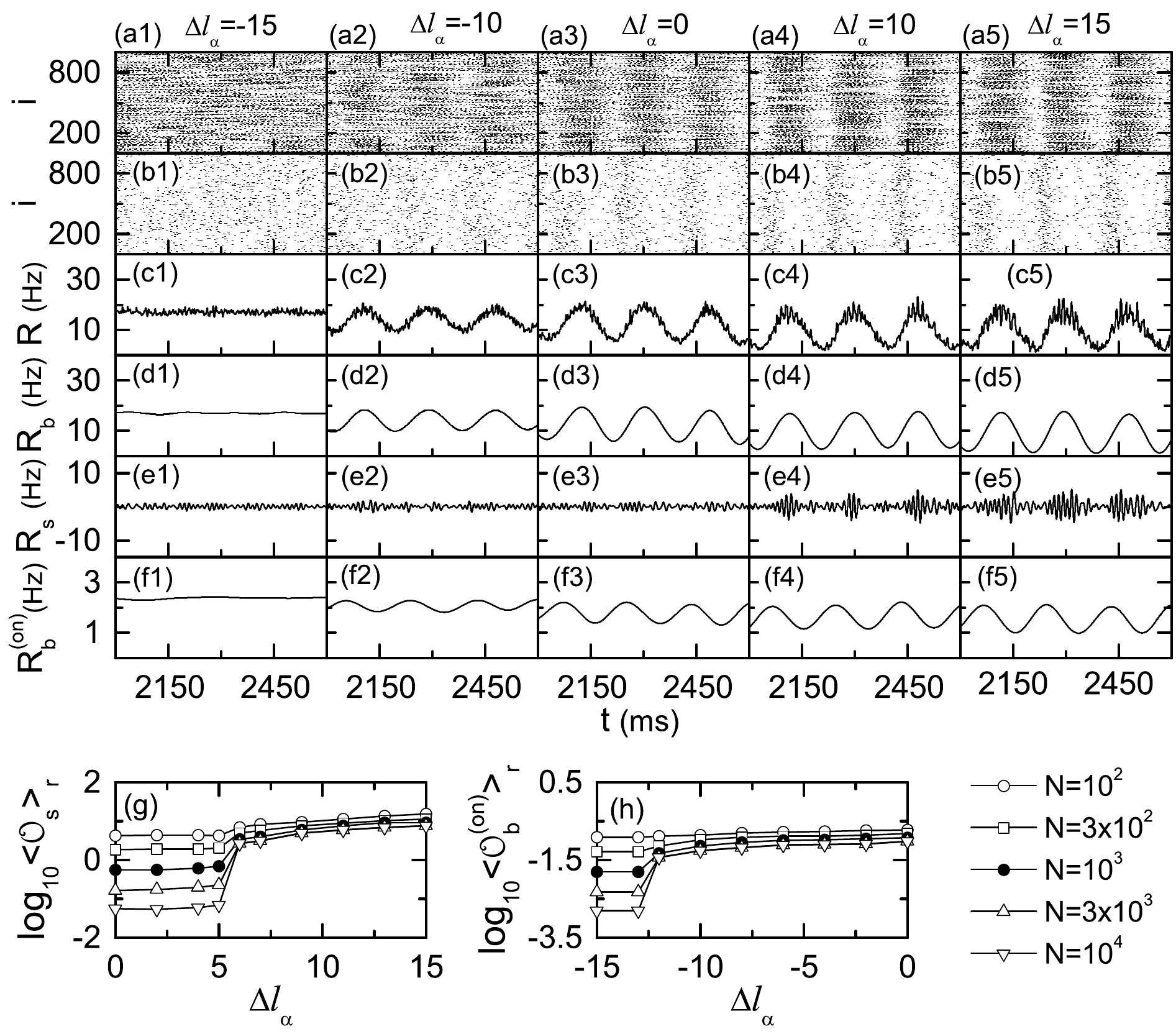}
\caption{
Emergence of burst and spike synchronization for various values of asymmetric parameter $\Delta l_{\alpha}$ in the pure $\alpha-$process ($\alpha=1$): unsynchronization for $\Delta l_{\alpha}=-15$, burst synchronization for $\Delta l_{\alpha}=0$ and 15, and complete synchronization (including both burst and intraburst spike synchronization) for  $\Delta l_{\alpha}=10$ and 15. (a1)-(a5) Raster plots of spikes, (b1)-(b5) raster plots of bursting onset times, (c1)-(c5) plots of IPFR kernel estimates $R(t)$, (d1)-(d5) plots of low-pass filtered IPBRs $R_b(t)$, (e1)-(e5) plots of band-pass filtered IPSRs $R_s(t)$, and (f1)-(f5) plots of IPBR kernel estimates $R_b^{(on)}(t)$ for various values of $\Delta l_{\alpha}$=-15, -10, 0, 10, and 15. Plots of (g) spiking order parameter $\langle {\cal{O}}_s \rangle_r$ and (h) bursting order parameter $\langle {\cal{O}}_b^{(on)} \rangle_r$  versus $\Delta l_{\alpha}$.
}
\label{fig:AA1}
\end{figure}

As the second case of network architecture, we consider the case of asymmetric preferential attachment $l_{\alpha}^{(in)} \neq l_{\alpha}^{(out)}$. We set $l_{\alpha}^{(in)}=  \widetilde{l_\alpha}+\Delta l_{\alpha}$ and ${l_\alpha}^{(out)}= \widetilde{l_\alpha} - \Delta l_{\alpha}$ such that $l_{\alpha}^{(in)} + l_{\alpha}^{(out)} = 2\,\widetilde{l_\alpha}=$ constant (i.e., the total number of in- and out-connections is fixed), and investigate the effect of asymmetric attachment on the burst  synchronization [which occurs for $J=4$, $D=0.06$, and $l_\alpha=\widetilde{l_\alpha} (=20)$] by varying the asymmetry parameter $\Delta l_{\alpha}$. Raster plots of spikes for $\Delta l_\alpha=-15$, -10, 0, 10, and 15 are shown in Figs.~\ref{fig:AA1}(a1)-\ref{fig:AA1}(a5), respectively.  For more direct visualization of bursting behavior, Figs.~\ref{fig:AA1}(b1)-\ref{fig:AA1}(b5) also show the raster plots of bursting onset times for the same values of $\Delta l_\alpha$. From the raster plots of spikes, we obtain IPFR kernel estimates $R(t)$ (showing the whole combined behaviors including both burst and spike synchronization) in Figs.~\ref{fig:AA1}(c1)-\ref{fig:AA1}(c5). Through frequency filtering, we separate the slow bursting and the fast spiking timescales, and decompose $R(t)$ into the IPBR $R_b(t)$ and the IPSR $R_s(t)$ which exhibit the bursting and the spiking behaviors separately. Figures \ref{fig:AA1}(d1)-\ref{fig:AA1}(d5) and Figs.~\ref{fig:AA1}(e1)-\ref{fig:AA1}(e5) show $R_b(t)$ and $R_s(t)$ for various values of $\Delta l_\alpha$, respectively. For bursting behavior, the IPBR kernel estimate $R_b^{(on)}(t)$ are also obtained from the raster plots of bursting onset times, and shown in Figs.~\ref{fig:AA1}(f1)-\ref{fig:AA1}(f5). As the asymmetry parameter $\Delta l_\alpha$ is increased from the symmetric case of $\Delta l_\alpha=0$, both bursting bands in the raster plots of spikes and bursting stripes in the raster plots of bursting onset times become more clear. Therefore, the amplitudes of $R_b(t)$ and $R_b^{(on)}(t)$ for $\Delta l_\alpha = 10$ and 15 become larger than those for $\Delta l_\alpha = 0$. Moreover, when passing a higher spiking threshold $\Delta l_{\alpha,h}^*$, a transition to intraburst spike-synchronization occurs. This type of spike-synchronized states are well shown in the fast-oscillating IPSRs $R_s(t)$ for $\Delta l_\alpha=10$ and 15. For these cases, the spiking amplitude of $R_s(t)$ (representing the degree of spike synchronization) for $\Delta l_\alpha=15$ is larger than that for $\Delta l_\alpha=10$. Consequently, as $\Delta l_\alpha$ is increased from $\Delta l_\alpha=0$, the burst and spike synchronization becomes better. In contrast, with decreasing $\Delta l_\alpha$ from 0 the amplitudes of $R_b(t)$ and $R_b^{(on)}(t)$ decrease, as shown in the case of $\Delta l_\alpha=-10$. Eventually, when passing a lower bursting threshold $\Delta l_{\alpha,l}^*$, unsynchronized states appear. As an example of unsynchronization, see the case of $\Delta l_\alpha=-15$ where $R_b(t)$ and $R_b^{(on)}(t)$ are nearly stationary. For these spiking and bursting transitions, we use both the spiking order parameter $\langle {\cal{O}}_s \rangle_r$ of Eq.~(\ref{eq:SO2}) and the bursting order parameter $\langle {\cal{O}}_b^{(on)} \rangle_r$ of Eq.~(\ref{eq:Border2}) to determine the higher and the lower thresholds, $l_{\alpha,h}^*$ and $l_{\alpha,l}^*$, respectively. Figures \ref{fig:AA1}(g) and \ref{fig:AA1}(h) show plots of $\langle {\cal{O}}_s \rangle_r$ and $\langle {\cal{O}}_b^{(on)} \rangle_r$ versus $\Delta l_\alpha$, respectively. As $\Delta l_\alpha$ is increased and passes the higher spiking threshold $\Delta l_{\alpha,h}^* (\simeq 6)$, a transition to intraburst spike synchronization takes place because ${\langle {\cal{O}}_s \rangle}_r$ saturates to non-zero limit values as $N$ is increased to $\infty$. As a result, for $\Delta l_\alpha > \Delta l_{\alpha,h}^*$ complete synchronization (including both burst and spike synchronization) emerges. On the other hand, as $l_\alpha$ is decreased and passes the lower threshold $\Delta l_{\alpha,l}^* (\simeq -12)$, a transition from burst synchronization to unsynchronization occurs because the values of $\langle {\cal{O}}_b^{(on)} \rangle_r$ tend to zero in the thermodynamic limit of $N \rightarrow \infty$.

\begin{figure}
\includegraphics[width=0.75\columnwidth]{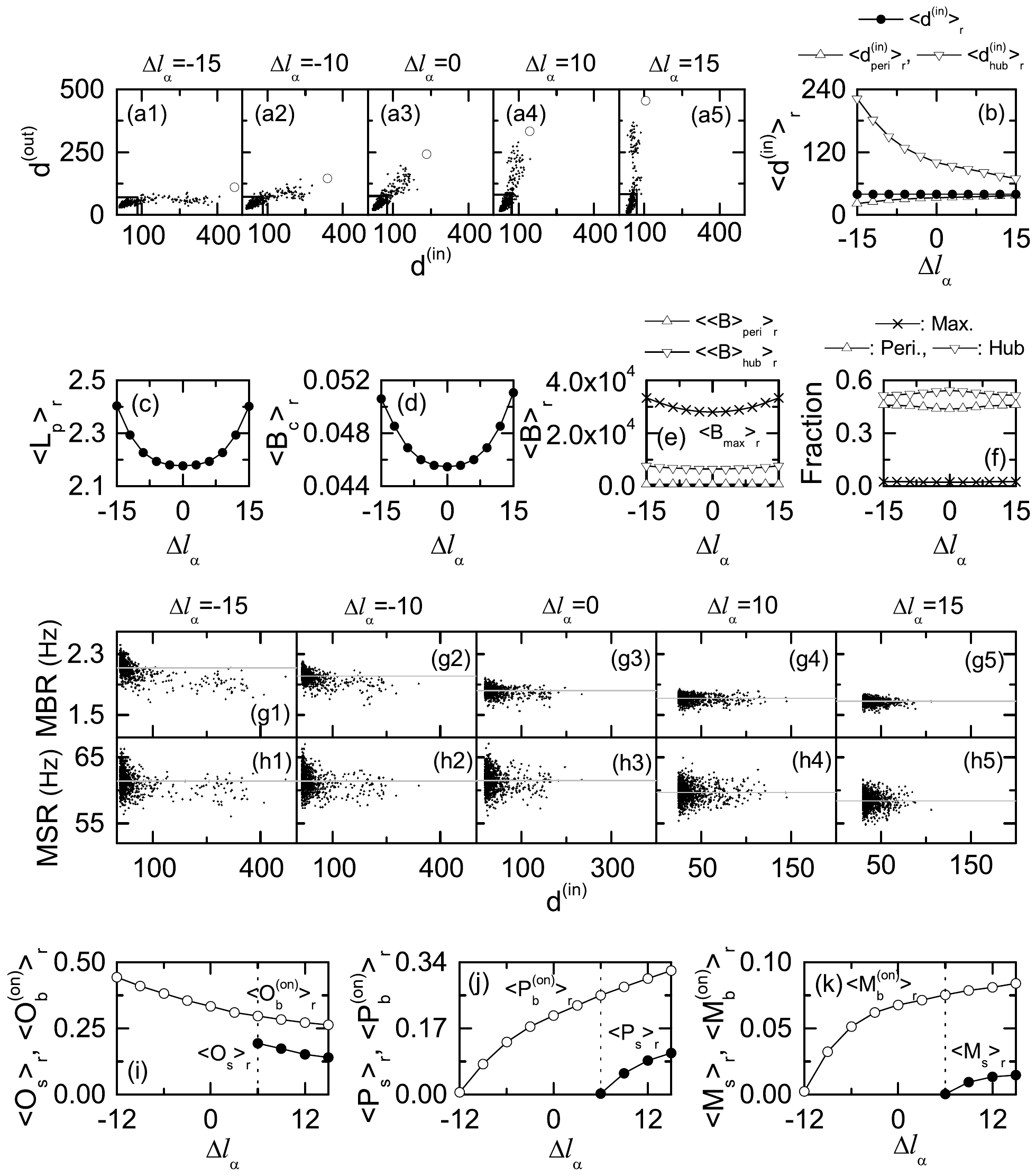}
\linespread{1}
\caption{
Effect of the asymmetric parameter $\Delta l_{\alpha}$ on the degree of burst and spike synchronization in the pure $\alpha-$process ($\alpha=1$). Plots of the out-degree $d^{(out)}$ versus the in-degree $d^{(in)}$ for $\Delta l_{\alpha}=$ (a1) -15, (a2) -10, (a3) 0, (a4) 10, and (a5) 15. (b) Plots of the ensemble-averaged in-degree $\langle d^{(in)} \rangle_r$ in the whole population, the average in-degree $\langle d^{(in)}_{peri} \rangle_r$ in the group of peripheral nodes, and the average in-degree $\langle d^{(in)}_{hub} \rangle_r$ in the group of secondary hubs versus $\Delta l_{\alpha}$. (c) Plot of the average path length $\langle L_p \rangle_r$ versus $\Delta l_{\alpha}$. (d) Plots of the maximum betweenness centrality $\langle B_{max} \rangle_r$, the average betweenness centrality $\langle \langle B \rangle_{hub} \rangle_r$ of secondary hubs, and the average betweenness centrality $\langle \langle B \rangle_{peri} \rangle_r$ of peripheral nodes versus $\Delta l_{\alpha}$. (e) Plot of  the betweenness centralization $\langle B_c \rangle_r$  versus  $\Delta l_{\alpha}$. (f) Plots of fractions $\langle B_{max} \rangle_r / \langle B_{tot} \rangle_r$, $\langle B_{tot}^{(hub)} \rangle_r / \langle B_{tot} \rangle_r$, and $\langle B_{tot}^{(peri)} \rangle_r / \langle B_{tot} \rangle_r$ versus $\Delta l_{\alpha}$. Here, quantities in (b)-(f) are obtained via 20 realizations. Plots of MBRs of individual neurons versus $\Delta l_{\alpha}$ for $\Delta l_{\alpha}=$ (g1) -15, (g2) -10, (g3) 0, (g4) 10, and (g5) 15. Plots of MSRs of individual neurons for $\Delta l_{\alpha}=$ (h1) -15, (h2) -10, (h3) 0, (h4) 10, and (h5) 15. Plots of (i) the average occupation degrees of bursting onset times $\langle O_b^{(on)} \rangle_r$  and spikings $\langle O_s \rangle_r$, (j) the average pacing degrees of bursting onset times $\langle P_b^{(on)} \rangle_r$ and spikings $\langle P_s \rangle_r$, and (k) the statistical-mechanical bursting measure $\langle M_b^{(on)} \rangle_r$ and spiking measure $\langle M_s \rangle_r$ versus $\Delta l_{\alpha}$. Data for $\langle O_b^{(on)} \rangle_r$, $\langle P_b^{(on)} \rangle_r$, and $\langle M_b^{(on)} \rangle_r$ are denoted by open circles, while those for $\langle O_s \rangle_r$, $\langle P_s \rangle_r$, and $\langle M_s \rangle_r$ are represented by solid circles. The vertical dotted line represents the higher spiking threshold $\Delta l_{\alpha,h}^* (\simeq 6)$.
}
\label{fig:AA2}
\end{figure}

To study the relation between network topology and population synchronization, we consider the effect of the asymmetry attachment on the network topology by varying $\Delta l_\alpha$. Plots of the out-degree $d^{(out)}$ versus the in-degree $d^{(in)}$ for $\Delta l_\alpha=-15$, -10, 0, 10, and 15 are shown in Figs.~\ref{fig:AA2}(a1)-\ref{fig:AA2}(a5), respectively. A majority of peripheral nodes with lower degrees are enclosed by rectangles, while hubs with higher degrees lie outside the rectangles (particularly, the head hub is denoted by the open circle). For the case of symmetric attachment with $\Delta l_\alpha=0$, the in- and out-degrees of the hubs and the peripheral nodes are distributed nearly symmetrically around the diagonal. On the other hand, the degree distributions vary significantly in the case of asymmetric attachment with $\Delta l_\alpha \neq 0$. For $\Delta l_\alpha >0$, the in-degrees of peripheral nodes are more than their out-degrees, while the out-degrees of hubs are much more than their in-degrees. The degree distributions for $\Delta l_\alpha >0$ seem to be similar to those obtained through counter-clockwise rotations of the symmetric distribution for $\Delta l_\alpha=0$ about a center, as shown in Figs.~\ref{fig:AA2}(a4)-\ref{fig:AA2}(a5). Thus, the distribution of in-degrees is narrow, while the distribution of out-degrees is wide, unlike the case of symmetric attachment. As a result, the ensemble-averaged in-degree $\langle d^{(in)} \rangle$ for $\Delta l_\alpha >0$, affecting distribution of individual MBRs and MSRs, becomes larger than that for $\Delta l_\alpha = 0$ due to increase in average in-degrees of the majority of peripheral nodes. In contrast, for $\Delta l_\alpha <0$, the in-degrees of peripheral nodes are less than their out-degrees, while the out-degrees of hubs are much less than their in-degrees. The degree distributions for $\Delta l_\alpha <0$ seem to be similar to those obtained through clockwise rotations of the symmetric distribution for $\Delta l_\alpha=0$ about a center, as shown in Figs.~\ref{fig:AA2}(a1)-\ref{fig:AA2}(a2). Hence, the distribution of in-degrees is wide, while the distribution of out-degrees is narrow. As a result of decreased average in-degrees of the majority of peripheral nodes, the ensemble-averaged in-degree $\langle d^{(in)} \rangle$ for $\Delta l_\alpha <0$, affecting distribution of individual MBRs and MSRs, becomes smaller than that for $\Delta l_\alpha = 0$. Figure \ref{fig:AA2}(b) shows plots of the average in-degree $\langle d^{(in)} \rangle_r$ (solid circles) in the whole population, the average in-degree $\langle d^{(in)}_{peri} \rangle_r$ (triangles) in the peripheral group, and the average in-degree $\langle d^{(in)}_{hub} \rangle_r$ (inverted triangles) in the hub group versus $\Delta l_\alpha$. With increasing $\Delta l_\alpha$, $\langle d^{(in)}_{peri} \rangle_r$ increases slowly, while $\langle d^{(in)}_{hub} \rangle_r$ decreases rapidly. Since the peripheral group is a majority one, $\langle d^{(in)} \rangle_r$ lies a little above $\langle d^{(in)}_{peri} \rangle_r$. Consequently, as $\Delta l_\alpha$ is increased, the distribution of in-degrees becomes narrower, which also implies that, with increasing $\Delta l_\alpha$ the distribution of out-degrees becomes wider because the total number of in- and out-connections is fixed. In this way, as $| \Delta l_\alpha |$ (magnitude of the asymmetry parameter $\Delta l_\alpha$) is increased, mismatching between in- and out-degrees of nodes increases, with keeping the total number of in- and out-connections constant, in contrast to the above 1st case of symmetric attachment where the total number of connections increases with increasing $l_\alpha$.

Next, we study the effect of $\Delta l_\alpha$ on the average path length $L_p$ of Eq.~(\ref{eq:APL}) (representing typical separation between two nodes) and the betweenness centralization $B_c$ of Eq.~(\ref{eq:BC2})
(denoting the relative degree of load of communication traffic concentrated to the head hub), both of which affect global communication between nodes. Figures \ref{fig:AA2}(c) and \ref{fig:AA2}(d) show the plots of $L_p$ and $B_c$ versus $\Delta l_{\alpha}$, respectively. As $| \Delta l_{\alpha} |$ is increased, both $L_p$ and $B_c$ increase symmetrically, independently of the sign of $\Delta l_{\alpha}$, due to increased mismatching between in- and out-degrees of nodes. Since both inward and outward links are involved equally in computation of $L_p$ and $B_c$, the values of $L_p$ and $B_c$ for both cases of different signs but the same magnitude (i.e., $\Delta l_{\alpha}$ and -$\Delta l_{\alpha}$) become the same. Such increase in $L_p$ implies enhancement of intermediate mediation of nodes controlling communication between nodes (i.e., enhancement in total betweenness $B_{tot}$).
As shown in Fig.~\ref{fig:AA2}(e), with increasing $| \Delta l_\alpha |$ the maximum betweenness $\langle B_{max} \rangle_r$ (crosses) of the head hub is much more enhanced than the average centralities of the secondary hubs and the peripheral nodes, $\langle {\langle B \rangle}_{hub} \rangle_r$ (inverted triangles) and $\langle {\langle B \rangle}_{peri} \rangle_r$ (triangles). Hence, as $| \Delta l_\alpha |$ is increased, differences between $B_{max}$ of the head hub and $B_i$ of other nodes are increased, which leads to increase in $B_c$. For large $B_c$, it is difficult to get efficient communication between nodes due to destructive interference between many signals passing through the head hub. Figure \ref{fig:AA2}(f) also shows plots of fractions $\langle B_{max} \rangle_r / \langle B_{tot} \rangle_r$ (crosses), $\langle B_{tot}^{(hub)} \rangle_r / \langle B_{tot} \rangle_r$ (inverted triangles), and $\langle B_{tot}^{(peri)} \rangle_r / \langle B_{tot} \rangle_r$ (triangles) versus $\Delta l_\alpha$. These fractions denote how the total centrality $B_{tot}$ is distributed in the head hub, the secondary hub group, and the peripheral group. As already shown in $\langle B_c \rangle_r$, with increasing $| \Delta l_\alpha |$, the fraction $\langle B_{max} \rangle_r / \langle B_{tot} \rangle_r$ for the head hub (i.e., relative load of communication traffic for the head hub) increases symmetrically. The secondary hub group has more than half load of total communication traffic due to their large average in-degree. However, with increasing $| \Delta l_\alpha |$, the fraction $\langle B_{tot}^{(hub)} \rangle_r / \langle B_{tot} \rangle_r$ for the secondary hub group decreases in a slow symmetrical way due to a slow slight increase in the fraction $\langle B_{tot}^{(peri)} \rangle_r / \langle B_{tot} \rangle_r$ for the peripheral group (resulting from a little increase in the number of peripheral nodes). As a result of effect of $\Delta l_\alpha$ on $L_p$ and $B_c$, with increasing $| \Delta l_\alpha |$ efficiency of global communication between nodes becomes worse. However, unlike the symmetric change in $L_p$ and $B_c$, population synchronization varies depending on the sign of $\Delta l_\alpha$. As an example, see both cases of $\Delta l_\alpha=10$ and -10 in Fig.~\ref{fig:AA1}. Their population synchronization is different because of distinctly different in-degree distributions affecting MBRs and MSRs of individual neurons, although their $L_p$ and $B_c$ are the same.

As shown in the above case of symmetric attachment in Fig.~\ref{fig:SA2}, the in-degree distributions affect MBRs and MSRs of individual neurons. Figures \ref{fig:AA2}(g1)-\ref{fig:AA2}(g5) [Figs.~\ref{fig:AA2}(h1)-\ref{fig:AA2}(h5)] show plots of MBRs (MSRs) of individual neurons versus $d^{(in)}$ for $\Delta l_\alpha=-15$, -10, 0, 10, and 15, respectively. As $\Delta l_\alpha$ is increased from 0, the distributions of in-degrees become narrowed, which results in decrease in variations in MBRs (MSRs). In addition, with increasing $\Delta l_\alpha$ from 0, the ensemble-averaged MBRs $\langle f_i^{(b)} \rangle$ [MSRs $\langle f_i^{(s)} \rangle$] (represented by horizontal gray lines) also decrease due to increase in average inhibition given to individual neurons (resulting from increased ensemble-averaged in-degrees). On the other hand, with decreasing $\Delta l_\alpha$ from 0, the distributions of in-degrees become broadened, and hence variations in MBRs (MSRs) increase. Moreover, as $\Delta l_\alpha$ is decreased from 0, the ensemble-averaged MBRs $\langle f_i^{(b)} \rangle$ [MSRs $\langle f_i^{(s)} \rangle$] increase because of decrease in average inhibition given to neurons (resulting from decreased ensemble-averaged in-degrees). This kind of changes occur distinctly for both cases of burst [$\Delta l_\alpha > \Delta l_{\alpha,l}^* (\simeq -12)$] and (intraburst) spike [$\Delta l_\alpha > \Delta l_{\alpha,h}^* (\simeq 6)$] synchronization because individual neurons receive more coherent inputs; in both cases of burst unsynchronization and intraburst spike unsynchronization only a little changes appear. Based on these changes in MBRs and MSRs, we discuss their effects on the pacing and occupation degrees for both cases of burst and spike synchronization. Figures \ref{fig:AA2}(i)-\ref{fig:AA2}(k) show the average occupation degrees of bursting onset times $\langle O_b^{(on)} \rangle_r$  and spikings $\langle O_s \rangle_r$, the average pacing degrees of bursting onset times $\langle P_b^{(on)} \rangle_r$  and spikings $\langle P_s \rangle_r$, and the statistical-mechanical bursting measure $\langle M_b^{(on)} \rangle_r$ and spiking measure $\langle M_s \rangle_r$ versus $\Delta l_\alpha$. Here, open circles denote data for burstings, while solid circles represent data for spikings. For the case of burst synchronization, with increasing $\Delta l_\alpha$ the variation in MBRs decreases, and hence $\langle P_b^{(on)} \rangle_r$ increases. On the other hand, $\langle O_b^{(on)} \rangle_r$ decreases due to decreased ensemble-averaged MBRs. However, $\langle P_b^{(on)} \rangle_r$ increases more rapidly than decrease in $\langle O_b^{(on)} \rangle_r$. Consequently, with increase in $\Delta l_\alpha$, $\langle M_b^{(on)} \rangle_r$ (given by the product of $\langle O_b^{(on)} \rangle_r$ and $\langle P_b^{(on)} \rangle_r$) also increases. For the case of intraburst spike synchronization, as $\Delta l_\alpha$ is increased, $\langle O_s \rangle_r$ also decreases because of decrease in ensemble-averaged MSR. Since it is less than $\langle O_b^{(on)} \rangle_r$, only a fraction of neurons exhibiting bursting activity fire spikings in the intraburst spiking cycles. On the other hand, with increasing $\Delta l_\alpha$ $\langle P_s \rangle_r$ makes an increase due to decreased variation in MSRs. But, it is much less than $\langle P_b^{(on)} \rangle_r$, because both the ensemble-averaged MSR $\langle f_i^{(s)} \rangle$ and the variation in MSRs from $\langle f_i^{(s)} \rangle$ are much larger than those for the bursting case. Like the bursting case, $\langle P_s \rangle_r$ also increases more rapidly than decrease in $\langle O_s \rangle_r$, and hence with increasing $\Delta l_\alpha$ $\langle M_s \rangle_r$ also increases. However, the degree of spike synchronization ${\langle M_s \rangle_r}$ is much less than the degree of burst synchronization $\langle M_b^{(on)} \rangle_r$.

\subsubsection{3rd Case of Network Architecture: Varying $\beta$}
\label{subsubsec:3rd}
\begin{figure}
\includegraphics[width=0.8\columnwidth]{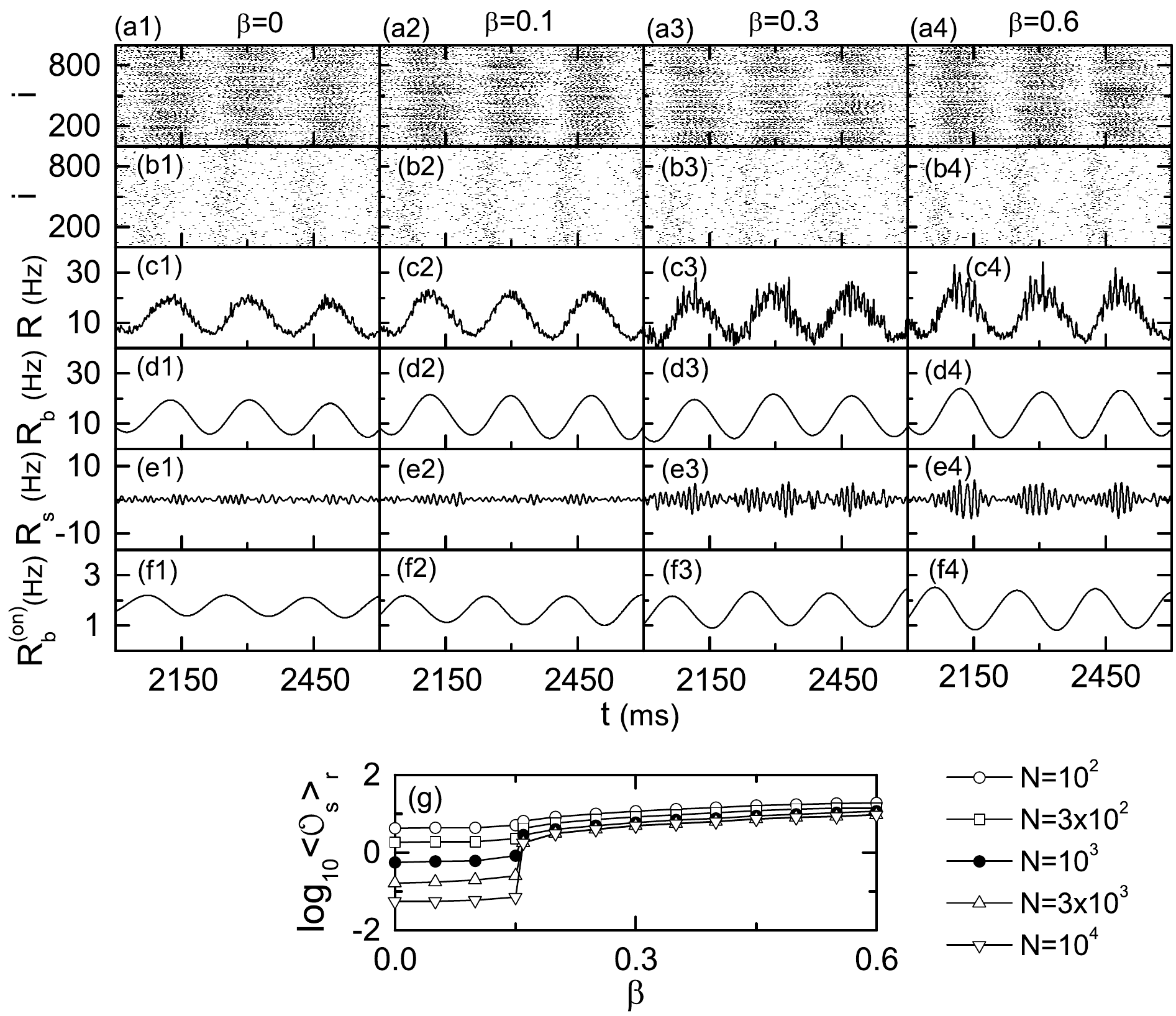}
\caption{
Emergence of burst and spike synchronization for various values of the probability $\beta$ in the $\beta-$process: burst synchronization for $\beta=0$ and 0.1, and complete synchronization (including both burst and intraburst spike synchronization) for $\beta=0.3$ and 0.6. (a1)-(a5) Raster plots of spikes, (b1)-(b5) raster plots of bursting onset times, (c1)-(c5) plots of IPFR kernel estimates $R(t)$, (d1)-(d5) plots of low-pass filtered IPBRs $R_b(t)$, (e1)-(e5) plots of band-pass filtered IPSRs $R_s(t)$, and (f1)-(f5) plots of IPBR kernel estimates $R_b^{(on)}(t)$ for various values of $\beta=0$, 0.1, 0.3, and 0.6. (g) Plots of spiking order parameter $\langle {\cal{O}}_s \rangle_r$ versus $\beta$.
}
\label{fig:beta1}
\end{figure}

As the third case of network architecture, along with the above $\alpha$-process (occurring with the probability $\alpha$) we consider the $\beta$-process (occurring with the probability $\beta$; $\alpha + \beta =1$). Unlike the case of $\alpha$-process, no new nodes are added, and symmetric preferential attachments with the same in- and out-degrees [$l_{\beta}^{(in)} = l_{\beta}^{(out)} (\equiv l_{\beta}$)] are made between $l_{\beta}$ pairs of (pre-existing) source and target nodes which are also preferentially chosen according to the attachment probabilities $\Pi_{source}(d_i^{(out)})$ and $\Pi_{target}(d_i^{(in)})$ of Eq.~(\ref{eq:AP}), respectively, such that self-connections (i.e., loops) and duplicate connections (i.e., multiple edges) are excluded. Here we set $l_\beta=5$ (see the fifth item in Table \ref{tab:Parm}). We investigate the effect of the $\beta$-process on the burst synchronization [which occurs $J=4$ and $D=0.06$ in the pure $\alpha-$process with symmetric preferential attachment ($l_\alpha^{(in)} = l_\alpha^{(out)} \equiv  l_\alpha = 20)]$ by varying $\beta$. Figures \ref{fig:beta1}(a1)-\ref{fig:beta1}(a4) [Figs.~\ref{fig:beta1}(b1)-\ref{fig:beta1}(b4)] show raster plots of spikes (raster plots of bursting onset times) for $\beta=0$, 0.1, 0.3, and 0.6, respectively. Their corresponding IPFR (IPBR) kernel estimates, $R(t)$ [$R_b^{(on)}(t)$] are also shown in Figs.~\ref{fig:beta1}(c1)-\ref{fig:beta1}(c4) [Figs.~\ref{fig:beta1}(f1)-\ref{fig:beta1}(f4)] for various values of $\beta$, respectively. Through frequency filtering, we decompose $R(t)$ into the IPBR $R_b(t)$ and the IPSR $R_s(t)$, which exhibit the bursting and the spiking behaviors separately and are shown in Figs.~\ref{fig:beta1}(d1)-\ref{fig:beta1}(d4) and Figs.~\ref{fig:beta1}(e1)-\ref{fig:beta1}(e4), respectively. As $\beta$ is increased, the bursting bands in the raster plots of spikes (the bursting stripes in the raster plots of bursting onset times) become more clear, and hence the amplitude of $R_b(t)$ [$R_b^{(on)}(t)$] increases. Furthermore, when passing a spiking threshold $\beta^*$, a transition to intraburst spike-synchronization occurs. This kind of spike synchronization is well shown in the fast-oscillating IPSR $R_s(t)$ for $\beta=0.3$ and 0.6. To determine $\beta^*$, we employ the spiking order parameter $\langle {\cal{O}}_s \rangle_r$ of Eq.~(\ref{eq:SO2}). Figure \ref{fig:beta1}(g) shows plots of $\langle {\cal{O}}_s \rangle_r$ versus $\beta$. As $\beta$ passes the spiking threshold $\beta^* (\simeq 0.16)$, a transition to intraburst spike synchronization occurs because ${\langle {\cal{O}}_s \rangle}_r$ saturates to non-zero limit values as $N$ is increased to $\infty$. Consequently, for $\beta > \beta^*$ complete synchronization (including both burst and spike synchronization) emerges.

\begin{figure}
\includegraphics[width=0.8\columnwidth]{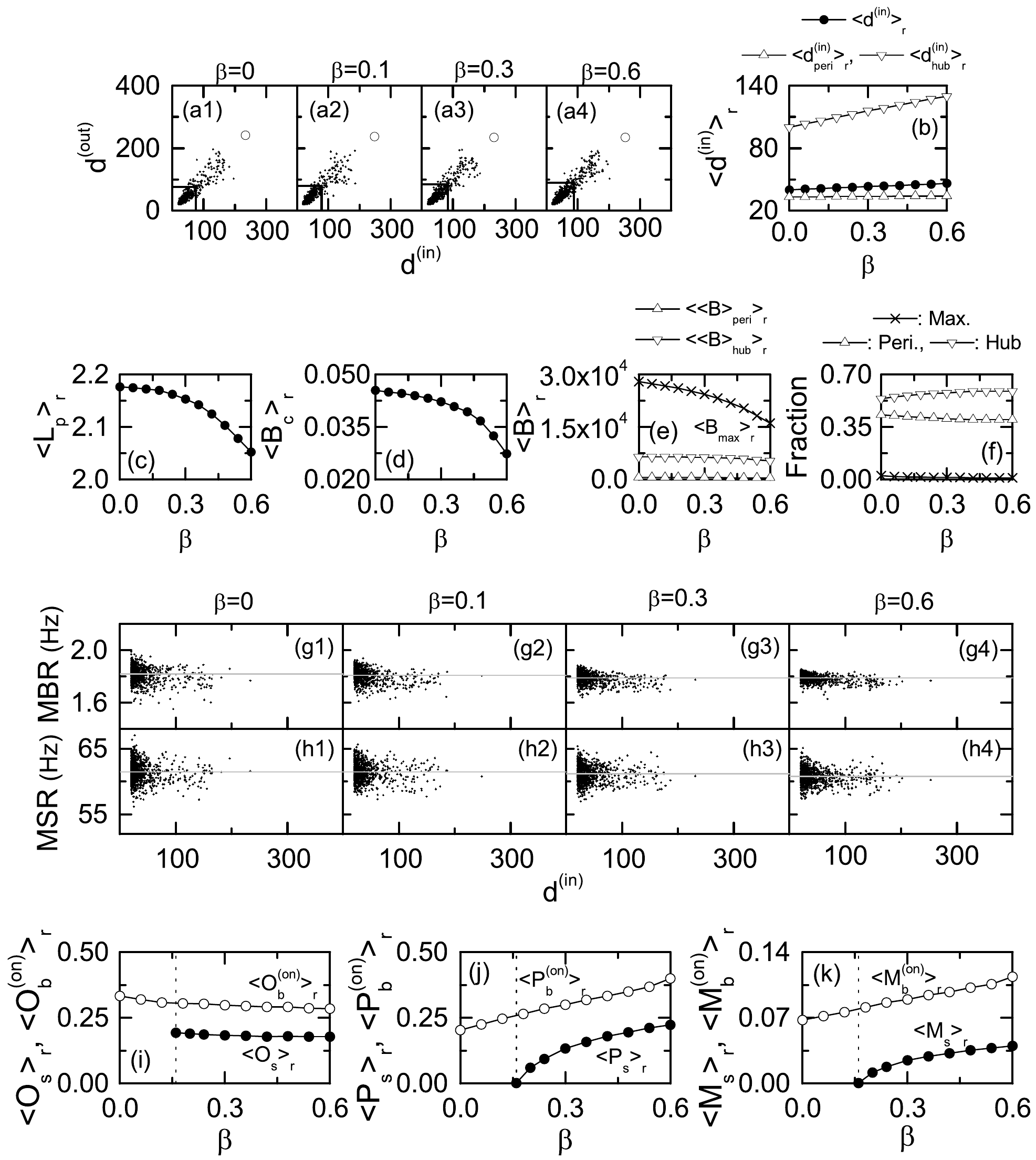}
\linespread{1.1}
\caption{
Effect of the $\beta-$process on the degree of burst and spike synchronization. Plots of the out-degree $d^{(out)}$ versus the in-degree $d^{(in)}$ for $\beta=$  (a1) 0, (a2) 0.1, (a3) 0.3, and (a4) 0.6. (b) Plots of the ensemble-averaged in-degree $\langle d^{(in)} \rangle_r$ in the whole population, the average in-degree $\langle d^{(in)}_{peri} \rangle_r$ in the group of peripheral nodes, and the average in-degree $\langle d^{(in)}_{hub} \rangle_r$ in the group of secondary hubs versus $\beta$. (c) Plot of the average path length $\langle L_p \rangle_r$ versus $\beta$. (d) Plots of the maximum betweenness centrality $\langle B_{max} \rangle_r$, the average betweenness centrality $\langle \langle B \rangle_{hub} \rangle_r$ of secondary hubs, and the average betweenness centrality $\langle \langle B \rangle_{peri} \rangle_r$ of peripheral nodes versus $\beta$. (e) Plot of the betweenness centralization $\langle B_c \rangle_r$  versus  $\beta$. (f) Plots of fractions $\langle B_{max} \rangle_r / \langle B_{tot} \rangle_r$, $\langle B_{tot}^{(hub)} \rangle_r / \langle B_{tot} \rangle_r$, and $\langle B_{tot}^{(peri)} \rangle_r / \langle B_{tot} \rangle_r$ versus $\beta$. Here, quantities in (b)-(f) are obtained via 20 realizations.
Plots of MBRs of individual neurons for $\beta=$ (g1) 0, (g2) 0.1, (g3) 0.3, and (g4) 0.6. Plots of MSRs of individual neurons for $\beta=$ (h1) 0, (h2) 0.1, (h3) 0.3, and (h4) 0.6. Plots of (i) the average occupation degrees of bursting onset times $\langle O_b^{(on)} \rangle_r$  and spikings $\langle O_s \rangle_r$, (j) the average pacing degrees of bursting onset times $\langle P_b^{(on)} \rangle_r$ and spikings $\langle P_s \rangle_r$, and (k) the statistical-mechanical bursting measure $\langle M_b^{(on)} \rangle_r$ and spiking measure $\langle M_s \rangle_r$ versus $\beta$. Data for $\langle O_b^{(on)} \rangle_r$, $\langle P_b^{(on)} \rangle_r$, and $\langle M_b^{(on)} \rangle_r$ are denoted by open circles, while those for $\langle O_s \rangle_r$, $\langle P_s \rangle_r$, and $\langle M_s \rangle_r$ are represented by solid circles. The vertical dotted line represents the spiking threshold $\beta^* (\simeq 0.16)$.
}
\label{fig:beta2}
\end{figure}

We also study the effect of the network topology on burst and spike synchronization by varying $\beta$. Figures \ref{fig:beta2}(a1)-\ref{fig:beta2}(a4) show ``comet-shaped'' plots of the out-degree $d^{(out)}$ versus the in-degree $d^{(in)}$ for $\beta=$0, 0.1, 0.3, and 0.6, respectively. For each $\beta$, peripheral nodes (corresponding to the coma part of the comet) are enclosed by the rectangle, while hubs (corresponding to the tail part of the comet) lie outside the rectangle and the head hub (node 1) with the highest degree is denoted by the open circle. In the $\beta-$process, the probability that the head hub with the highest degree may be chosen as a source and/or a target node is low because self-connections and duplicate connections are excluded. Such probability for the peripheral node with the lowest degree is also low because its degree is lowest. Hence, in the
$\beta-$process, there is no particular change in the degrees at both ends (with the highest and the lowest degrees) of the comet-shaped degree distribution, in contrast to the first case of symmetric attachment in the $\alpha-$process (see Fig.~\ref{fig:SA2}). On the other hand, there is distinct increase in the degrees of some (pre-existing) peripheral nodes and secondary hubs through the $\beta-$process, which leads to the immigration of some peripheral nodes into the secondary hub group. As a result, with increasing $\beta$ the tail part of the comet (i.e. corresponding to the secondary group) is particularly intensified than the coma part of the comet (corresponding to the peripheral group), because the number of links of secondary hubs is much more increased than those of peripheral nodes. Figure \ref{fig:beta2}(b) shows plots of the average in-degree $\langle d^{(in)} \rangle_r$ (solid circles) in the whole population, the average in-degree $\langle d^{(in)}_{peri} \rangle_r$ (triangles) in the peripheral group, and the average in-degree $\langle d^{(in)}_{hub} \rangle_r$ (inverted triangles) in the hub group   versus $\beta$. With increasing $\beta$, $\langle d^{(in)}_{hub} \rangle_r$ increases rapidly in comparison to the slow increase in $\langle d^{(in)}_{peri} \rangle_r$, in contrast to the case of the 1st case of network architecture where both the hub group and the peripheral group are intensified in a similar rate with increasing $l_\alpha$ [compare with Fig.~\ref{fig:SA2}(b)]. In this way, the secondary hub group is much intensified than the peripheral group through the $\beta-$process. Hence, as $\beta$ is increased, the ensemble-averaged degree $\langle d^{(in)} \rangle_r$ in the whole population also increases slowly because of slow increase in the average in-degree in the majority group of peripheral nodes. Consequently, with increasing $\beta$, the total number of connections increases slowly.

In addition to the degree distribution, we also measure the average path length $L_p$ and the betweenness centralization $B_c$ by varying $\beta$. Figures \ref{fig:beta2}(c) and \ref{fig:beta2}(d) show the plots of $L_p$ and $B_c$ versus $\beta$, respectively. As $\beta$ is increased, both $L_p$ and $B_c$ decrease in a slow monotonic way due to slow increase in the total number of connections. As explained above, decrease in $L_p$ leads to reduction in total centrality $B_{tot}$ (i.e., the sum of centralities of the head hub, the secondary hubs, and the peripheral nodes). How $B_{tot}$ decreases with increase in $\beta$ can be seen explicitly in Fig.~\ref{fig:beta2}(e). We note that the maximum betweenness $\langle B_{max} \rangle_r$ (crosses) of the head hub is much more reduced than the average centralities of the secondary hubs and the peripheral nodes, $\langle {\langle B \rangle}_{hub} \rangle_r$ (inverted triangles) and $\langle {\langle B \rangle}_{peri} \rangle_r$ (triangles), which results in decrease in differences between $B_{max}$ of the head hub and $B_i$ of other nodes (i.e., decrease in $B_c$). Thus, with increasing $\beta$ relative load of communication traffic for the head hub decreases. How the total centrality $B_{tot}$ is distributed in the head hub, the secondary hub group, and the peripheral group may also been seen in the plots of fractions $\langle B_{max} \rangle_r / \langle B_{tot} \rangle_r$ (crosses), $\langle B_{tot}^{(hub)} \rangle_r / \langle B_{tot} \rangle_r$ (inverted triangles), and $\langle B_{tot}^{(peri)} \rangle_r / \langle B_{tot} \rangle_r$ (triangles) versus $l_\alpha$ in Fig.~\ref{fig:beta2}(f), where $B_{tot}^{(hub)}$ ($B_{tot}^{(peri)}$) is the total centrality in the group of secondary hubs (peripheral nodes). As already shown in $\langle B_c \rangle_r$, with increasing $\beta$ the fraction $\langle B_{max} \rangle_r / \langle B_{tot} \rangle_r$ for the head hub decreases. However, unlike the 1st case of varying $l_\alpha$, as $\beta$ is increased, the fraction $\langle B_{tot}^{(hub)} \rangle_r / \langle B_{tot} \rangle_r$ for the secondary hub group increases, while the fraction $\langle B_{tot}^{(peri)} \rangle_r / \langle B_{tot} \rangle_r$ for the peripheral group decreases, because the secondary hub group is much more intensified in the $\beta-$process. Thanks to the effect of $\beta$ on $L_p$ and $B_c$, as $\beta$ is increased, typical separation between two nodes in the network becomes shorter and load of communication traffic becomes less concentrated on the head hub, mainly due to intensified role of the secondary hub group in the $\beta-$process. As a result, with increasing $\beta$, efficiency of global communication between nodes becomes better, which may lead to increase in the degree of burst and spike synchronization.

Finally, based on the above change in the in-degree distribution, we study the effect of the $\beta-$process on the MBRs (MSRs) of individual neurons. Figures \ref{fig:beta2}(g1)-\ref{fig:beta2}(g4) [Figs.~\ref{fig:beta2}(h1)-\ref{fig:beta2}(h4)] show the distribution of MBRs (MSRs) of individual neurons versus the in-degree $d^{(in)}$ for $\beta=$0, 0.1, 0.3, and 0.6, respectively. For the case of $\beta=0$, pre-synaptic neurons of a peripheral neuron belong to a small subset of the whole population because its in-degree is small, while those of a hub neuron with higher degrees belong to a relatively large sub-population. As a result, MBRs (MSRs) of peripheral neurons are somewhat broadly distributed around the ensemble-averaged horizontal gray line, while the distribution of MBRs (MSRs) of hub neurons is a little reduced. As $\beta$ is increased from 0, the ensemble-averaged in-degree $\langle d^{(in)} \rangle_r$ increases, and then the size of the subset of pre-synaptic neurons for a typical neuron becomes larger. Consequently, with increasing $\beta$ the variation in MBRs (MSRs) of individual neurons decreases. Due to these variations in distributions of MBRs (MSRs), the ensemble-averaged MBR (MSR) in the whole population decreases slightly because of a little decrease in the average MBR (MSR) in the majority group of peripheral nodes. Based on these changes in MBRs and MSRs, we also discuss their effects on the occupation and pacing degrees for both cases of burst and spike synchronization. Figures \ref{fig:beta2}(i)-\ref{fig:beta2}(k) show the average occupation degrees of bursting onset times $\langle O_b^{(on)} \rangle_r$  and spikings $\langle O_s \rangle_r$, the average pacing degrees of bursting onset times $\langle P_b^{(on)} \rangle_r$  and spikings $\langle P_s \rangle_r$, and the statistical-mechanical bursting measure $\langle M_b^{(on)} \rangle_r$ and spiking measure $\langle M_s \rangle_r$ versus $\beta$; open circles denote data for burstings, while solid circles represent data for spikings. For the case of burst synchronization, with increasing $\beta$ the variation in MBRs decreases, and hence $\langle P_b^{(on)} \rangle_r$ increases. On the other hand, $\langle O_b^{(on)} \rangle_r$ decreases only a little due to slight decrease in the ensemble-averaged MBRs. However, $\langle P_b^{(on)} \rangle_r$ increases much more rapidly than decrease in $\langle O_b^{(on)} \rangle_r$. As a result, with increase in $\beta$, $\langle M_b^{(on)} \rangle_r$ (given by the product of $\langle O_b^{(on)} \rangle_r$ and $\langle P_b^{(on)} \rangle_r$) also increases. For the case of intraburst spike synchronization, as $\beta$ is increased, $\langle O_s \rangle_r$ also decreases very little because of slight decrease in ensemble-averaged MSRs. Since it is less than $\langle O_b^{(on)} \rangle_r$, only a fraction of neurons exhibiting bursting activity fire spikings in the intraburst spiking cycles. On the other hand, with increase in $\beta$ $\langle P_s \rangle_r$ makes an increase because of decreased variation in MSRs. But, it is much less than $\langle P_b^{(on)} \rangle_r$, because both the ensemble-averaged MSR $\langle f_i^{(s)} \rangle$ and the variation in MSRs from $\langle f_i^{(s)} \rangle$ are much larger than those for the bursting case. As in the bursting case, $\langle P_s \rangle_r$ also increases more rapidly than decrease in $\langle O_s \rangle_r$, and hence with increasing $\beta$ $\langle M_s \rangle_r$ also increases. However, the degree of spike synchronization ${\langle M_s \rangle_r}$ is much less than the degree of burst synchronization $\langle M_b^{(on)} \rangle_r$.

\section{Summary} \label{sec:SUM}
Brain networks have been found to exhibit scale-free property in the rat hippocampal networks and the human cortical functional network. This kind of SFNs are inhomogeneous ones with a few hubs (with an exceptionally large number of connections), unlike statistically homogeneous networks such as random graphs and small-world networks. As our SFN, we considered a directed variant of the Barab\'{a}si-Albert SFN model, evolved via two independent $\alpha-$ and $\beta$-processes which occur with the probabilities $\alpha$ and $\beta$, respectively, and investigated the effect of scale-free connectivity on burst and spike synchronization of bursting neurons. In addition to the $\alpha$-process (i.e., standard Barab\'{a}si-Albert SFN model with growth and preferential directed attachment), the $\beta$-process, intensifying the internal connections between pre-existing nodes without addition of new nodes, is incorporated. Our SFN is composed of suprathreshold bursting HR neurons, interacting via inhibitory GABAergic synapses. We first considered the case of ``pure'' $\alpha$-process (i.e., $\alpha=1$) with symmetric preferential attachment with the same in- and out-degrees ($l_\alpha^{(in)} = l_\alpha^{(out)} \equiv l_\alpha$), and studied emergence of burst and spike synchronization by varying the coupling strength $J$ and the noise intensity $D$ for a fixed attachment degree $\widetilde{l_\alpha} (=20)$. Thus, complete synchronization (including both burst and spike synchronization) has been found to occur within a part of the region of burst synchronization in the $J-D$ plane. For an intensive study we fixed the value of $J$ at $J=4.0$, and investigated the evolution of population states by increasing $D$. For small $D$, complete synchronization emerges. However, as $D$ passes a lower threshold $D_l^*$ the intraburst spike synchronization breaks up, and then only the burst synchronization persists. Eventually when passing a higher thershold $D_h^*$, a transition to unsynchronization occurs due to a destructive effect of noise to spoil the synchronization. For characterization of burst and spike synchronization, we employed realistic order parameters and statistical-mechanical measures, based on the IPBR and IPSR. Then, the lower and the higher thresholds, $D_l^*$ and $D_h^*$, for the spike and the burst transitions were determined in terms of the spiking and bursting order parameters, respectively. Furthermore, the degrees of burst and spike synchronization were also measured in terms of the statistical-mechanical bursting and spiking measures, respectively. Next, we also fixed the value of $D$ at $D=0.06$ where only the burst synchronization occurs in the above pure $\alpha$-process with symmetric attachment of $l_\alpha = \widetilde{l_\alpha} (=20)$, and studied the effect of scale-free connectivity on the burst synchronization by varying (1) the degree $l_\alpha$ of symmetric attachment and (2) the asymmetry parameter $\Delta l_\alpha$ of asymmetric preferential attachment of new nodes. As the degree $l_\alpha$ for the first case of symmetric preferential attachment is increased from $\widetilde{l_\alpha}$, both the average path length $L_p$ and the betweenness centralization $B_c$ have been found to decrease due to increase in the total number of connection in the SFN (resulting from increase in the average in-degrees of hubs and peripheral nodes in a similar rate). Hence, typical separation between two nodes in the network becomes shorter and load of communication traffic becomes less concentrated on the head hub. Consequently, with increasing $l_\alpha$ from $\widetilde{l_\alpha}$ the pacing degree of the burst synchronization was found to increase due to increased efficiency of global communication between nodes, and eventually complete synchronization emerged when passing a higher threshold $l_{\alpha,h}^*$. On the other hand, as $l_\alpha$ is decreased from $\widetilde{l_\alpha}$, the pacing degree of burst synchronization was found to decrease due to the increase in $L_p$  and $B_c$, and when passing a lower threshold $l_{\alpha,l}^*$, a transition to unsynchronization occurred. For the second case of asymmetric attachment, as the magnitude $| \Delta l_\alpha |$ of the asymmetry parameter is increased from the symmetric case of $\Delta l_\alpha =0$, mismatching between inward and outward edges of nodes increases, along with keeping the total number of connections to be constant. As a result, both $L_p$ and $B_c$ have been found to increase symmetrically, independently of the sign of $\Delta l_\alpha$, because the inward and the outward links are equally involved in calculation of $L_p$ and $B_c$. Hence, with increasing $| \Delta l_\alpha |$, the efficiency of global communication between nodes becomes worse. However, the population synchronization has been found to vary depending on the sign of $\Delta l_\alpha$ because of distinctly different in-degree distributions affecting MBRs and MSRs of individual neurons. With increasing $\Delta l_\alpha$ from 0 (i.e., corresponding to the case of positive $\Delta l_\alpha$), the pacing degree of burst synchronization has been found to increase due to decrease in the variation in individual MBRs, resulting from the narrowed distribution of the in-degrees. Eventually, complete synchronization emerged when passing a higher threshold $\Delta l_{\alpha,h}^*$. In contrast, with decreasing $\Delta l_\alpha$ from 0 (i.e., corresponding to the case of negative $\Delta l_\alpha$), the pacing degree of the burst synchronization has been found to decrease because of increase in the variation in individual MBRs, resulting from the broadened distribution of the in-degrees. Eventually, when passing a lower threshold $\Delta l_{\alpha,l}^*$ a transition to unsynchronization occurred. As the third case of network architecture, we considered the $\beta$-process. With increasing $\beta$, internal links between pre-existing nodes are intensified. Particularly, the secondary hub group is much intensified than the peripheral group, in contrast to the 1st case of symmetric attachment in the $\alpha$-process where both the secondary hub and the peripheral groups are intensified in a similar rate with increasing $l_\alpha$. In this way, the total number of connections in the SFN increases slowly in the $\beta-$process. Hence, as $\beta$ is increased, both $L_p$ and $B_c$ decrease slowly, which results in increase in the effectiveness of global communication between neurons. Consequently, with increasing $\beta$ the degree of burst synchronization has been found to increase, and when passing a critical value $\beta^*$, complete synchronization emerged. From the results of these three cases, it follows that not only $L_p$ and $B_c$ (affecting global communication between nodes) but also the in-degree distribution (affecting local dynamics of individual neurons) are important network factors determining the pacing degree of population synchronization of bursting neurons in SFNs. Hence, a harmony between these network factors (affecting global communication and local dynamics) seems to be essential for effective population synchronization.

Finally, we expect that our results might provide important insights on emergence of burst and spike synchronization of bursting neurons, associated with neural information processes in health and disease, in real brain networks with scale-free property. As is well known, the real brain is considered as one of the most complex systems \cite{Sporns}. Particularly, the mammalian brain (e.g., cat and macaque) has been revealed to have a modular structure composed of sparsely linked clusters with spatial localization \cite{SF11,MN1,MN2,MN3}. The connection structure in each module of the real brain reveals complex topology which is neither regular nor random \cite{Sporns,Buz2,CN1,CN2,CN3,CN4,CN5,CN6,CN7} (e.g., scale-freeness and small-worldness). Hence, real brain networks are far more complex than minimal models such as scale-free and small-world networks. In the clustered network composed of scale-free sub-networks, we expect that ``modular'' burst and spike synchronization (where intra-dynamics of sub-populations make some mismatching) may also emerge, in addition to the global burst and spike synchronization (where the population behavior is globally identical). However, explicit study on the effect of modular structure on burst and spike synchronization in clustered SFNs is beyond our present subject, and it is left as a future work.

\begin{acknowledgments}
This research was supported by Basic Science Research Program through the National Research Foundation of Korea (NRF) funded by the Ministry of Education (Grant No. 2013057789).
\end{acknowledgments}

\newpage
\begin{table}
\caption{Parameter values used in our computations.}
\label{tab:Parm}
\begin{ruledtabular}
\begin{tabular}{lllllll}
(1) & \multicolumn{5}{l}{Single HR Bursting Neurons \cite{Longtin}} \\
& $a = 1$ & $b = 3$ & $c = 1$ & $d = 5$ & $r = 0.001$\\
& $s = 4$ & $x_o = −1.6$ \\
\hline
(2) & \multicolumn{5}{l}{External Stimulus to HR Bursting Neurons} \\
& $I_{DC} = 1.4$ \\
\hline
(3) & \multicolumn{5}{l}{Inhibitory GABAergic Synapse \cite{Sparse}} \\
& $\tau_l=1$ & $\tau_r=0.5$ & $\tau_d=5$ & $X_{syn}=-2$ \\
\hline
(4) & \multicolumn{5}{l}{Pure $\alpha$-process ($\alpha=1$) with Symmetric Attachment} \\
& \multicolumn{2}{l}{$l_{\alpha}^{(in)} = l_{\alpha}^{(out)} \equiv l_{\alpha} (=20)$} & $J$: Varying & $D$: Varying \\
\hline
(5) & \multicolumn{4}{l}{Effect of Scale-free Connectivity (3 Cases)} \\
& $J=4$ & $D=0.06$ \\
& \multicolumn{5}{l}{Pure $\alpha$-process: $l_{\alpha}$: Varying (1st case of symmetric attachment)} \\
& & \multicolumn{4}{l}{$\Delta l_{\alpha}$: Varying (2nd case of asymmetric attachment)} \\
& \multicolumn{5}{l}{Combined $\alpha$- and $\beta$-processes (with symmetric attachment):} \\
& & $l_\alpha=20$ & $l_\beta=5$ & \multicolumn{2}{l}{$\beta$: Varying (3rd case)} \\
\end{tabular}
\end{ruledtabular}
\end{table}

\newpage

\newpage

\newpage

\newpage

\newpage

\newpage

\newpage

\newpage

\newpage

\newpage

\newpage

\newpage

\newpage

\newpage


\begin{thebibliography}{}
\bibitem{Izhi1} E. M. Izhikevich, Int. J. Bif.  Chaos {\bf 10}, 1171 (2000).
\bibitem{Burst1} {\it Bursting: The Genesis of Rhythm in the Nervous System,} edited by S. Coombes and P. C. Bressloff (World Scientific, Singapore, 2005).
\bibitem{Burst2} E. M. Izhikevich, Scholarpedia {\bf {1(3)}}, 1300 (2006).
\bibitem{Rinzel1} J. Rinzel, in {\em Ordinary and Partial Differential Equations}, edited by B. D. Sleeman and R. J. Jarvis, Lecture Notes in Mathematics Vol. 1151 (Springer, Berlin, 1985), pp.~304-316.
\bibitem{Rinzel2} J. Rinzel, in {\em Mathematical Topics in Population Biology, Morphogenesis, and Neurosciences}, edited by E. Teramoto and M. Yamaguti, Lecture Notes in Biomathematics Vol. 71 (Springer, Berlin, 1987), pp.~267-281.
\bibitem{Burst3} E. M. Izhikevich, {\it Dynamical Systems in Neuroscience} (MIT Press, Cambridge, 2007).
\bibitem{Izhi2} E. M. Izhikevich, IEEE Trans. Neural Networks {\bf 15}, 1063 (2004).
\bibitem{Burst4} R. Krahe and F. Gabbian, Nature Rev. Neurosci. {\bf 5}, 13 (2004).
\bibitem{Burst5} J. Lisman, Trends Neurosci. {\bf 20}, 38 (1997).
\bibitem{Burst6} E. N. Izhikevich, N. S. Desai, E. C. Walcott, and F. C. Hoppensteadt, Trends Neurosci. {\bf 26}, 161 (2003).
\bibitem{CT1} B. W. Connors and M. J. Gutnick, Trends Neurosci. {\bf 13}, 99 (1990).
\bibitem{CT2} C. M. Gray and D. A. McCormick, Science {\bf 274}, 109 (1996).
\bibitem{TRN1} R. L. Llin\'{a}s and H. Jahnsen, Nature {\bf 297}, 406 (1982).
\bibitem{TRN2} D. A. McCormick and J. R. Huguenard, J. Neurophysiol. {\bf 8}, 1384 (1992).
\bibitem{TR} S. H. Lee, G. Govindaiah, and C. L. Cox, J. Physiol. {\bf 582}, 195 (2007).
\bibitem{HP} H. Su, G. Alroy, E. D. Kirson, and Y. Yaari, J. Neurosci. {\bf 21}, 4173 (2001).
\bibitem{PC} M. D. Womack and K. Khodakhah, J. Neurosci. {\bf 22}, 10603 (2002).
\bibitem{PBC1} T. R. Chay and J. Keizer, Biophys. J. {\bf 42}, 181 (1983).
\bibitem{PBC2} T. A. Kinard, G. de Vries, and A. Sherman, Biophys. J. {\bf 76}, 1423 (1999).
\bibitem{PBC3} M. Pernarowski, R. M. Miura, and J. Kevorkian, SIAM J. Appl. Math. {\bf 52}, 1627 (1992).
\bibitem{BC1} C. A. Del Negro, C.-F. Hsiao, S. H. Chandler, and A. Garfinkel, Biophys. J. {\bf 75}, 174 (1998).
\bibitem{BC2} R. J. Butera, J. Rinzel, and J. C. Smith, J. Neurophysiol. {\bf 82}, 382 (1999).
\bibitem{Burstsync1} J. E. Rubin, Scholarpedia {\bf {2(10)}}, 1666 (2007).
\bibitem{Burstsync2} I. Omelchenko, M. Rosenblum, and A. Pikovsky, Eur. Phys. J. {\bf 191}, 3 (2010).
\bibitem{Spindle1} M. Steriade, D. A. McCormick, and T. J. Sejnowski, Science {\bf 262}, 679 (1993).
\bibitem{Spindle2} M. Bazhenov and I. Timofevv, Scholarpedia {\bf {1(6)}}, 1319 (2006).
\bibitem{Spindle3} S. Gais, W. Plihal, U. Wagner, and J. Born, Nat. Neurosci. {\bf 3}, 1335 (2000).
\bibitem{Spindle4} T. J. Sejnowski and A. Destexhe, Brain Res. {\bf 886}, 208 (2000).
\bibitem{PD1} M. Bevan, P. Magill, D. Terman, J. Bolam, and C. Wilson, Trends Neurosci. {\bf 25}, 525 (2002).
\bibitem{PD2} P. Brown, Cur. Opin. Neurobiol. {\bf 17}, 656. (2007).
\bibitem{PD3} C. Park, R. M. Worth, and L. L. Rubchinsky, J. Neurophysiol. {\bf 103}, 2703 (2010).
\bibitem{PD4} C. Hammond, H. Bergman, and P. Brown, Trends Neurosci. {\bf 30}, 357 (2007).
\bibitem{PD5} P. J. Uhlhaas and W. Singer, Neuron {\bf 52}, 155 (2006).
\bibitem{Epilepsy} R. Fisher, W. van Emde Boas, W. Blume, C. Elger, P. Genton, P. Lee, and J. Engel, Epilepsia {\bf 46}, 470 (2005).
\bibitem{Sporns} O. Sporns, {\it Networks of the Brain} (MIT Press, Cambridge, 2011).
\bibitem{Buz2} G. Buzs$\acute{\rm a}$ki, C. Geisler, D. A. Henze, and X.-J. Wang, Trends Neurosci. {\bf 27}, 186 (2004).
\bibitem{CN1} D. B. Chklovskii, B. W. Mel, and K. Svoboda, Nature {\bf 431}, 782 (2004).
\bibitem{CN2} S. Song, P. J. Sj$\ddot{\rm o}$str$\ddot{\rm o}$m, M. Reigl, S. Nelson, and D. B. Chklovskii, PLoS Biol. {\bf 3}, e68 (2005).
\bibitem{CN3} O. Sporns and C. J. Honey, Proc. Natl. Acad. Sci. USA {\bf 103}, 19219 (2006).
\bibitem{CN4} P. Larimer and B. W. Strowbridge, J. Neurosci. {\bf 28}, 12212 (2008).
\bibitem{CN5} E. Bullmore and O. Sporns, Nat. Rev. Neurosci. {\bf 10}, 186 (2009).
\bibitem{CN6} O. Sporns, G. Tononi, and G. M. Edelman, Cereb. Cortex {\bf 10}, 127 (2000).
\bibitem{CN7} D. S. Bassett and E. Bullmore, The Neuroscientist {\bf 12}, 512 (2006).
\bibitem{SF1} P. Bonifazi, M. Goldin, M. A. Picardo, I. Jorquera, A. Cattani, G. Bianconi, A. Represa, Y. Ben-Ari, and R. Cossart, Science {\bf 326}, 1419 (2009).
\bibitem{SF2} C. Wiedemann, Nature Rev. Neurosci. {\bf 11}, 74 (2010).
\bibitem{SF3} X. Li, G. Ouyang, A. Usami, Y. Ikegaya, and A. Sik, Biophys. J. {\bf 98}, 1733 (2010)
\bibitem{SF4} R. J. Morgan and I. Soltesz, Proc. Natl. Acad. Sci. USA {\bf 105}, 6179 (2008).
\bibitem{SF5} V. M. Egu{\'{i}}luz, D. R. Chialvo, G. A. Cecchi, M. Baliki, and A. V. Apkarian, Phys. Rev. Lett. {\bf 94}, 018102 (2005).
\bibitem{SF6} M. P. Young, Philos. Trans. R. Soc. {\bf 252}, 13 (1993).
\bibitem{SF7} M. P. Young, J. W. Scannell, G. A. Burns, and C. Blakemore, Rev. Neurosci. {\bf 5}, 227 (1994).
\bibitem{SF8} J. W. Scannell, C. Blakemore, and M. P. Young, J. Neurosci. {\bf 15}, 1463 (1995).
\bibitem{SF9} D. J. Felleman and D. C. Van Essen, Cereb. Cortex {\bf 1}, 1 (1991).
\bibitem{SF10} J. W. Scannell, G. A. P. C. Burns, C. C. Hilgetag, M. A. O'Neill, and M. P. Young, Cereb. Cortex {\bf 9}, 277 (1999).
\bibitem{SF11} O. Sporns, D. R. Chialvo, M. Kaiser, and C. C. Hilgetag, Trends Cogn. Sci. {\bf 8}, 418 (2004).
\bibitem{SF12} M. Kaiser, R. Martin, P. Andras, and M. P. Young, Eur. J. Neurosci. {\bf 25}, 3185 (2007).
\bibitem{BA1} A.-L. Barab\'{a}si and R. Albert, Science {\bf 286}, 509 (1999).
\bibitem{BA2} R. Albert and A.-L. Barab\'{a}si, Rev. Mod. Phys. {\bf 74}, 47 (2002).
\bibitem{SF13} C. A. S. Batista, A. M. Batista, J. A. C. de Pontes, R. L. Viana, and S. R. Lopes, Phys. Rev. E {\bf 76}, 016218 (2007).
\bibitem{SF14} Q. Wang, M. Perc, Z. Duan, and G. Chen, Phys. Rev. E {\bf 80}, 026206 (2009).
\bibitem{SF15} C. A. S. Batisa, A. M. Batisa, J. C. A. de Pontes, S. R. Lopes, and R. L. Viana, Chaos, Solitons, and Fractals {\bf 41}, 2220 (2009).
\bibitem{SF16} C. A. S. Batisa, S. R. Lopes, R. L. Viana, and A. M. Batisa, Neural Networks {\bf 23}, 114 (2010).
\bibitem{SF17} Q. Wang, G. Chen, and M. Perc, PLoS ONE {\bf 6}, e15851 (2011).
\bibitem{SF18} F. A. S. Ferrari, R. L. Viana, S. R. Lopes, and R. Stoop, Neural Networks {\bf 66}, 107 (2015).
\bibitem{Bollobas} B. Bollob{\'{a}}s, C. Borgs, J. Chayes, and O. Riordan, ``Directed Scale-free Graph,'' Proc. 14th ACM-SIAM Symposium on Discrete Algorithms, pp. 132-139 (2003).
\bibitem{beta2} R. Albert and A.-L. Barab\'{a}si, Phys. Rev. Lett. {\bf 85}, 5234 (2000).
\bibitem{beta3} S. N. Dorogovtsev and J. F. F. Mendes, Europhys. Lett. {\bf 52}, 33 (2000).
\bibitem{beta4} A.-L. Barab\'{a}si, H. Jeong, E. Ravasz, Z. N\'{e}da, A, Schubert, and T. Vicsek, Physica A {\bf 311}, 590 (2002).
\bibitem{Kim2} S.-Y. Kim and W. Lim, Phys. Rev. E {\bf 92}, 022717 (2015).
\bibitem{plastic} A. Pascual-Leone, C. Freitas, L. Oberman, J. C. Horvath, M. Halko, M. Eldaief, S. Bashir, M. Vernet, M. Shafi, B. Westover, A. M. Vahabzadeh-Hagh, and A. Rotenberg, Brain Topography {\bf 24}, 302 (2011).
\bibitem{HR1} J. L. Hindmarsh and R. M. Rose, Nature {\bf 296}, 162 (1982).
\bibitem{HR2} J. L. Hindmarsh and R. M. Rose, Proc. R. Soc. London, Ser. B {\bf 221}, 87 (1984).
\bibitem{HR3} R. M. Rose and J. L. Hindmarsh, Proc. R. Soc. London, Ser. B {\bf 225}, 161 (1985).
\bibitem{IPFR1} X. J. Wang, Physiol. Rev. {\bf 90}, 1195 (2010).
\bibitem{IPFR2} N. Brunel and V. Hakim, Chaos {\bf 18}, 015113 (2008).
\bibitem{Kim1} S.-Y. Kim and W. Lim, Physica A {\bf 438}, 544 (2015).
\bibitem{Longtin} A. Longtin, Phys. Rev. E {\bf 55}, 868 (1997).
\bibitem{Sparse} N. Brunel and X.-J. Wang, J. Neurophysiol. {\bf 90}, 415 (2003).
\bibitem{SDE} M. San Miguel and R. Toral, in {\it Instabilities and Nonequilibrium Structures VI}, edited by J. Martinez, R. Tiemann,
and E. Tirapegui (Kluwer Academic Publisher, Dordrecht, 2000), pp.~35-130.
\bibitem{EP1} R. J. Staba, C. L. Wilson, A. Bragin, I. Fried, and J. Engel Jr., J. Neurosci. {\bf 22}, 5694 (2002).
\bibitem{EP2} C. Alvarado-Rojas, K. Lehongre, J. Bagdasaryan, A. Bragin, R. Staba, J. Engel Jr., V. Navarro, and M. Le Van Quyen, Front. Comput. Neurosci. {\bf 7}, 00140 (2013).
\bibitem{EP3} W. Truccolo, O. J. Ahmed, M. T. Harrison, E. N. Eskandar, G. R. Cosgrove, J. R. Madsen, A. S. Blum, N. S. Potter, L. R. Hochberg, and S. S. Cash, J. Neurosci. {\bf 34}, 9927(2014).
\bibitem{EP4} M. R. Bower and P. S. Buckmaster, J. Neurophysio. {\bf 99}, 2431 (2008).
\bibitem{Kernel} H. Shimazaki and S. Shinomoto, J. Comput. Neurosci. {\bf 29}, 171 (2010).
\bibitem{Kim3} S.-Y. Kim and W. Lim, Physica A {\bf 421}, 109 (2015).
\bibitem{BTC1} L. C. Freeman, Sociometry {\bf 40}, 35 (1977).
\bibitem{BTC2} L. C. Freeman, Soc. Net. {\bf 1}, 215 (1978).
\bibitem{Kim4} W. Lim and S.-Y. Kim, J. Korean Phys. Soc. {\bf 57}, 1290 (2010).
\bibitem{BTC3} T. Nishikawa, A. E. Motter, Y.-C. Lai, and F. C. Hoppensteadt, Phys. Rev. Lett. {\bf 91}, 014101 (2003).
\bibitem{MN1} C. C. Hilgetag, G. A. P. C. Burns, M. A. O'Neill, J. W. Scannell, and M. P. Young, Phil. Trans. R. Soc. Lond. B {\bf 355}, 91 (2000).
\bibitem{MN2} C. C. Hilgetag and M. Kaiser, Neuroinformatics {\bf 2}, 353 (2004).
\bibitem{MN3} S.-J. Wang, C. C. Hilgetag, and C. S. Zhou, Front. Comput. Neurosci. {\bf 5}, 30 (2011).
\end{thebibliography}
\end{document}